\documentclass[final]{IEEEtran}
\usepackage{amsthm,amssymb,graphicx,multirow,amsmath,color,amsfonts}%,ulem}
\usepackage[update,prepend]{epstopdf}
\usepackage[noadjust]{cite}
\usepackage[latin1]{inputenc}
\usepackage{tikz}
\usepackage{bm}
\usepackage{bbm} % for \mathbbm{1}
\usepackage[nolist]{acronym}
\usepackage{pdfpages}
\usepackage{multirow}
\usepackage{subfig}

\usepackage{comment}
\usepackage{fancyhdr}
\def\nb0{{\mathbf{0}}}
\def\nb1{{\mathbf{1}}}

\newtheorem{prop}{Proposition}

\newtheorem{remark}{Remark}

\allowdisplaybreaks % Allows breaking of eqnarray over multiple pages (avoids unnecessary blanks in the document before eqnarray)
\usepackage{algorithm}
\usepackage{algorithmic}
\usepackage{setspace}	% Remove in double column version. Also search for \setstretch in the body of the paper and comment these commands for double column

\begin{acronym}
	
	\acro{5G-NR}{5G New Radio}
	\acro{3GPP}{3rd Generation Partnership Project}
	\acro{AC}{address coding}
	\acro{ACF}{autocorrelation function}
	\acro{ACR}{autocorrelation receiver}
	\acro{ADC}{analog-to-digital conversion}
	\acrodef{aic}[AIC]{Analog-to-Information Converter}     
	\acro{AIC}[AIC]{Akaike information criterion}
	\acro{aric}[ARIC]{asymmetric restricted isometry constant}
	\acro{arip}[ARIP]{asymmetric restricted isometry property}
	\acro{AoD}{angle of departure}
	\acro{ARQ}{automatic repeat request}
	\acro{AUB}{asymptotic union bound}
	\acrodef{awgn}[AWGN]{Additive White Gaussian Noise}     
	\acro{AWGN}{additive white Gaussian noise}

	\acro{APSK}[PSK]{asymmetric PSK} 
	
	\acro{waric}[AWRICs]{asymmetric weak restricted isometry constants}
	\acro{warip}[AWRIP]{asymmetric weak restricted isometry property}
	\acro{BCH}{Bose, Chaudhuri, and Hocquenghem}        
	\acro{BCHC}[BCHSC]{BCH based source coding}
	\acro{BEP}{bit error probability}
	\acro{BFC}{block fading channel}
	\acro{BG}[BG]{Bernoulli-Gaussian}
	\acro{BGG}{Bernoulli-Generalized Gaussian}
	\acro{BPAM}{binary pulse amplitude modulation}
	\acro{BPDN}{Basis Pursuit Denoising}
	\acro{BPPM}{binary pulse position modulation}
	\acro{BPSK}{binary phase shift keying}
	\acro{BPZF}{bandpass zonal filter}
	\acro{BSC}{binary symmetric channels}              
	\acro{BU}[BU]{Bernoulli-uniform}
	\acro{BER}{bit error rate}
	\acro{BS}{base station}
	\acro{MS}{mobile station}
	\acro{BM}{beamspace modulation}

	\acro{CP}{Cyclic Prefix}
	\acrodef{cdf}[CDF]{cumulative distribution function}   
	\acro{CDF}{cumulative distribution function}
	\acrodef{c.d.f.}[CDF]{cumulative distribution function}
	\acro{CCDF}{complementary cumulative distribution function}
	\acrodef{ccdf}[CCDF]{complementary CDF}               
	\acrodef{c.c.d.f.}[CCDF]{complementary cumulative distribution function}
	\acro{CD}{cooperative diversity}
	
	\acro{CDMA}{Code Division Multiple Access}
	\acro{ch.f.}{characteristic function}
	\acro{CIR}{channel impulse response}
	\acro{cosamp}[CoSaMP]{compressive sampling matching pursuit}
	\acro{CR}{cognitive radio}
    \acro{CRB}{Cram{\'e}r-Rao bound}
\acro{BCRB}{Bayesian Cram{\'e}r-Rao bound}
	\acro{cs}[CS]{compressed sensing}                   
	\acrodef{cscapital}[CS]{Compressed sensing}
	\acrodef{CS}[CS]{compressed sensing}
	\acro{CSI}{channel state information}
	
	\acro{CCSDS}{consultative committee for space data systems}
	\acro{CC}{convolutional coding}

	\acro{DAA}{detect and avoid}
	\acro{DAB}{digital audio broadcasting}
	\acro{DCT}{discrete cosine transform}
	\acro{DFT}{discrete Fourier transform}
	\acro{DR}{distortion-rate}
	\acro{DS}{direct sequence}
	\acro{DS-SS}{direct-sequence spread-spectrum}
	\acro{DTR}{differential transmitted-reference}
	\acro{DVB-H}{digital video broadcasting\,--\,handheld}
	\acro{DVB-T}{digital video broadcasting\,--\,terrestrial}
	\acro{DL}{downlink}
	\acro{DSSS}{Direct Sequence Spread Spectrum}
	\acro{DFT-s-OFDM}{Discrete Fourier Transform-spread-Orthogonal Frequency Division Multiplexing}
	\acro{DAS}{distributed antenna system}
	\acro{DNA}{Deoxyribonucleic Acid}
	\acro{DoA}{direction of arrival}

	\acro{EC}{European Commission}
	\acro{EED}[EED]{exact eigenvalues distribution}
	\acro{EIRP}{Equivalent Isotropically Radiated Power}
	\acro{ELP}{equivalent low-pass}
	\acro{eMBB}{Enhanced Mobile Broadband}
	\acro{EMF}{electric and magnetic fields}
	\acro{EU}{European union}
	\acro{EMR}{effective multiplexing region}

	\acro{FC}[FC]{fusion center}
	\acro{FCC}{Federal Communications Commission}
	\acro{FEC}{forward error correction}
	\acro{FFT}{fast Fourier transform}
	\acro{FH}{frequency-hopping}
	\acro{FH-SS}{frequency-hopping spread-spectrum}
	\acrodef{FS}{Frame synchronization}
	\acro{FSsmall}[FS]{frame synchronization}  
	\acro{FDMA}{Frequency Division Multiple Access}    
	\acro{FT}{Fourier transform}
	\acro{FIM}{Fisher information matrix} 
	
	\acro{GA}{Gaussian approximation}
	\acro{GF}{Galois field }
	\acro{GG}{Generalized-Gaussian}
	\acro{GIC}[GIC]{generalized information criterion}
	\acro{GLRT}{generalized likelihood ratio test}
	\acro{GPS}{Global Positioning System}
	\acro{GMSK}{Gaussian minimum shift keying}
	\acro{GSMA}{Global System for Mobile communications Association}
	
	\acro{HAP}{high altitude platform}

	\acro{IDR}{information distortion-rate}
	\acro{IFFT}{inverse fast Fourier transform}
	\acro{iht}[IHT]{iterative hard thresholding}
	\acro{i.i.d.}{independent, identically distributed}
	\acro{IoT}{Internet of Things}                      
	\acro{IR}{impulse radio}
	\acro{lric}[LRIC]{lower restricted isometry constant}
	\acro{lrict}[LRICt]{lower restricted isometry constant threshold}
	\acro{ISI}{intersymbol interference}
	\acro{ITU}{International Telecommunication Union}
	\acro{ICNIRP}{International Commission on Non-Ionizing Radiation Protection}
	\acro{IEEE}{Institute of Electrical and Electronics Engineers}
	\acro{ICES}{IEEE international committee on electromagnetic safety}
	\acro{IEC}{international Electrotechnical Commission}
	\acro{IARC}{International Agency on Research on Cancer}
	\acro{IS-95}{Interim Standard 95}
	\acro{ISAC}{integrated sensing and communication}
	\acro{I/Q}{in-phase and quadrature}
	
	\acro{KPI}{Key Performance Indicator}
	
	\acro{LEO}{low earth orbit}
	\acro{LF}{likelihood function}
	\acro{LLF}{log-likelihood function}
	\acro{LLR}{log-likelihood ratio}
	\acro{LLRT}{log-likelihood ratio test}
	\acro{LoS}{line-of-sight}
	\acro{NLoS}{non-line-of-sight}
	\acro{LRT}{likelihood ratio test}
	\acro{wlric}[LWRIC]{lower weak restricted isometry constant}
	\acro{wlrict}[LWRICt]{LWRIC threshold}
	\acro{LPWAN}{low power wide area network}
	\acro{LoRaWAN}{Low power long Range Wide Area Network}

	\acro{MB}{multiband}
	\acro{MC}{multicarrier}
	\acro{MDS}{mixed distributed source}
	\acro{MF}{matched filter}
	\acro{m.g.f.}{moment generating function}
	\acro{MI}{mutual information}
	\acro{MIMO}{multiple-input multiple-output}
	\acro{MISO}{multiple-input single-output}
	\acrodef{maxs}[MJSO]{maximum joint support cardinality}                       
	\acro{ML}[ML]{maximum likelihood}
	\acro{MMSE}{minimum mean-square error}
        \acro{MSE}{mean-square error}
	\acro{MMV}{multiple measurement vectors}
	\acrodef{MOS}{model order selection}
	\acro{M-PSK}[${M}$-PSK]{$M$-ary phase shift keying}                       
	\acro{M-APSK}[${M}$-PSK]{$M$-ary asymmetric PSK} 
	
	\acro{M-QAM}[$M$-QAM]{$M$-ary quadrature amplitude modulation}
	\acro{MRC}{maximal ratio combiner}                  
	\acro{maxs}[MSO]{maximum sparsity order}                                      
	\acro{M2M}{machine to machine}                                                
	\acro{MUI}{multi-user interference}
	\acro{mMTC}{massive Machine Type Communications}      
	\acro{mm-Wave}[mm-Wave]{millimeter-wave}
	\acro{MP}{mobile phone}
	\acro{MPE}{maximum permissible exposure}
	\acro{MAC}{media access control}
	\acro{NB}{narrowband}
	\acro{NBI}{narrowband interference}
	\acro{NLA}{nonlinear sparse approximation}
	\acro{NTIA}{National Telecommunications and Information Administration}
	\acro{NTP}{National Toxicology Program}
	\acro{NF}{near-field}
	\acro{FF}{far-field}

	\acro{OC}{optimum combining}                             
	\acro{OC}{optimum combining}
	\acro{ODE}{operational distortion-energy}
	\acro{ODR}{operational distortion-rate}
	\acro{OFDM}{orthogonal frequency-division multiplexing}
	\acro{omp}[OMP]{orthogonal matching pursuit}
	\acro{OSMP}[OSMP]{orthogonal subspace matching pursuit}
	\acro{OQAM}{offset quadrature amplitude modulation}
	\acro{OQPSK}{offset QPSK}
	\acro{OFDMA}{Orthogonal Frequency-division Multiple Access}
	
	\acro{OQPSK/PM}{OQPSK with phase modulation}

	\acro{PAM}{pulse amplitude modulation}
	\acro{PAR}{peak-to-average ratio}
	\acrodef{pdf}[PDF]{probability density function}                      
	\acro{PDF}{probability density function}
	\acrodef{p.d.f.}[PDF]{probability distribution function}
	\acro{PDP}{power dispersion profile}
	\acro{PMF}{probability mass function}                             
	\acrodef{p.m.f.}[PMF]{probability mass function}
	\acro{PN}{pseudo-noise}
	\acro{PPM}{pulse position modulation}
	\acro{PRake}{Partial Rake}
	\acro{PSD}{power spectral density}
	\acro{PSEP}{pairwise synchronization error probability}
	\acro{PSK}{phase shift keying}
	\acro{PD}{power density}
	\acro{8-PSK}[$8$-PSK]{$8$-phase shift keying}
	\acro{PHP}{Poisson hole process}
	\acro{PPP}{Poisson point process}
	\acrodefplural{PPP}{Poisson point processes}
	\acro{FSK}{frequency shift keying}

	\acro{QAM}{Quadrature Amplitude Modulation}
	\acro{QPSK}{quadrature phase shift keying}
	\acro{OQPSK/PM}{OQPSK with phase modulator }
	
	\acro{RD}[RD]{raw data}
	\acro{RDL}{"random data limit"}
	\acro{ric}[RIC]{restricted isometry constant}
	\acro{rict}[RICt]{restricted isometry constant threshold}
	\acro{rip}[RIP]{restricted isometry property}
	\acro{ROC}{receiver operating characteristic}
	\acro{rq}[RQ]{Raleigh quotient}
	\acro{RS}[RS]{Reed-Solomon}
	\acro{RSC}[RSSC]{RS based source coding}
	\acro{r.v.}{random variable}                               
	\acro{R.V.}{random vector}
	\acro{RMS}{root mean square}
	\acro{RFR}{radiofrequency radiation}
	\acro{RIS}{reconfigurable intelligent surface}
	\acro{SA}[SA-Music]{subspace-augmented MUSIC with OSMP}
	\acro{SCBSES}[SCBSES]{Source Compression Based Syndrome Encoding Scheme}
	\acro{SCM}{sample covariance matrix}
	\acro{SEP}{symbol error probability}
	\acro{SER}{symbol error rate}
	\acro{BER}{bit error rate}
	\acro{SED}{symbol Euclidean distance}
	\acro{SG}[SG]{sparse-land Gaussian model}
	\acro{SIMO}{single-input multiple-output}
	\acro{SINR}{signal-to-interference plus noise ratio}
	\acro{SIR}{signal-to-interference ratio}
	\acro{SISO}{single-input single-output}
	\acro{SMV}{single measurement vector}
	\acro{SNR}[\textrm{SNR}]{signal-to-noise ratio} 
	\acro{sp}[SP]{subspace pursuit}
	\acro{SS}{spread spectrum}
	\acro{SW}{sync word}
	\acro{SAR}{specific absorption rate}
	\acro{SSB}{synchronization signal block}
	\acro{SDG}{Sustainable Development Goal}
	\acro{SBF}{spatial bandpass filtering}
	\acro{SVD}{singular value decomposition}

	\acro{TH}{time-hopping}
	\acro{ToA}{time-of-arrival}
	\acro{TR}{transmitted-reference}
	\acro{TW}{Tracy-Widom}
	\acro{TWDT}{TW Distribution Tail}
	\acro{TCM}{trellis coded modulation}
	\acro{TDD}{time-division duplexing}
	\acro{TDMA}{Time Division Multiple Access}
	
	\acro{UAV}{unmanned aerial vehicle}
	\acro{uric}[URIC]{upper restricted isometry constant}
	\acro{urict}[URICt]{upper restricted isometry constant threshold}
	\acro{UWB}{ultrawide band}
	\acro{UWBcap}[UWB]{Ultrawide band}   
	\acro{URLLC}{Ultra Reliable Low Latency Communications}
	
	\acro{wuric}[UWRIC]{upper weak restricted isometry constant}
	\acro{wurict}[UWRICt]{UWRIC threshold}                
	\acro{UE}{user equipment}
	\acro{UL}{uplink}
	
	\acro{ULA}{uniform linear array}
        \acro{URA}{uniform rectangular array}
	\acro{UPA}{uniform planar array}
	\acro{UCA}{uniform circular array}

	\acro{WiM}[WiM]{weigh-in-motion}
	\acro{WLAN}{wireless local area network}
	
	\acro{wm}[WM]{Wishart matrix}                               
	\acroplural{wm}[WM]{Wishart matrices}
	\acro{WMAN}{wireless metropolitan area network}
	\acro{WPAN}{wireless personal area network}
	\acro{wric}[WRIC]{weak restricted isometry constant}
	\acro{wrict}[WRICt]{weak restricted isometry constant thresholds}
	\acro{wrip}[WRIP]{weak restricted isometry property}
	\acro{WSN}{wireless sensor network}                        
	\acro{WSS}{wide-sense stationary}
	\acro{WHO}{World Health Organization}
	\acro{WP}{work package}
	
	\acro{sss}[SpaSoSEnc]{sparse source syndrome encoding}
	\acro{SO}{strategic objective}
	
	\acro{VLC}{visible light communication}
	\acro{RF}{radio frequency}
	\acro{FSO}{free space optics}
	\acro{IoST}{Internet of space things}
	
	\acro{GSM}{Global System for Mobile Communications}
	\acro{2G}{second-generation cellular network}
	\acro{3G}{third-generation cellular network}
	\acro{4G}{fourth-generation cellular network}
	\acro{5G}{5th-generation cellular network}	
	\acro{gNB}{next generation node B base station}
	\acro{NR}{New Radio}
	\acro{UN}{United Nations}
	\acro{UMTS}{Universal Mobile Telecommunications Service}
	\acro{LTE}{Long Term Evolution}
	
	\acro{QO}{quasi-orthogonal}
	\acro{QoS}{Quality of Service}
	\acro{Tx}{transmitter}
	\acro{Rx}{receiver}
	\acro{DoF}{degrees-of-freedom}
	\acro{EDoF}{effective degrees-of-freedom}
	\acro{QOM}{quasi-orthogonal mode}
	\acro{BBS}{best beam selection}
        \acro{TD}{time delayer}
        \acro{PS}{phase shifter}
        \acro{ELAA}{extremely large antenna array}
        \acro{SU}{single-user}
        \acro{MU}{multi-user}

\acro{CTRV}{constant turn rate and velocity}
\acro{KF}{Kalman filter}
\acro{EKF}{extended Kalman filter}
\acro{RMSE}{root mean-square error}    
\acro{NMSE}{normalized mean-square error}
\acro{w.r.t}{with respect to}
\acro{THz}{terahertz}
\acro{mmWave}{millimeter wave}
\end{acronym}

\begin{document}
%\pagenumbering{gobble}
\graphicspath{{./Figure/}}
\title{
Near-Field Position and Orientation Tracking With Hybrid ELAA Architecture
} 
\author{Lin Chen, \IEEEmembership{Graduate Student Member, IEEE}, Xiaojun Yuan, {\em Fellow, IEEE}, \\and Ying-Jun Angela Zhang, {\em Fellow, IEEE}
		\thanks{  
			Lin Chen and Ying-Jun Angela Zhang are with the Department of Information Engineering, The Chinese University of Hong Kong (CUHK), Hong Kong (e-mail: \{cl022,~yjzhang\}@ie.cuhk.edu.hk).
			
			Xiaojun Yuan is with the National Key Laboratory of Wireless Communications, University of Electronic Science and Technology of China (UESTC), Chengdu 611731, China (e-mail: xjyuan@uestc.edu.cn).
		} % remove the date for conference drafts
		% \vspace{-4mm}
\vspace{-4mm}}

\maketitle
%%%%%%%%%%%%%%%%%%%%%%%%%%%%%%%%%%%%%%%%%%%%%%%%%%%%%%%%%%%%%%%%%%%%%
% \maketitle
% \thispagestyle{fancy}  
% \fancyhead{}   
% \lhead{\today}  
% \cfoot{\quad}     
% \renewcommand{\headrulewidth}{0pt} 
% \renewcommand{\footrulewidth}{0pt}
%%%%%%%%%%%%%%%%%%%%%%%%%%%%%%%%%%%%%%%%%%%%%%%%%%%%%%%%%%%%%%%%%%%%% 

\begin{abstract} 
This paper investigates \ac{NF} \textcolor{black}{position and orientation (i.e., pose)} tracking of a multi-antenna \ac{MS} using an \ac{ELAA}-equipped \ac{BS} with a limited number of \ac{RF} chains. 
Under this hybrid array architecture, the received uplink pilot signal at the BS is first combined by analog phase shifters, producing a low-dimensional observation before digital processing.
Such analog compression provides only partial access to the ELAA measurement,  
making it essential to design an analog combiner that can preserve pose-relevant signal components despite channel uncertainty and unit-modulus hardware constraints. 
To address this, we propose a predictive analog combining-assisted extended Kalman filter (PAC-EKF) framework, where the analog combiner leverages the temporal correlation in the MS pose variation to capture the most informative signal components in a predictive manner.
We then analyze fundamental performance limits via the Bayesian Cram{\'e}r-Rao bound and the Fisher information matrix, explicitly quantifying how the analog combiner, array size, and signal-to-noise ratio influence the pose-relevant information contained in the uplink observation. 
% Bayesian \ac{CRB} and \ac{FIM}, \ac{SNR}
%%%%%%%%
Building on these insights,  we develop two methods for designing a low-complexity analog combiner. 
% Numerical results confirm the potential of \ac{NF} position and orientation tracking and show that the proposed predictive analog combining approach significantly improves tracking accuracy, even with fewer RF chains and lower transmit power.
Numerical results show that the proposed predictive analog combining approach significantly improves tracking accuracy, even with fewer RF chains and lower transmit power, \textcolor{black}{and remains effective under practical phase quantization.}
% We also discuss the potential impact of prediction error on the predictive analog combiner design and the tracking performance under the PAC-EKF framework.
\end{abstract}


\acresetall

\vspace{-3mm}

\section{Introduction}

To meet the demand for ultra-high data rates, future wireless systems are expected to exploit both high frequencies and \acp{ELAA}~\cite{6G_freq_antenna,6G_freq,NFtutorial_Dai}. This paradigm shifts the propagation regime from the conventional \ac{FF}, i.e., Fraunhofer region, into the radiative \ac{NF}, i.e., Fresnel region~\cite{selvan2017fraunhofer}.
\textcolor{black}{In the \ac{FF} regime, the propagation from a radiating source point can be well approximated by a planar-wave model, where the phase differences across the receive array depend only on the angle of arrival.
%%%%%
In contrast, in the \ac{NF} regime, this approximation no longer holds. The propagation from a radiating source point is more accurately described by a spherical-wave model, where the phase variations across the receive array depend jointly on both the angle and distance of the source point~\cite{NFtutorial,mismatch,NFlocalization}.}  
 
The richer spatial structure of the NF signals enables an ELAA-equipped \ac{BS} to estimate not only the {\em angle} but also the {\em distance} of a radiating source point, effectively localizing a {\em single-antenna} user from an uplink \ac{SIMO} channel snapshot (obtained from the received pilot signal)~\cite{mismatch,NFlocalization,yuan2024scalable}.
For a multi-antenna user, \textcolor{black}{its {\em position and orientation}, i.e., pose,} jointly determine the \ac{MIMO} channel response, making the user pose recoverable from the channel snapshot~\cite{yuan-MIMO}.
%%%%%%%%%
For a moving user, e.g., a multi-antenna \ac{MS} with a frequently changing pose, the classic estimation, which re-estimates the pose from scratch each time, is highly inefficient.  
Instead, the BS can perform {\em pose tracking} by leveraging the temporal correlation inherent in the MS's movement and rotation. 
Specifically, the MS pose is modeled as a time-varying state that evolves via a {\em state transition model}, and each new channel snapshot provides a noisy {\em observation/measurement} of the current state~\cite{guerra2021near}. By combining the previous state estimate with the current observation, the BS can achieve high-accuracy pose tracking with low pilot overhead.

Despite the promise of using the ELAA-equipped BS for pose tracking, the practical implementation faces a significant challenge in the array architecture. A {\em fully digital} array, with one dedicated \ac{RF} chain per antenna, is prohibitively costly and power-consuming due to the large number of antennas. 
Practical ELAAs therefore adopt {\em hybrid analog-digital} architectures, where a small number of RF chains are connected to massive antennas through low-cost analog phase shifters~\cite{AltMin,hybridMMSE,el2014spatially}. 
However, this hardware-efficient architecture introduces a critical constraint: The received signal at the BS must pass through the analog phase shifters before \ac{ADC}~\cite{mendez2015channel}, yielding only a {\em compressed observation} of the NF channel snapshot. 

Consequently, the configuration of the analog phase shifters, known as analog beamforming or, more specifically, {\em analog combining}, is essential. 
It determines how the high-dimensional NF signal is projected onto a low-dimensional subspace, directly impacting whether the subtle wavefront features needed for accurate pose tracking are captured or not. 
This gives rise to a fundamental {\em chicken-and-egg problem} in analog combiner design.
On one hand, accurate pose estimation requires an analog combiner that preserves informative signal components in the channel snapshot.
On the other hand, 
identifying which components are informative requires \ac{CSI}, which depends on the MS pose (the unknown that needs to be estimated).
This motivates the study of NF tracking under a hybrid ELAA architecture, leveraging NF propagation characteristics, temporal pose correlation, and a principled analog combiner design to achieve efficient and accurate pose tracking.

 \vspace{-2mm}

\subsection{Related Work}\label{subsec:related}
 
Classical Bayesian filters, such as \ac{KF} and its variants including \ac{EKF}, provide a principled framework for dynamic state tracking~\cite{simon2006optimal,salmi2008detection}. These filters recursively {\em predict} the state evolution via the state transition model and {\em update} the state estimate based on noisy observations~\cite {clock1,clock2,koivisto2021channel}. 
This framework has recently been adapted to leverage NF characteristics for tracking the position of a {\em single-antenna} user with an ELAA-equipped BS~\cite{guerra2021near,LiuISAC2}. 
Beyond position tracking, Ref.~\cite{dai2025attitude} studied the NF pose tracking for a rotary-wing UAV with several antennas.  

However, the above works~\cite{guerra2021near,LiuISAC2,dai2025attitude} assume that the BS employs a {\em fully digital} ELAA architecture with massive RF chains.
The potential of NF pose tracking using a practical {\em hybrid} ELAA architecture with a limited number of RF chains remains largely unexplored.  
Unlike the fully digital architecture that allows direct access to the {\em full}-array observation, under the hybrid architecture, only a {\em partial}-array observation, i.e., a low-dimensional compressed observation, is available for pose estimation due to analog combining~\cite{mendez2015channel}.
This makes the analog combiner design crucial for accurate tracking.

\color{black}
Although the analog combiner design has been studied in \ac{FF} systems primarily for {\em data communication}, such designs are generally unsuitable for \ac{NF} tracking.
This is because their design objectives are fundamentally different, despite sharing the same hardware constraints, e.g., the {\em unit-modulus constraint} imposed by the analog phase shifters and the need for {\em low-complexity real-time} processing. 
Specifically, communication-oriented designs~\cite{AltMin,hybridMMSE,el2014spatially} are typically guided by criteria such as data rate maximization or \ac{MMSE} of symbol detection, under the assumption of a {\em known} and slowly varying channel.  
Tracking, in contrast, deals with a channel that depends on the unknown and dynamically evolving state of the MS. In this case, the analog combiners must handle {\em channel uncertainty}, with the primary goal of preserving as much information about the MS state in the compressed observations.
 
Random analog combiners~\cite{cui2022channel,mixed,mendez2015channel} can be easily implemented without the knowledge of the instantaneous channel or MS state and are widely employed in classic estimation problems, where each measurement is treated as {\em independent}. 
However, when applied to tracking scenarios, such combiners that ignore the {temporal pose correlation} may degrade tracking performance, particularly under {low \ac{SNR}} conditions where the useful signal is easily swamped by noise. 
A key question arises: {How can we leverage the temporal pose correlation to design analog combiners that achieve high tracking accuracy at the BS despite having limited RF chains and low \ac{SNR}?}

\color{black} 
 
\vspace{-2mm}

%%%%%%%%%%%%%
\subsection{Contribution}\label{sec:contri} 
This paper investigates the NF position and orientation tracking of a multi-antenna MS using a practical hybrid ELAA architecture at a BS.
The MS pose evolves according to a state transition model, and the NF-MIMO channel is parameterized by the MS pose. 
At each tracking interval, an uplink pilot transmitted by the MS passes through the NF channel and an analog combiner, producing a compressed observation at the BS.
Using an \ac{EKF}, the BS first predicts the MS pose via the state transition model and designs the analog combiner; the BS then updates the pose estimate from the compressed observation. 
% Within this framework, we analyze the fundamental tracking performance and develop practical analog combiners.
The main contributions are listed below. 
\begin{itemize}

    \item  We propose a predictive analog combining-assisted \ac{EKF} (PAC-EKF) framework for \ac{NF} pose tracking, where the analog combiner is configured during the EKF prediction stage using the predicted MS pose.
    This enables the combiner to anticipate the informative signal components before the pilot arrives. 
    By explicitly accounting for the channel's unknown but time-correlated nature, predictive analog combining enriches the EKF update with more informative compressed observations, enabling accurate tracking even with a limited number of RF chains. 
    
    \item To quantify the fundamental performance limits of \ac{NF} pose tracking, we derive the Bayesian \ac{CRB} and the associated \ac{FIM}. Our analysis shows how the NF channel structure and the analog combiner design influence the Fisher information related to the MS pose contained in the uplink observation. Furthermore,  we derive asymptotic expressions to reveal how this information scales with the array size, \ac{SNR}, and MS pose. We also validate these scaling behaviors through numerical simulations.

    \item Within the PAC-EKF framework, we design two low-complexity predictive analog combiners using the predicted MS pose, 
    i.e., a \ac{FIM}-based design that preserves the expected informative components in the observation, 
    and a geometry-inspired design that exploits NF beamfocusing to collect \textcolor{black}{pose-relevant information} from expected dominant spatial viewpoints. Simulations show that the proposed analog combiners nearly achieve fully digital tracking accuracy while requiring fewer RF chains, outperform the random analog combiner while requiring lower transmit power, \textcolor{black}{and remain effective under practical phase quantization. We also show that the impact of the prediction error on the PAC-EKF framework is fundamentally governed by the EKF linearization.}
    % affecting both the analog combiner design and the EKF update.

\end{itemize}

% \subsection{Outline and Notations}
		{\em Organization:}  The remainder of this paper is organized as follows. Sec.~\ref{sec:sys} introduces the system model. Sec.~\ref{sec:EKF} develops the tracking framework. Sec.~\ref{sec:performance} analyzes the fundamental performance of \ac{NF} tracking. Sec.~\ref{sec:AnalogCombiner} proposes two low-complexity designs of predictive analog combiners. \textcolor{black}{Sec.~\ref{sec:ErrorProp} discusses the impact of the prediction error in the proposed framework.}
        Sec.~\ref{sec:results} provides simulation results. \textcolor{black}{Sec.~\ref{sec:conclusion} concludes this work and presents several directions for future work.}
        
		{\em Notations:} The upper- and lower-case bold letters denote matrices and vectors, respectively. 
		For a matrix, $\mathbf{(\cdot)}^{\mathsf{T}}$, $\mathbf{(\cdot)}^{*}$, and $\mathbf{(\cdot)}^{\mathsf{H}}$ represent transpose, conjugate, and conjugate transpose operators, respectively. $\left\|\cdot\right\|_2$ and $\left\|\cdot\right\|_{\mathsf{F}}$ denote the $l_2$ norm and the Frobenius norm. $\lambda_i(\mathbf{A})$ represents the $i$-th singular value of a matrix $\mathbf{A}$.  
        $\mathbf{I}$ denotes the identity matrix of an appropriate size. $\mathcal{U}^{N_1 \times N_2}$ denotes the set of all ${N_1 \times N_2}$ matrices that have unit modulus entries. For a scalar-valued function $f:\mathbb{R}^{N_1}\rightarrow\mathbb{R}$, its gradient at $\mathbf{x}\in\mathbb{R}^{N_1}$ is a column vector $\nabla_{\mathbf{x}} f(\mathbf{x})=\left[\frac{\partial f}{\partial [\mathbf{x}]_{1}},..., \frac{\partial f}{\partial [\mathbf{x}]_{n_1}},...,\frac{\partial f}{\partial [\mathbf{x}]_{N_1}}\right]^\mathsf{T}$, where $[\mathbf{x}]_{n_1}$ is the $n_1$-th element of $\mathbf{x}$. For a vector-valued function $\bm{f}:\mathbb{R}^{N_1}\rightarrow\mathbb{R}^{N_2}$, its Jacobian at $\mathbf{x}\in \mathbb{R}^{N_1}$  is $\nabla_{\mathbf{x}} \bm{f}(\mathbf{x})=\frac{\partial \mathbf{f}(\mathbf{x})}{ \partial \mathbf{x}^{\mathsf{T}}}$, which is a $N_2\times N_1$ matrix with the $(n_2,n_1)$-th element being $[\nabla_{\mathbf{x}} \bm{f}(\mathbf{x})]_{n_2,n_1}=\frac{\partial [\bm{f}(\mathbf{x})]_{n_2}}{\partial [\mathbf{x}]_{n_1}}$.
        %%%%%%%%%%%%%%%%
        For real numbers $a,b\in \mathbb{R}$, 
		$\left \lfloor a \right \rfloor $ represents the floor function, 
		$\left \lceil a \right \rceil$ represents the ceiling function, 
		and ${\rm mod} (a,b)$ is the remainder of $a$ divided by $b$. For a positive integer $N\in \mathbb{Z}_+$, $<\!\!N\!\!>$ represents a set of integers, i.e., $<\!\!N\!\!>=\{ - \frac{N-1}{2},...,\frac{N-1}{2}\} \subset \mathbb{Z}$ if $N$ is odd, or {$ <\!\!N\!\!>=\{-\frac{N}{2},...,\frac{N}{2}-1\} \subset \mathbb{Z}$ if $N$ is even.}
        Moreover, ${\rm j}$ is the imaginary unit and $\log(\cdot)$ is the natural logarithm.

\begin{figure*}[ht]
% \vspace{-4mm}
    \centering
    \subfloat[\label{fig:NFchannel}]{
        \includegraphics[width=0.7\linewidth]{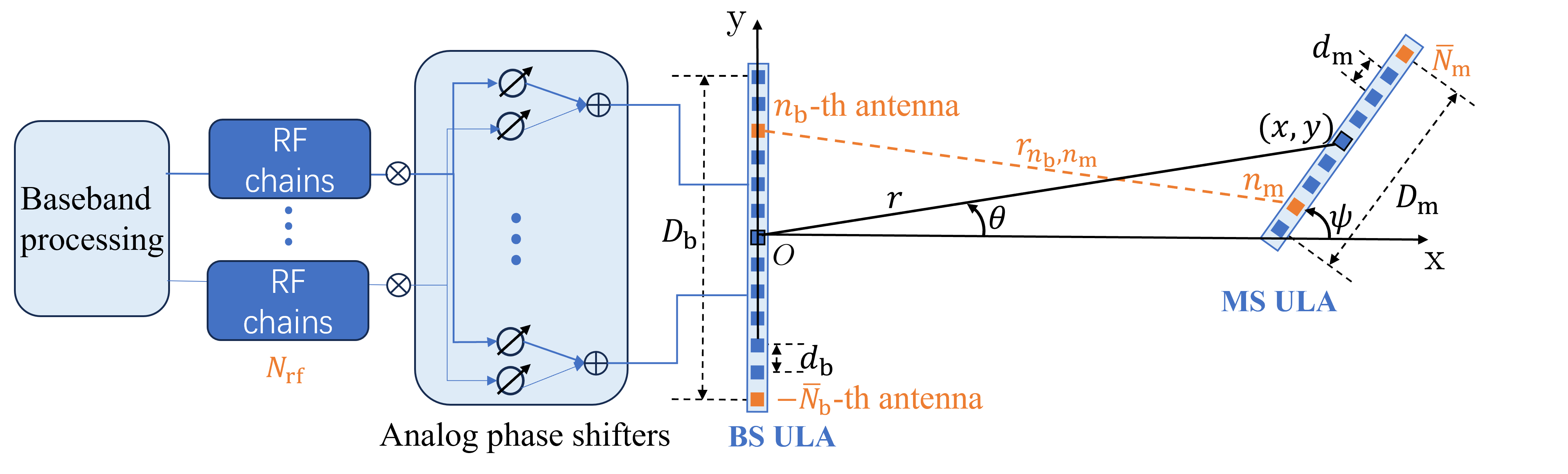}
    } 
    \subfloat[\label{fig:CTRV}]{
        \includegraphics[width=0.23\linewidth]{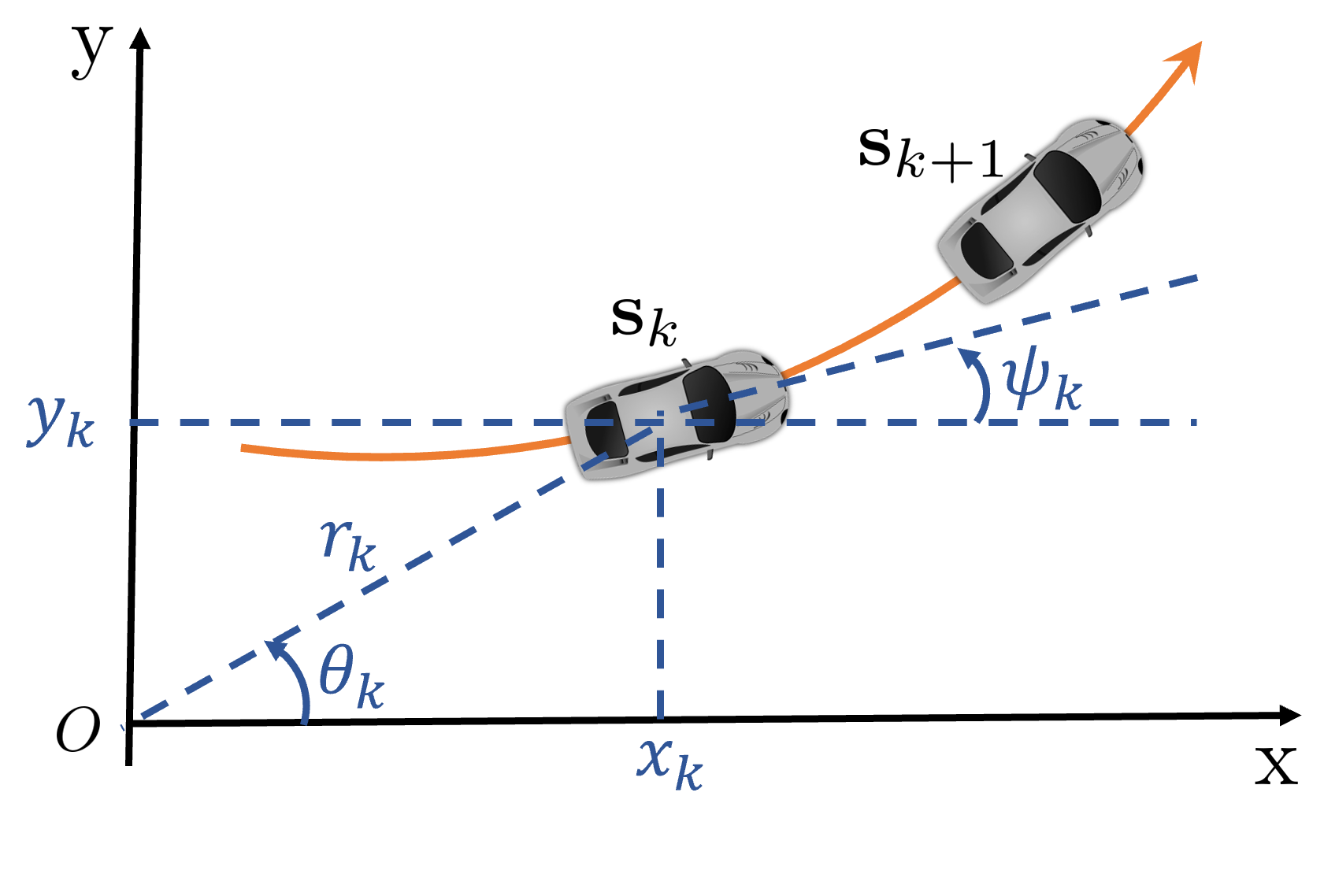}
    }  
    \vspace{-2mm}
    \caption{System Model. (a) A geometric model of the LoS MIMO channel between ULAs with a hybrid array architecture at the BS. (b) A state transition model of a vehicle-type MS from the time instant $k$ to $k+1$. } %, where the received signal at the BS must be combined by the analog phase shifters before baseband processing.
    % \vspace{-2mm}
    \label{fig:sys}
\end{figure*}

% \vspace{-2mm}

%%%%%%%%%%%%%%%%%%
\section{System Model}\label{sec:sys}

We consider a point-to-point narrowband massive \ac{MIMO} system operating at a high carrier frequency $f$ with a corresponding wavelength $\lambda$, e.g., in \ac{mmWave} and \ac{THz} bands. As shown in Fig.~\ref{fig:NFchannel}, the \ac{BS} is equipped with a \ac{ULA} of $N_{\rm b}$ antennas with antenna spacing $d_{\rm b}=\frac{\lambda}{2}$. The \ac{MS} is equipped with a \ac{ULA} of $N_{\rm m}$ antennas with antenna spacing $d_{\rm m}=\frac{\lambda}{2}$.
The corresponding array apertures are $D_{\rm b}=(N_{\rm b}-1)d_{\rm b}$ and $D_{\rm m}=(N_{\rm m}-1)d_{\rm m}$, respectively.
Moreover, the \ac{BS} employs a hybrid array architecture with only $N_{\rm rf}$ \ac{RF} chains, where $ N_{\rm rf} \ll N_{\rm b}$. We assume that the \ac{MS} employs a fully digital array architecture, as the MS typically has a smaller number of antennas than the BS, i.e., $N_{\rm m}< N_{\rm b}$.
The MS-BS distance is assumed to be larger than $d_{\rm Fn}$, where $d_{\rm Fn}=0.62\sqrt{D_{\rm b}^3/\lambda}$ is the {\em Fresnel distance} separating the reactive and radiative \ac{NF} regions~\cite{selvan2017fraunhofer}. 
Accordingly, the MS primarily operates in the radiative NF region of the BS array~\cite{kosasih2024finite,tang2023line}. % reactiveNF
For sufficiently large distances, the MS may also enter the FF region of the BS array. 

%%%%%%%%%%%%%
In the following, we present the NF channel model, which reveals how the instantaneous channel response depends on the MS's position and orientation. 
To enable the BS to track channel variation induced by MS mobility, we further describe the state transition model that characterizes the time-varying trajectory of the MS pose, and the uplink observation model under the BS's hybrid array architecture.

 \vspace{-2mm}
 
%%%%%%%%%%%
\subsection{Channel Model}\label{subsec:channel}

We denote the BS antenna index by $n_{\rm b} \in {<\!N_{\rm b}\!>} \, \triangleq \{-\bar N_{\rm b},...,\bar N_{\rm b}\}$ and the MS antenna index by $n_{\rm m} \in\, {<\!N_{\rm m}\!>}\, \triangleq \{-\bar N_{\rm m},...,\bar N_{\rm m}\}$, where $N_{\rm b}$ and $N_{\rm m}$ are assumed to be odd numbers with $\bar N_{\rm b}=(N_{\rm b}-1)/2$ and $\bar N_{\rm m}=(N_{\rm m}-1)/2$ for notational simplicity.\footnote{The model can be generalized to the ULA with an even number of antennas, i.e.,  $N_{\rm b}=2\bar N_{\rm b}$ or $N_{\rm m}=2\bar N_{\rm m}$. In this case, $n_{\rm b} \in \,<\!\!N_{\rm b}\!> \,\triangleq \{-\bar N_{\rm b},...,\bar N_{\rm b}-1\}$ or $n_{\rm m} \in\, <\!\!N_{\rm m}\!\!>\,\triangleq \{-\bar N_{\rm m},...,\bar N_{\rm m}-1\}$.} The ULA center corresponds to index $0$ in each array.
The coordinate origin $O$ is set at the BS ULA center, and the $\rm y$-axis is aligned with the BS ULA.
We denote the MS pose as $\mathbf{p}=[x,y,\psi]^{\mathsf{T}}$, where $[x,y]^{\mathsf{T}}$ is the position of the MS ULA center and $\psi$ is the orientation angle relative to the $\rm x$-axis. Under this model, the position of the $n_{\rm b}$-th BS antenna is $\mathbf{r}_{n_{\rm b}}=[0,n_{\rm b} d_{\rm b}]^{\mathsf{T}}$, the position of the $n_{\rm m}$-th MS antenna is $\mathbf{r}_{n_{\rm m}}=[x+n_{\rm m} d_{\rm m}\cos\psi, y+n_{\rm m} d_{\rm m} \sin\psi]^{\mathsf{T}}$, and their distance is
\begin{align}\label{eq:r(nb,nm)}
        &r_{n_{\rm b},n_{\rm m}}=\left\|\mathbf{r}_{n_{\rm b}}\!-\mathbf{r}_{n_{\rm m}}\right\|
        \nonumber\\&=\sqrt{(x+n_{\rm m} d_{\rm m}\cos\psi)^2+(y+n_{\rm m} d_{\rm m} \sin\psi-n_{\rm b} d_{\rm b} )^2}.
\end{align}
Particularly, $r_{0,0}\!=\!\sqrt{x^2+y^2}\! \triangleq \! r$ is the MS-BS distance. 
%%%%
Due to the severe attenuation of \ac{NLoS} components in high-frequency communications, the \ac{LoS} path typically dominates the propagation channel~\cite{indoorSM,chen2025quasi,tang2023line}. We therefore focus on modeling the strong LoS component.
Using the ray-tracing technique, the \ac{LoS} channel from the MS to the BS is parameterized by $\mathbf{p}$ as~\cite{mixed,chen2025quasi}
\begin{align}\label{eq:Hexact}
				\mathbf{H}(\mathbf{p})&=\left[\mathbf{h}_{-\bar N_{\rm m}},...,\mathbf{h}_0,...,\mathbf{h}_{\bar N_{\rm m}}\right]
                \nonumber\\&=\!\begin{bmatrix}
					h_{-\bar N_{\rm b},-\bar N_{\rm m}} & \dots  &  h_{-\bar N_{\rm b},\bar N_{\rm m}}\\
					\vdots  & \ddots  & \vdots\\
					h_{\bar N_{\rm b},-\bar N_{\rm m}} & \dots &  h_{\bar N_{\rm b},\bar N_{\rm m}}
				\end{bmatrix} \in \mathbb{C}^{N_{\rm b} \times N_{\rm m}},
		\end{align}
        where $\mathbf{h}_{ n_{\rm m}}$ is the $(n_{\rm m}+\bar N_{\rm m}+1)$-th column of $\mathbf{H}(\mathbf{p})$, and $h_{n_{\rm b},n_{\rm m}} = \frac{\lambda}{4\pi r_{n_{\rm b},n_{\rm m}}}\exp \left(-{\rm j}\frac{2\pi}{\lambda}r_{n_{\rm b},n_{\rm m}} \right)$ is determined by $\mathbf{p}$.
	% \begin{align}\label{eq:h}
 %        h_{n_{\rm b},n_{\rm m}} &= \frac{\lambda}{4\pi r_{n_{\rm b},n_{\rm m}}}\exp \!\left(\!-{\rm j}\frac{2\pi}{\lambda}r_{n_{\rm b},n_{\rm m}}\!\!\right). 
	% 	\end{align}
 
 \vspace{-2mm}

\subsection{State Transition Model}\label{subsec:state}

\textcolor{black}{We assume that the MS antenna array is mounted on a vehicle, such that the ULA center represents the vehicle position and the ULA orientation represents the vehicle heading. We model the MS state evolution in discrete time with a sampling interval $\tau$.}
As shown in Fig.~\ref{fig:CTRV}, at the $k$-th time instant, the MS state is characterized by $\mathbf{s}_k = [\mathbf{p}_k^{\mathsf{T}},\mathbf{v}_k^{\mathsf{T}}]^{\mathsf{T}} \in \mathbb{R}^{5}$, where $\mathbf{p}_k=[x_k,y_k,\psi_k]^{\mathsf{T}}\in\mathbb{R}^{3}$ describes the position and orientation of the MS, and $\mathbf{v}_k=[v_k,\omega_k]^{\mathsf{T}}\in \mathbb{R}^{2}$ consists of the linear velocity $v_k$ and the angular velocity (turn rate) $\omega_k$, where $k\in\{0,...,K\}$ and $K$ is the total number of time instants.
The state transition follows  
\begin{align}
    \mathbf{s}_{k+1} = \mathbf{a}(\mathbf{s}_k)+\mathbf{n}_{{\rm s},k} \in \mathbb{R}^{5},
\end{align}
where $\mathbf{a}(\mathbf{s}_k)$ is the state transition function (modeling the deterministic dynamics of the \ac{MS}) and $\mathbf{n}_{{\rm s},k}$ is the process noise. 
\textcolor{black}{For the considered vehicle-mounted scenario, we adopt the \ac{CTRV} model~\cite{CTRV+CV,CTRV+diag} for $\mathbf{a}(\mathbf{s}_k)$, which assumes that the linear velocity and turn rate are constant over one sampling interval.
This model captures the coupled evolution of the vehicle position and heading, and is widely used in vehicle tracking systems with non-zero turn rate~\cite{CTRV+CV,CTRV+diag}.}
% \textcolor{red}{For the considered vehicle-mounted scenario,} the translational motion of the vehicle centroid and the rotational evolution of its heading are coupled, especially during turning. We therefore adopt the \ac{CTRV} model for $\mathbf{a}(\mathbf{s}_k)$, which provides \textbf{a standard and tractable description} of planar rigid-body motion with constant linear velocity and constant turn rate over one sampling interval~\cite{CTRV+CV,CTRV+diag}. 
% In particular, during turning, the translational motion of the vehicle centroid and the rotational evolution of its heading are coupled. 
Under the \ac{CTRV}, $\mathbf{a}(\mathbf{s}_k)$ is given by
\begin{align}\label{eq:fcn_a}
    \begin{split}
        \mathbf{a}(\mathbf{s}_k)=\begin{bmatrix}
x_k + \frac{v_k}{\omega_k} \left[ \sin(\psi_k + \omega_k \tau) - \sin(\psi_k) \right]  \\
y_k + \frac{v_k}{\omega_k} \left[ -\cos(\psi_k + \omega_k \tau) + \cos(\psi_k) \right]  \\
\psi_k + \omega_k \tau\\
v_k  \\
\omega_k 
\end{bmatrix}.
    \end{split}
\end{align} 
The process noise $\mathbf{n}_{{\rm s},k}$ accounts for random linear and angular accelerations. We assume
$\mathbf{n}_{{\rm s},k}\sim\mathcal{N}(\mathbf{0},\mathbf{N}_{{\rm s}})$ with $\mathbf{N}_{{\rm s}}={\rm diag}\{0,0,0,(\tau\sigma_v)^2,(\tau\sigma_\omega)^2\}$, where $\sigma_v^2$ and $\sigma_\omega^2$ are the variances of the linear and angular acceleration noises, respectively~\cite{CTRV+CV,CTRV+diag}.
 
\vspace{-2mm}

\subsection{Uplink Observation Model}\label{subsec:uplink}  

Eqs.~\eqref{eq:r(nb,nm)}-\eqref{eq:Hexact} show that the channel is highly sensitive to the instantaneous MS pose.
% As the MS moves, its pose changes, causing rapid channel variations. 
% Reliable communication thus requires frequently estimating the MS state so that the transmission strategy can be adapted in real time. 
Consequently, MS mobility causes rapid channel variations, thereby necessitating frequent MS state estimation to adapt the transmission strategy in real time. 
To achieve this, we consider a \ac{TDD} system with uplink-downlink channel reciprocity, which includes the following three stages.
%%%%%%%%%%
At the {\em initial access} stage, the BS and MS establish a LoS link and obtain the initial state and transition model parameters via beam training, channel estimation, and/or auxiliary sensors. Then, at the {\em tracking} stage, the MS transmits an uplink pilot signal. Based on the uplink observation and the state transition model, the BS estimates the MS state and updates the channel estimate. At the subsequent {\em data transmission} stage, the BS transmits data in the downlink or receives data in the uplink using the updated channel. By iteratively executing tracking and transmission, the system maintains reliable connectivity. Sudden channel changes (e.g., LoS blockage) trigger re-initialization~\cite{GDN2016tracking}.
In this paper, we focus on uplink tracking. Initial estimation and data transmission schemes are available in~\cite{mixed,tang2023line,chen2025quasi}. 
The uplink observation model is provided below.

Denote by $P_{\rm m}$ the MS transmit power. At time instant $k$, the MS transmits an uplink pilot $\mathbf{x}_{k}\in \mathbb{C}^{N_{\rm m}\times 1}$, satisfying 
\begin{align}\label{eq:xk}
    \mathbb{E}[\mathbf{x}_{k}\mathbf{x}_{k}^{\mathsf{H}}]=\frac{P_{\rm m}}{N_{\rm m}}\mathbf{I}.
\end{align} 
Let $\mathbf{n}_{{\rm o},k}\sim\mathcal{CN}(\mathbf{0},\mathbf{N}_{\rm o})$ be the channel noise, where $\mathbf{N}_{\rm o}=\sigma_{\rm o}^2\mathbf{I}\in \mathbb{C}^{N_{\rm b}\times N_{\rm b}}$, and $\sigma_{\rm o}^2$ is the noise power.
The BS receives
    \begin{align} 
		\mathbf{y}_k 
		&=   \mathbf{H}_k \mathbf{x}_{k}+ \mathbf{n}_{{\rm o},k} \in \mathbb{C}^{N_{\rm b}\times 1},
\end{align}
where $\mathbf{H}_k\triangleq\mathbf{H}(\mathbf{p}_k)$.
This model neglects Doppler shifts as their effect within a single pilot transmission is negligible~\cite{salmi2008detection,LiuISAC2}.
Besides, we assume the BS and MS are clock-synchronized after initial access. Any residual offset can be handled by extending the model to incorporate a clock-transition model~\cite{clock1,clock2}.

%%%%%%%%%%%
As illustrated in Fig.~\ref{fig:NFchannel}, the BS employs a hybrid array architecture in which analog combining is performed in the RF domain before \ac{ADC}. 
% $\mathbf{y}_k$ represents the (uncompressed) uplink NF channel snapshot, i.e., the full-array observation available before analog combining.
Note that $\mathbf{y}_k$ is the (uncompressed) channel snapshot, i.e., the full-array observation.
% Under the BS's hybrid array architecture, analog combining is performed in the RF domain before \ac{ADC}. 
% Under the BS's hybrid array architecture, 
In practice, only a compressed snapshot is accessible at the BS, given by
\begin{align} \label{eq:zk-observe}
        \mathbf{z}_k 
		&=  \mathbf{Q}_k \mathbf{y}_k=  \mathbf{Q}_k\underbrace{\mathbf{H}_k \mathbf{x}_{k}}_{\mathbf{b}(\mathbf{s}_k)}+\mathbf{Q}_k\mathbf{n}_{{\rm o},k} \in \mathbb{C}^{N_{\rm rf}\times 1},
\end{align}
where $\mathbf{Q}_k \in \mathcal{U}^{ N_{\rm rf}\times N_{\rm b} }$ is the analog combining matrix with unit-modulus entries, $\mathbf{b}(\mathbf{s}_k)\triangleq\mathbf{H}(\mathbf{p}_k) \mathbf{x}_{k}$ is the observation function (before compression).
The compressed observation $\mathbf{z}_k$ serves as the key information source for pose estimation. Its informativeness depends on (i) {\em the NF-LoS channel structure}, which encodes the MS pose $\mathbf{p}_k$ into $\mathbf{H}_k$ and (ii) {\em the analog combiner design} $\mathbf{Q}_k$, which determines how much of the available information is preserved after RF-domain compression.  
Random analog combiners~\cite{cui2022channel,mixed,mendez2015channel}, such as matrices with $\{\pm 1\}$ entries, generally fail to preserve the informative signal components, leading to degraded estimation accuracy. 
 
% With limited \ac{RF} chains, 
Designing an analog combiner for tracking is fundamentally more challenging than for data transmission. In conventional data transmission, the channel is typically assumed to be known or slowly varying, so the analog combiner can be designed to maximize the communication rate or minimize the detection error, under the unit-modulus constraint~\cite{AltMin,hybridMMSE,el2014spatially}. In contrast, during tracking, the channel depends on the unknown time-varying MS state, rendering instantaneous \ac{CSI} unavailable. Consequently, the analog combiner should be designed to anticipate which signal components are most informative for tracking, an approach we refer to as {\em predictive analog combining}. In this context, the design objective becomes preserving pose-relevant information, which requires principled measures to quantify informativeness. 
Furthermore, the predictive analog combining must remain computationally lightweight, as tracking generally operates under strict real-time constraints and short update intervals.

The {\em unit-modulus constraint}, {\em unavailable CSI}, {\em unclear design objective}, and {\em computational efficiency} together make analog combiner design for NF pose tracking particularly challenging.     
To address these challenges, we proceed in three steps. First, in Sec.~\ref{sec:EKF}, we develop a recursive tracking framework that integrates analog combiner design with pose estimation, leveraging temporal pose correlation to overcome the lack of instantaneous CSI.
% effectively addressing the challenge of the lack of instantaneous CSI by leveraging temporal correlation in the MS pose evolution.
Second, in Sec.~\ref{sec:performance}, we derive fundamental performance limits for NF pose tracking, explicitly characterizing the impact of the NF channel structure and the analog combiner design on tracking accuracy.
%%%%%%%%%%%%%%%%%% 
These theoretical insights then guide the predictive analog combiner designs in Sec.~\ref{sec:AnalogCombiner}, where we propose two low-complexity designs under the unit-modulus constraint.

\vspace{-2mm}

\section{Tracking Framework}\label{sec:EKF}
\color{black}
In this section, we present the \ac{EKF}-based framework for tracking the time-varying MS state under the hybrid ELAA architecture. 
The \ac{EKF} is widely adopted for recursive state estimation with nonlinear state transition and observation models~\cite{simon2006optimal}.
However, the EKF assumes a given observation model and does not account for the analog combiner design.  
In the following, we first introduce the standard \ac{EKF} framework under a given analog combiner. Building upon the EKF framework, we propose the PAC-EKF framework that incorporates the analog combiner design to enhance the informativeness of the compressed observation for state estimation.  
\color{black}

\vspace{-3mm}

\subsection{Extended Kalman Filter (EKF)}

\textcolor{black}{The EKF  operates through recursive prediction and update steps, where the nonlinear models are locally linearized around the current estimate.
Through this recursive procedure, the EKF effectively fuses prior state information with newly observed data to achieve accurate tracking.}
Specifically, at each time instant $k$, the EKF maintains a {\em prior estimate} ${\mathbf{s}}_{k|k-1}$ with covariance ${\mathbf{P}}_{k|k-1}=\mathbb{E} [(\mathbf{s}_k-{\mathbf{s}}_{k|k-1})(\mathbf{s}_k-{\mathbf{s}}_{k|k-1})^{\mathsf{T}}]$, and a {\em posterior estimate} ${\mathbf{s}}_{k|k}$ with covariance ${\mathbf{P}}_{k|k}=\mathbb{E} [(\mathbf{s}_k-{\mathbf{s}}_{k|k})(\mathbf{s}_k-{\mathbf{s}}_{k|k})^{\mathsf{T}}]$. These quantities are recursively computed through the following prediction and update stages~\cite{simon2006optimal}.

\subsubsection{EKF Prediction Stage} 
The prior estimate ${\mathbf{s}}_{k|k-1}$ is obtained by propagating the previous posterior estimate ${\mathbf{s}}_{k-1|k-1}$ through the state transition function in \eqref{eq:fcn_a}, i.e.,
\begin{align}
    {\mathbf{s}}_{k|k-1}&=\mathbf{a} ({\mathbf{s}}_{k-1|k-1}) \in \mathbb{R}^{5}.
\end{align}
The corresponding prior covariance at time instant $k$ is
\begin{align}
    {\mathbf{P}}_{k|k-1} =\mathbf{A}_{k-1}{\mathbf{P}}_{k-1|k-1}\mathbf{A}_{k-1}^{\mathsf{T}}+\mathbf{N}_{\rm s} \in \mathbb{R}^{5\times 5},
\end{align}
where $\mathbf{A}_{k-1}=\nabla_{\mathbf{s}} \mathbf{a}(\mathbf{s})\big|_{{\mathbf{s}}_{k-1|k-1}}$ is the Jacobian of the state transition function evaluated at ${\mathbf{s}}_{k-1|k-1}$, \textcolor{black}{which provides the first-order local linearization of the nonlinear state evolution}, and $\mathbf{N}_{\rm s}$ is the process noise covariance.

\subsubsection{EKF Update Stage}\label{subsec:EKFupdate} 
The new compressed observation $\mathbf{z}_k=\mathbf{Q}_k\mathbf{y}_k$ under the chosen analog combiner $\mathbf{Q}_k$ is then used to refine the prior estimate.  
Note that the system involves a complex-valued observation model and a real-valued state transition model.
The Kalman-gain form of the EKF~\cite{simon2006optimal} is typically employed for purely real-valued systems. 
Therefore, we adopt an information form of the EKF~\cite{simon2006optimal,salmi2008detection,koivisto2021channel,clock1}, which handles the complex-valued observation through the following real-valued score function and \ac{FIM}. 
%%%%%%%%%%%%%%
\begin{prop} \label{prop:FIMdata}
At each time instant, the score function of the compressed observation $\mathbf{z}_k$ in \eqref{eq:zk-observe}, defined as the gradient of the log-likelihood function $\log p(\mathbf{z}_k|\mathbf{s}_k)$ with respect to (w.r.t.) the state $\mathbf{s}_k$ is 
\begin{align}\label{eq:score}
    \!\!\mathbf{g}_k & = \nabla_{\mathbf{s}_k} \log p(\mathbf{z}_k|\mathbf{s}_k)
    \nonumber\\&=\frac{2}{\sigma^2_{\rm o}} {\rm Re} \left\{   \mathbf{B}_k^{\mathsf{H}}\mathbf{Q}_k^{\mathsf{H}} (\mathbf{Q}_k \mathbf{Q}_k^{\mathsf{H}})^{-1}(\mathbf{z}_k -\!\mathbf{Q}_k\mathbf{b}(\mathbf{s}_k))\right\}\!\in\!\mathbb{R}^{5}.\!
\end{align}
Here, $\sigma_{\rm o}^2$ is the channel noise power, and $\mathbf{B}_k=\nabla_{\mathbf{s}_k} \mathbf{b}(\mathbf{s}_k)$ is the observation Jacobian (before analog combining), \textcolor{black}{which provides the first-order local linearization of the nonlinear observation model.}
Specifically, $\mathbf{B}_k$ is given by 
\begin{align}\label{eq:Bk}
 \mathbf{B}_k&= 
\big[ \mathbf{J}_{x,k}\mathbf{x}_k, 
\mathbf{J}_{y,k}\mathbf{x}_k, 
\mathbf{J}_{\psi,k} \mathbf{x}_k, 
\mathbf{0},\mathbf{0}\big] \in \mathbb{C}^{N_{\rm b}\times5},
\end{align}
where $\mathbf{x}_k$ is the known uplink pilot signal, and  $\mathbf{J}_{\mu,k}\triangleq\frac{\partial \mathbf{H}(\mathbf{p}_k)}{\partial \mu_k}\in \mathbb{C}^{N_{\rm b}\times N_{\rm m}}$ is the derivative of the channel w.r.t. $\mu\in\{x,y,\psi\}$.
The \ac{FIM}, defined as the covariance of the score function, is 
 \begin{align}\label{eq:Fdata1}
    \mathbf{F}_{k}& 
    =\mathbb{E}_{\mathbf{z}_k|\mathbf{s}_{k}}[ \mathbf{g}_k \mathbf{g}_k^{\mathsf{T}} ]
    =\frac{2}{\sigma_{\rm o}^2}{\rm Re}\{ \mathbf{B}_k^{\mathsf{H}} \mathbf{P}_{\mathbf{Q}_k}\mathbf{B}_k\}\in \mathbb{R}^{5\times 5},
\end{align}
where $\mathbf{P}_{\mathbf{Q}_k} \triangleq \mathbf{Q}_k^{\mathsf{H}} (\mathbf{Q}_k\mathbf{Q}_k^{\mathsf{H}})^{-1} \mathbf{Q}_k$ is the orthogonal projection onto the row space of $\mathbf{Q}_k$.
\begin{proof}
 See Appendix~\ref{app:dataFIM}. 
\end{proof}
\end{prop}

With Proposition~\ref{prop:FIMdata}, the posterior covariance is updated as
\begin{align}\label{eq:P(k|k)}
      {\mathbf{P}}_{k|k} =({\mathbf{P}}_{k|k-1}^{-1}+\mathbf{F}_{k})^{-1} \in\mathbb{R}^{5\times 5},
\end{align} 
where $\mathbf{F}_{k}\in \mathbb{R}^{5\times5}$ is given in \eqref{eq:Fdata1} and is evaluated at the predicted state ${\mathbf{s}}_{k|k-1}$, \textcolor{black}{i.e., based on the local linearization of the observation model around ${\mathbf{s}}_{k|k-1}$.}
The posterior estimate is then updated as
\begin{align}\label{eq:s(k|k)}
{\mathbf{s}}_{k|k}&= {\mathbf{s}}_{k|k-1}+ {\mathbf{P}}_{k|k}\mathbf{g}_k \in \mathbb{R}^{5},
\end{align}
where $\mathbf{g}_k\in\mathbb{R}^5$ is given in \eqref{eq:score} and is also evaluated at ${\mathbf{s}}_{k|k-1}$.

 \vspace{-2mm}
 
\subsection{Predictive Analog Combining-Assisted EKF (PAC-EKF)}\label{subsec:PAC-EKF}

\begin{figure}
% \vspace{-4mm}
    \centering
\includegraphics[width=0.95\linewidth]{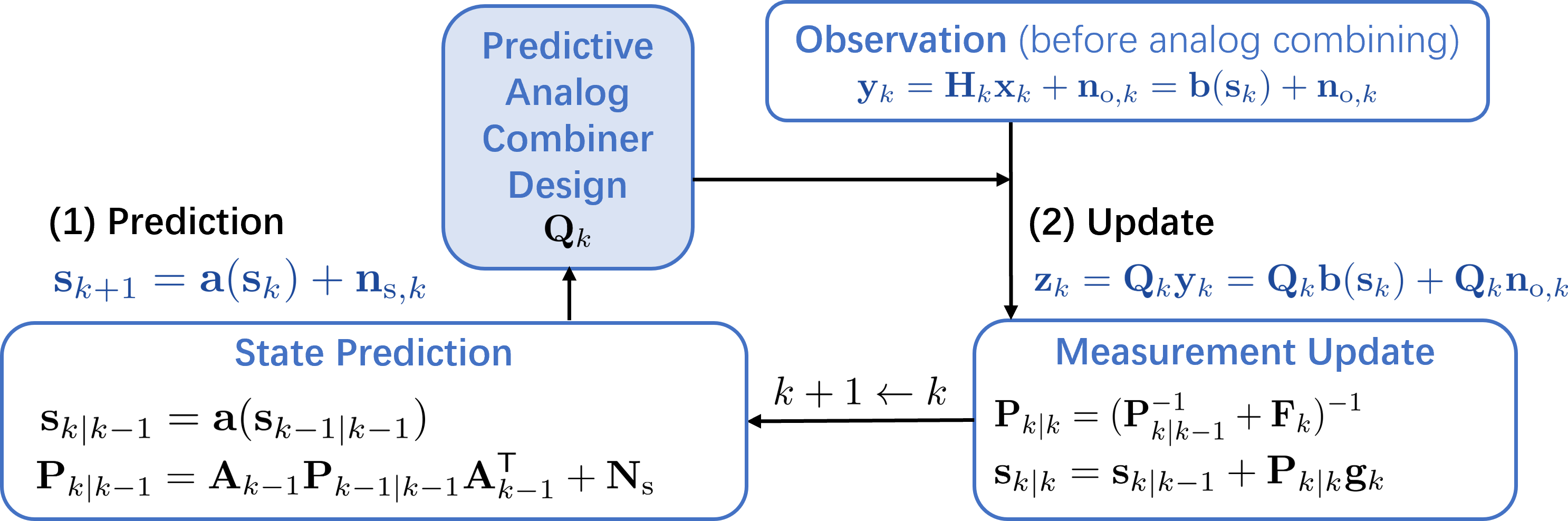}
% \vspace{-1mm}
    \caption{Schematic diagram of PAC-EKF.}
    \label{fig:PAC-EKF}
\end{figure}

From \eqref{eq:P(k|k)}-\eqref{eq:s(k|k)}, the effectiveness of the EKF update depends critically on the analog combiner $\mathbf{Q}_k$. 
Since analog combining is performed before the received signal $\mathbf{y}_k$ is converted to baseband, $\mathbf{Q}_k$ must be determined before observing $\mathbf{y}_k$, i.e., during the prediction stage. 
This motivates the {\em PAC-EKF} framework in Fig.~\ref{fig:PAC-EKF}. 
%%%%
As discussed in Sec.~\ref{subsec:uplink}, the true state $\mathbf{s}_k$ is unknown, and thus the instantaneous CSI is unavailable at the time of designing $\mathbf{Q}_k$. 
The PAC-EKF addresses this challenge by leveraging the temporal correlation in the MS pose evolution through the predicted state ${\mathbf{s}}_{k|k-1}$ and its covariance ${\mathbf{P}}_{k|k-1}$. These priors, obtained during the EKF prediction stage, capture both the expected MS pose and its uncertainty.  
By utilizing these priors, $\mathbf{Q}_k$ can be designed in a predictive manner to extract the signal components that are expected to be most informative in the upcoming observation.

Intuitively, a well-designed analog combiner should minimize the expected pose estimation error. 
Under the PAC-EKF framework, this leads to the following \ac{MMSE} design problem
\begin{align}\label{opt:mmse}
\mathbf{Q}^{\rm mse}_k&=\arg \min_{\mathbf{Q}_k \in \mathcal{U}^{N_{\rm rf}\times N_{\rm b}}} {\rm Tr}( \mathbf{D} \mathbb{E} [(\mathbf{s}_k-{\mathbf{s}}_{k|k})(\mathbf{s}_k-{\mathbf{s}}_{k|k})^{\mathsf{T}}] )
\nonumber\\&=\arg \min_{\mathbf{Q}_k \in \mathcal{U}^{N_{\rm rf}\times N_{\rm b}}} {\rm Tr}( \mathbf{D}{\mathbf{P}}_{k|k}), 
% \nonumber\\&\overset{(a)}{=} \arg \min_{\mathbf{Q}_k \in \mathcal{U}^{N_{\rm rf}\times N_{\rm b}}} {\rm Tr}(  \textcolor{red}{\mathbf{D}}({\mathbf{P}}_{k|k-1}^{-1}+\mathbf{F}_{k})^{-1}),
\end{align}
where $\mathbf{D}={\rm diag}([1,1,1,0,0])$ extracts the pose parameters, i.e., $x_k,y_k,\psi_k$.
% , \textcolor{red}{and (a) is from \eqref{eq:P(k|k)}.}
The optimization problem in \eqref{opt:mmse}, subject to the unit-modulus constraint, can be addressed via manifold optimization~\cite{hybridMMSE, AltMin}. 
\color{black}
A key component in manifold optimization is the conjugate gradient of ${\rm Tr}( \mathbf{D}{\mathbf{P}}_{k|k})$ w.r.t. $\mathbf{Q}_k$, denoted by $\frac{\partial{\rm Tr}( \mathbf{D}{\mathbf{P}}_{k|k})}{\partial \mathbf{Q}_k^*}$.
Following the derivation in Appendix~\ref{app:conjugate}, 
$\frac{\partial{\rm Tr}( \mathbf{D}{\mathbf{P}}_{k|k})}{\partial \mathbf{Q}_k^*}=\frac{2}{\sigma_{\rm o}^2} (\mathbf{Q}_k\mathbf{Q}_k^{\mathsf{H}})^{-1} \mathbf{Q}_k \mathbf{B}_k {\mathbf{P}}_{k|k} \mathbf{D} {\mathbf{P}}_{k|k}  \mathbf{B}_k ^{\mathsf{H}} (\mathbf{P}_{\mathbf{Q}_k}-\mathbf{I}).$ 
\color{black}
However, this approach requires iterative updates at each time instant, which is computationally demanding and sensitive to initialization~\cite{hybridMMSE}.
This motivates the development of efficient analog combining strategies, 
which will be presented in Sec.~\ref{sec:AnalogCombiner}.

 \vspace{-2mm}
 
%%%%%%%%%%%%%%%%%%
\section{Fundamental Performance Limits}\label{sec:performance}

This section analyzes the fundamental performance limits of NF pose tracking via the Bayesian \ac{CRB} and the Fisher information.
% To understand the fundamental performance limits of NF pose tracking, we derive the Bayesian \ac{CRB} and the associated \acp{FIM} in this section.
 
\vspace{-2mm}

\subsection{Bayesian CRB}

Denote by $p_{0:k}=p(\mathbf{s}_{0:k},\mathbf{z}_{0:k})$ the joint distribution of the states and observations up to time instant $k$, where $\mathbf{s}_{0:k}=\{\mathbf{s}_0,...,\mathbf{s}_k\}$ and $\mathbf{z}_{0:k}=\{\mathbf{z}_{0},...,{\mathbf{z}_{k}}\}$. 
The Bayesian (or posterior) \ac{CRB} provides a lower bound on the mean-square error of estimating ${\mathbf{s}}_k$, which can be expressed as~\cite{tichavsky1998posterior}
\begin{align}
\mathbb{E}_{p_{0:k}}\left[(\mathbf{s}_k-\mathbf{s}_{k|k})(\mathbf{s}_k-\mathbf{s}_{k|k})^{\rm T}\right] \succeq \mathbf{V}_{{\rm b},k}  \triangleq  \mathbf{F}_{{\rm b},k}^{-1},
\end{align}
where $\mathbf{s}_{k|k}$ is the estimate of ${\mathbf{s}}_k$, $\mathbf{V}_{{\rm b},k}$ is the Bayesian \ac{CRB} matrix, and $\mathbf{F}_{{\rm b},k}$ is the {\em Bayesian \ac{FIM}}. 
Specifically, $\mathbf{F}_{{\rm b},0}$
% The initial Bayesian FIM 
is derived from the prior distribution of the state $p(\mathbf{s}_0)$ as~\cite{guerra2021near} 
\begin{align}\label{eq:FB0}
    \mathbf{F}_{{\rm b},0}=\mathbb{E}_{p(\mathbf{s}_0)}[\nabla_{\mathbf{s}_{0}}\log p(\mathbf{s}_0) \nabla_{\mathbf{s}_0}\log p(\mathbf{s}_0)^{\mathsf{T}} ].
\end{align} 
For $k>0$, $\mathbf{F}_{{\rm b},k}$ combines the information propagated from the past states with new data information from the current observation, which can be expressed recursively as~\cite{tichavsky1998posterior,koivisto2021channel} 
\begin{align}\label{eq:Fbayesian}
    \mathbf{F}_{{\rm b},k}=\mathbf{F}_{{\rm p},k}+\mathbf{F}_{{\rm d},k},  \forall k\in[1,K].%\forall k\in\{1,...,K\}.
\end{align} 
%%%%%%%%%%
$\mathbf{F}_{{\rm p},k}$ is the {\em prior \ac{FIM}}, propagated from the previous Bayesian FIM $\mathbf{F}_{{\rm b},k-1}$ through the state transition model as~\cite{tichavsky1998posterior,koivisto2021channel} 
\begin{align}\label{eq:Fprior}
    \mathbf{F}_{{\rm p},k}\!= (\mathbf{A}_k \mathbf{F}_{{\rm b},k-1}^{-1} \mathbf{A}_k^{\mathsf{T}}  + \mathbf{N}_{\rm s})^{-1},  \forall k\in[1,K],
\end{align} 
where $\mathbf{A}_k=\nabla_{\mathbf{s}_k} \mathbf{a}(\mathbf{s}_k)$ is the state transition Jacobian. This term measures how well prior information predicts the current state.
%%%%%%%%%
$\mathbf{F}_{{\rm d},k}$ is the {\em expected data \ac{FIM}} associated with newly observed data, defined as the expectation of the standard \ac{FIM} $\mathbf{F}_{k}$ over the prior distribution of the state, i.e.,~\cite{tichavsky1998posterior,guerra2021near}
\begin{align}\label{eq:FDk}
    \mathbf{F}_{{\rm d},k} = \mathbb{E}_{\mathbf{s}_k|\mathbf{s}_{k-1}}[ \mathbf{F}_{k} ], \forall k\in[1,K],
\end{align}
where $\mathbf{F}_{k}=\mathbb{E}_{\mathbf{z}_k|\mathbf{s}_{k}}[\nabla_{\mathbf{s}_{k}}\log p(\mathbf{z}_k|\mathbf{s}_{k})\nabla_{\mathbf{s}_{k}}\log p(\mathbf{z}_k|\mathbf{s}_{k})^{\mathsf{T}}] $ quantifies the information that the new observation  $\mathbf{z}_k$ provides about the current state $\mathbf{s}_{k}$.  
The expectation $\mathbb{E}_{\mathbf{s}_k|\mathbf{s}_{k-1}}[\cdot]$ in \eqref{eq:FDk} generally lacks a closed form and is typically evaluated via Monte Carlo methods~\cite{koohifar2018autonomous,guerra2021near}.
% In general, the expectation $\mathbb{E}_{\mathbf{s}_k|\mathbf{s}_{k-1}}[\cdot]$ in \eqref{eq:FDk} does not admit a closed-form expression and is often approximated using Monte Carlo integration~\cite{koohifar2018autonomous,guerra2021near}.  
Since $\mathbf{F}_{k}$ admits a closed-form expression as in~\eqref{eq:Fdata1}, it enables a clear interpretation of how the NF channel structure and the analog combiner design jointly influence the information carried by the current observation $\mathbf{z}_k$. 
We therefore proceed to analyze $\mathbf{F}_{k}$ in detail.

 \vspace{-2mm}

\subsection{Fisher Information From Uplink Observation}\label{subsec:DataFIM}

% As the following analysis applies to any time instant, we simplify the notation by omitting the subscript $k$.
%%%%
% As we focus on the analysis at any one time instant, the subscript corresponding to the time instant index $k$ is omitted hereafter for notational simplicity.
%%%%%
% Given that the form of $\mathbf{F}_k$ is identical for all time instances $k$, we drop the subscript in what follows to simplify the notation for our general analysis.
%%%%%
% Because this analysis concerns the information contributed by a single observation snapshot, the time index $k$ does not play a role and is omitted hereafter for notational simplicity.

From Proposition~\ref{prop:FIMdata}, the diagonal entries of $\mathbf{F}_{k}$ are 
\begin{align}
    \operatorname{diag}(\mathbf{F}_{k})=[F_{x,k},F_{y,k},F_{\psi,k},0,0] ,  \forall k\in[1,K],
\end{align}
where ${F}_{\mu,k}$ denotes the Fisher information associated with each pose parameter $\mu\in\{x,y,\psi\}$ contained in the compressed observation $\mathbf{z}_k$, which is given by
 \begin{align}\label{eq:F_mu_k}
         {F}_{\mu,k}  
         =\frac{2}{\sigma_{\rm o}^2} \mathbf{x}_k^{\mathsf{H}} \mathbf{J}_{\mu,k}^{\mathsf{H}}\mathbf{P}_{\mathbf{Q}_k}\mathbf{J}_{\mu,k} \mathbf{x}_k \in\mathbb{R}, \forall k\in[1,K].
    \end{align}
Note that ${F}_{\mu,k}$ depends on the specific pilot realization $\mathbf{x}_k$. 
To analyze pilot-independent performance, we next quantify the average Fisher information about $\mu$ over the pilot distribution.

\begin{prop}\label{prop:F_mu}
Under the pilot statistics \eqref{eq:xk},
the average Fisher information for estimating each pose parameter $\mu\in\{x,y,\psi\}$ from the uplink observation at each time instant is 
\begin{align}\label{eq:barFmu}
\!\!\bar{F}_{\mu,k}\!=\!\frac{2 P_{\rm m}}{\sigma_{\rm o}^2 N_{\rm m}}\!\left\| \mathbf{P}_{\mathbf{Q}_k} \mathbf{J}_{\mu,k} \right\|_{\mathsf{F}}^2 \!\le\! \!\frac{2 P_{\rm m}}{\sigma_{\rm o}^2 N_{\rm m}} \!\left\| \mathbf{J}_{\mu,k} \right\| _{\mathsf{F}}^2, \!\forall k\!\in\![1,K]. \!
\end{align} 
%%%%%%
\begin{proof}
Taking the expectation of ${F}_{\mu,k}$ over $\mathbf{x}_k$, we obtain
 \begin{align}\label{eq:barFmu0}
\bar{F}_{\mu,k}=\mathbb{E}_{\mathbf{x}_k}[{F}_{\mu,k}] &\overset{(a)}{=} 
\mathbb{E}_{\mathbf{x}_k}\left[{\rm Tr}\left(\frac{2}{\sigma_{\rm o}^2} \mathbf{x}_k^{\mathsf{H}} \mathbf{J}_{\mu,k}^{\mathsf{H}}\mathbf{P}_{\mathbf{Q}_k}\mathbf{J}_{\mu,k} \mathbf{x}_k\right)\right]
  \nonumber\\& \overset{(b)}{=}   \frac{2}{\sigma_{\rm o}^2}  {\rm Tr} \left(\mathbb{E}_{\mathbf{x}_k} \left[\mathbf{x}_k\mathbf{x}_k^{\mathsf{H}} \right]\mathbf{J}_{\mu,k}^{\mathsf{H}}\mathbf{P}_{\mathbf{Q}_k} \mathbf{J}_{\mu,k} \right)
  \nonumber\\& \overset{(c)}{=}  \frac{2  P_{\rm m}}{\sigma_{\rm o}^2 N_{\rm m}}  {\rm Tr} \left(\mathbf{J}_{\mu,k}^{\mathsf{H}}\mathbf{P}_{\mathbf{Q}_k}\mathbf{J}_{\mu,k} \right),
\end{align}
where (a) is from \eqref{eq:F_mu_k} and ${\rm Tr}(t)=t\in\mathbb{R}$,  (b) is from the cyclic property of the trace, and (c) is from \eqref{eq:xk}.
Following the properties of projection matrices (i.e., $\mathbf{P}_{\mathbf{Q}_k}^2=\mathbf{P}_{\mathbf{Q}_k}$ and $\mathbf{P}_{\mathbf{Q}_k}=\mathbf{P}_{\mathbf{Q}_k}^{\mathsf{H}}$), we obtain
\begin{align}\label{eq:trace_J}
     {\rm Tr} \left(\mathbf{J}_{\mu,k}^{\mathsf{H}}\mathbf{P}_{\mathbf{Q}_k}\mathbf{J}_{\mu,k} \right)&={\rm Tr} \left(\mathbf{J}_{\mu,k}^{\mathsf{H}}\mathbf{P}_{\mathbf{Q}_k}^{\mathsf{H}}\mathbf{P}_{\mathbf{Q}_k}\mathbf{J}_{\mu,k} \right)
      \nonumber\\&\overset{(a)}{=} \left\| \mathbf{P}_{\mathbf{Q}_k} \mathbf{J}_{\mu,k} \right\|_{\mathsf{F}}^2 \overset{(b)}{\le}  \left\| \mathbf{J}_{\mu,k} \right\| _{\mathsf{F}}^2, 
\end{align}
where (a) is based on the definition of the Frobenius norm, and (b) is from $\left\|\mathbf{P}_{\mathbf{Q}_k}\mathbf{J}_{\mu,k}\right\|_{\mathsf{F}}^2\le \left\|\mathbf{P}_{\mathbf{Q}_k}\right\|^2_2\left\|\mathbf{J}_{\mu,k}\right\|_{\mathsf{F}}^2$ and $\left\|\mathbf{P}_{\mathbf{Q}_k}\right\|^2_2=1$.
Substituting \eqref{eq:trace_J} into \eqref{eq:barFmu0}, we obtain \eqref{eq:barFmu}.
\end{proof}
\end{prop}

Proposition~\ref{prop:F_mu} explicitly reveals that at each time instant $k\in[1,K]$, the average Fisher information $\bar{F}_{\mu,k}$ contained in the observation $\mathbf{z}_k$ depends jointly on the NF channel structure (through $\mathbf{J}_{\mu,k}$) and the analog combiner (through $\mathbf{P}_{\mathbf{Q}_k}$). Specifically, $\mathbf{J}_{\mu,k}$ reflects the sensitivity of the NF channel to changes in pose parameter $\mu$, while $\mathbf{P}_{\mathbf{Q}_k}$ determines how much of this sensitivity is preserved after analog combining.
% This characterization establishes a clear connection between the physical propagation, array architecture, and the achievable Fisher information.
 
\vspace{-2mm}

\subsection{Scaling Behavior of Fisher Information}\label{subsec:FIMapprox}
 
From Proposition~\ref{prop:F_mu}, the upper bound of $\bar{F}_{\mu,k}$ in~\eqref{eq:barFmu} represents the maximum achievable Fisher information when analog combining introduces no distortion to the uplink observation at each time instant $k\in[1,K]$. This bound can be achieved in the fully digital architecture with $N_{\rm rf}=N_{\rm b}$ and $\mathbf{Q}_k=\mathbf{Q}^{\rm fd}\triangleq\mathbf{I}\in \mathbb{C}^{N_{\rm b}\times N_{\rm b}}$. Moreover, this bound depends explicitly on the \ac{SNR} $\frac{ P_{\rm m}}{\sigma_{\rm o}^2}$ and the channel derivatives $\mathbf{J}_{\mu,k}$, $\mu\in \{x,y,\psi\}$, whose structure varies with the MS pose and the antenna number. 
To analytically characterize how the fundamental performance limit, i.e., the maximum of $\bar{F}_{\mu,k}$, scales with these factors, we next derive asymptotic forms of  $\mathbf{J}_{\mu,k}$ and $\left\|\mathbf{J}_{\mu,k}\right\|_{\mathsf{F}}^2$, based on the following widely adopted NF assumptions (A1)-(A3)~\cite{cui2022channel,tang2023line,chen2025quasi,kosasih2024finite}.
% \footnote{These NF assumptions are used for theoretical analysis. Simulations are conducted based on the ray-tracing model in \eqref{eq:Hexact}.} 
Since the derivation holds for any $k\in[1,K]$, we omit the subscript $k$ in the remainder of this subsection for notational simplicity. 

\begin{itemize}
    \item[(A1)] The wavefront is spherical with uniform amplitude across the antenna array, i.e., uniform spherical wave. 
    \item[(A2)] The BS employs a large array with $N_{\rm b}\gg 1$, while the MS uses a smaller array with $N_{\rm m} < N_{\rm b}$ and $D_{\rm m}<D_{\rm b}$.
    \item[(A3)] The MS is located beyond the Fresnel distance $d_{\rm Fn}$, i.e., $|x| > d_{\rm Fn}$ and/or $|y| > d_{\rm Fn}$.
    We consider two representative cases based on whether the MS lies on the axis perpendicular to the BS ULA: 
    \begin{itemize}
        \item[(A3.1)] {Off-Axis Case: $|x| > d_{\rm Fn}$ and $|y| > d_{\rm Fn}$.}
        \item[(A3.2)] {On-Axis Case: $|x| > d_{\rm Fn}$ and $|y| = 0$.} 
    \end{itemize} 
\end{itemize}

\begin{prop}\label{prop:Jacobian}
Under assumptions (A1)-(A3.1), as $N_{\rm b}\to\infty$, the channel derivative $\mathbf{J}_{\mu}$ w.r.t. the position parameter $\mu \in\{x,y\}$ converges in the Frobenius norm to a scaled version of the channel $\mathbf{H}$, given by
\begin{align}\label{eq:Jmu-H} 
\tilde{\mathbf{J}}_{\mu}\triangleq \eta\frac{\mu}{r} \mathbf{H}, \quad \mu \in\{x,y\},
\end{align}
where $\eta =-\left(\frac{1}{r}+{\rm j} \frac{2\pi}{\lambda}  \right)$, $r=\sqrt{x^2+y^2}$, and the relative approximation error vanishes with rate $\frac{\|\mathbf{J}_{\mu}- \tilde{\mathbf{J}}_{\mu}\|_{\mathsf{F}}^2}{ \|\tilde{\mathbf{J}}_{\mu} \|_{\mathsf{F}}^2} = \mathcal{O}(N_{\rm b}^{-1})$.
%%%%%%%%%%%%
For the orientation parameter $\psi$, $\mathbf{J}_{\psi}$ converges to a column-scaled version of $\mathbf{H}$, given by 
\begin{align}\label{eq:Jpsi-H}
\tilde{\mathbf{J}}_{\psi}\triangleq\eta d_{\rm m} \sin(\theta -\psi)\mathbf{H}\mathbf{D}_{N_{\rm m}},
\end{align} 
where $\mathbf{D}_{N_{\rm m}}={\rm diag}[-\bar N_{\rm m}, ..., 0,..., \bar N_{\rm m}]$ is a diagonal weighting matrix, and $[\theta,r]^{\mathsf{T}}$ is the polar coordinate of the MS.
When $|\sin(\theta-\psi)|>c>0$ for some non-zero constant $c$, the relative error satisfies $   \frac{\|\mathbf{J}_{\psi}- \tilde{\mathbf{J}}_{\psi}\|_{\mathsf{F}}^2}{ \|\tilde{\mathbf{J}}_{\psi} \|_{\mathsf{F}}^2} = \mathcal{O}\left(N_{\rm b}^{-1}\right)$.
\vspace{-2mm}

\begin{proof}
    See Appendix~\ref{app:SVD_Hjacobian}.
\end{proof}
\end{prop}

\begin{figure}
% \vspace{-1mm}
    \centering
    \includegraphics[width=0.8\linewidth]{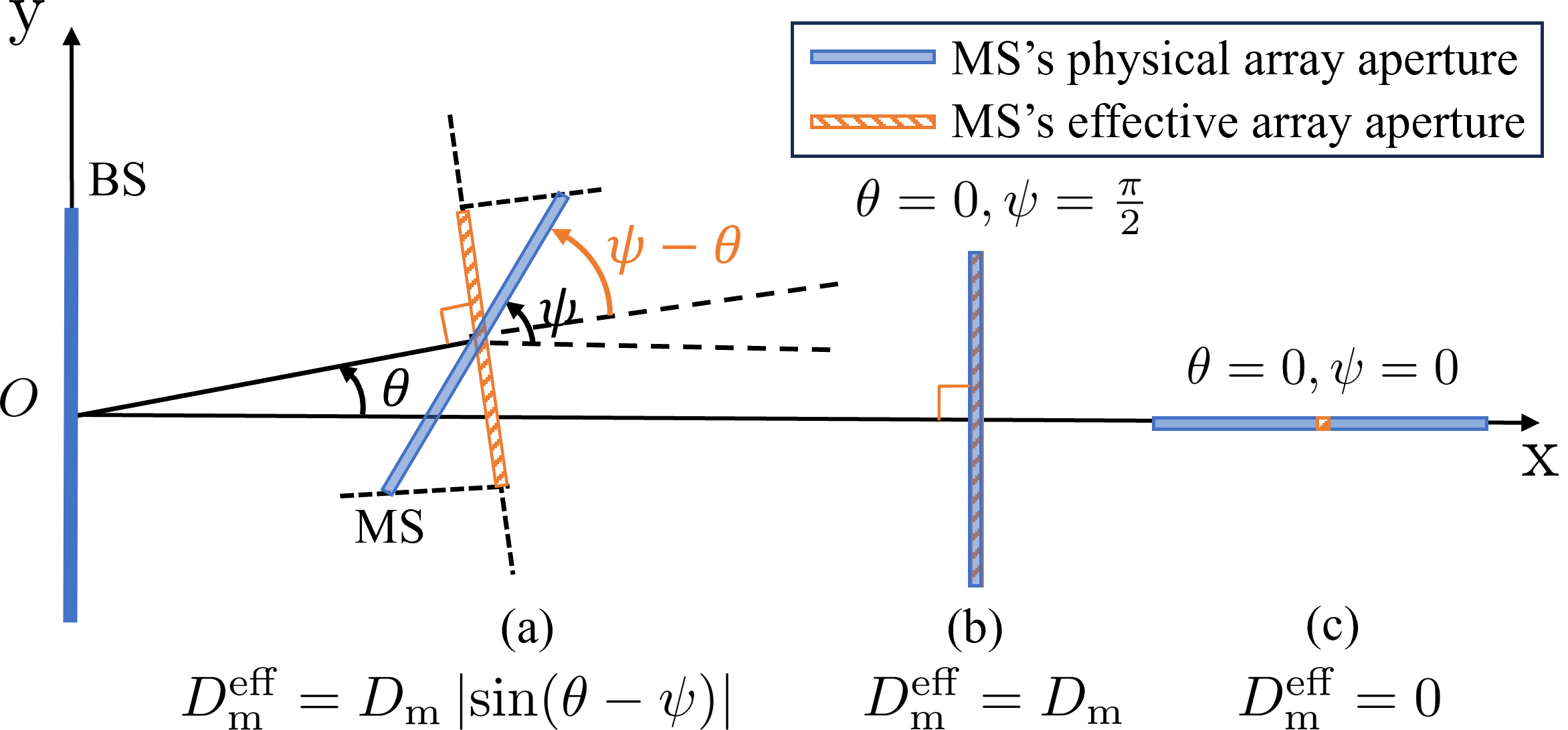}
    \vspace{-1mm}
    \caption{MS's physical and effective array apertures ($D_{\rm m}$ and $D_{{\rm m}}^{\rm eff}$).}
    \label{fig:Deff}
\end{figure}

\begin{figure}
\vspace{-3mm}
    \centering
    \includegraphics[width=0.85\linewidth]{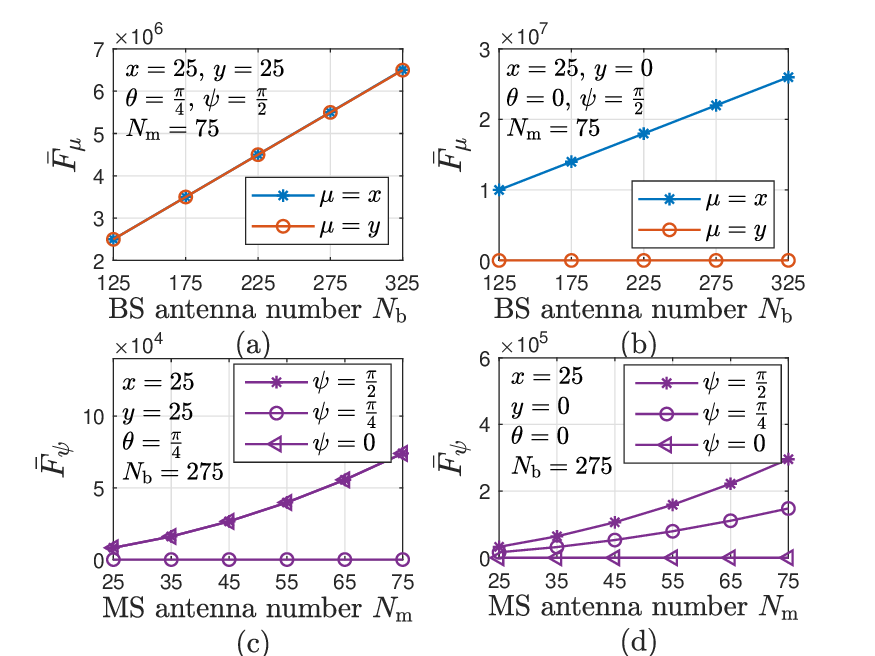}
    \vspace{-2mm}
    \caption{Impact of system parameters on $\bar{F}_{\mu}$ under a fully digital BS
array architecture, where $f=28 ~\rm GHz$, $P_{\rm m}=10~\rm dBm$, and $\sigma^2_{\rm o}=-70~\rm dBm$. }
    \label{fig:FIM-norm-Nb-Nm}
\end{figure}

Note that the term $\left|\sin(\theta-\psi)\right|$ acts as
a projection factor that determines the MS's effective array aperture, denoted by $D_{{\rm m}}^{\rm eff}\triangleq D_{\rm m}\left|\sin(\theta-\psi)\right|$. This corresponds to projecting the MS ULA onto the direction orthogonal to the line connecting the BS's and MS's ULA centers (referred to as the MS-BS line), as illustrated in Fig.~\ref{fig:Deff}. Proposition~\ref{prop:Jacobian} reveals that the approximations $\mathbf{J}_{\mu} \approx \tilde{\mathbf{J}}_{\mu}$, $\mu\in\{x,y,\psi\}$, hold when the BS aperture is sufficiently large and the MS has a non-zero effective aperture, which we will verify later via numerical results in Figs.~\ref{fig:FIM-norm-Nb-Nm} and \ref{fig:QOM-comparison}. 
Building upon Propositions~\ref{prop:F_mu} and~\ref{prop:Jacobian}, we next characterize the asymptotic behavior of the average Fisher information $\bar{F}_{\mu}$, $\mu\in\{x,y,\psi\}$.

\begin{prop}\label{prop:Fnorm} 
Under assumptions (A1)-(A3), the average Fisher information about the MS position in the uplink observation, abbreviated as position information, satisfies  
\begin{align}\label{eq:Fxy-upper}
    \!\!\bar{F}_{x} \!+ \!\bar{F}_{y} & 
  \!\le  \! \!\frac{2  P_{\rm m}}{\sigma_{\rm o}^2 N_{\rm m}}  (\left\|\mathbf{J}_{x}\right\|_{\mathsf{F}}^2 \!\! + \!\left\| \mathbf{J}_{y}\right\|_{\mathsf{F}}^2 )
 \!\!= \!\!\frac{ P_{\rm m} N_{\rm b} }{2\sigma_{\rm o}^2r^2}  \!\left(1 \!+ \!\mathcal{O}(N_{\rm b}^{-1})\right)\!. \!\!
\end{align} 
In the off-axis case (A3.1), both $x$- and $y$-related components contribute significantly to the position information, whereas in the on-axis case (A3.2), the $y$-related contribution becomes negligible.
% the MS's effective array aperture is non-zero, 
% i.e.,
When $|\sin(\theta-\psi)|>c>0$, the average Fisher information about the MS orientation (abbreviated as orientation information) satisfies  
\begin{align}\label{eq:Fpsi-upper}
   \!\bar{F}_{\psi} \!\!\le \! \!  \frac{2  P_{\rm m}}{\sigma_{\rm o}^2 N_{\rm m}}   \!\left\|\mathbf{J}_{\psi}\right\|_{\mathsf{F}}^2 
  \! =  \!\frac{ P_{\rm m} N_{\rm b} (D_{{\rm m}}^{\rm eff})^2 }{24\sigma_{\rm o}^2r^2} \!\left(\! \!1 \!+ \!\frac{2}{N_{\rm m} \!\!- \!1} \!\right)\!  \left(1 \!+ \!\mathcal{O}(N_{\rm b}^{-1})\right)\!. \!
\end{align} 

\vspace{-4mm}

\begin{proof}
    See Appendix~\ref{app:Hjacobian_norm}.
\end{proof} 
\end{prop}

\vspace{-2mm}

%%%%%%%%%%%%%%%%%%%%%%%
Eqs.~\eqref{eq:Fxy-upper}-\eqref{eq:Fpsi-upper} intuitively reveal how the position and orientation information scales with system parameters under the fully digital BS array architecture and NF assumptions.
\begin{itemize}
    \item Ratio of MS transmit power to noise power $\frac{P_{\rm m}}{\sigma_{\rm o}^2}$: Both position and orientation information increase with $P_{\rm m}$ and decrease with $\sigma_{\rm o}^2$, i.e., proportional to the \ac{SNR}.
    \item BS antenna number $N_{\rm b}$: Position and orientation information scale linearly with $N_{\rm b}$ since adding more BS antennas can provide higher-dimensional observations.  
    \item MS-BS distance $r$: Position and orientation information decay proportionally to $1/r^2$, primarily due to the reduction in received signal power from higher path loss. 
    \item MS's effective array aperture $D_{{\rm m}}^{\rm eff}$: Orientation information scales quadratically with $ D_{{\rm m}}^{\rm eff}$,  determined by both the MS's physical array aperture $D_{\rm m}$ and the projection factor $\left|\sin(\theta-\psi)\right|$.
    Intuitively, a sufficiently long MS array provides a spatial baseline that allows the BS to infer the MS orientation from relative path differences across antennas. 
    Conversely, if the MS array is too short (approximating a source point) or aligned with the MS-BS line, the effective spatial separation between antennas vanishes from the BS's viewpoint, making orientation unidentifiable. 
\end{itemize}
 
% \begin{itemize}
% \item[$\frac{P_{\rm m}}{\sigma_{\rm o}^2}$](Ratio of MS transmit power to noise power):  
% Both position and orientation information increase with $P_{\rm m}$ and decrease with $\sigma_{\rm o}^2$, i.e., proportional to the \ac{SNR}.
% \item[$N_{\rm b}$] (BS antenna number):
% Position and orientation information scale linearly with $N_{\rm b}$ since adding more BS antennas can provide higher-dimensional observations.  
% \item[$r$](MS-BS distance): Position and orientation information decay proportionally to $1/r^2$, primarily due to the reduction in received signal power from higher path loss. 
% \item[$D_{{\rm m}}^{\rm eff}$] (MS effective array aperture): 
% Orientation information scales quadratically with $ D_{{\rm m}}^{\rm eff}$, determined by both the physical array aperture $D_{\rm m}$ and the projection factor $\left|\sin(\theta-\psi)\right|$.
% Intuitively, a sufficiently long MS array provides a spatial baseline that allows the BS to infer the MS orientation from relative path differences across antennas. 
% Conversely, if the MS array is too short (approximating a source point) or aligned with the MS-BS line, the effective spatial separation between antennas vanishes from the BS's viewpoint, making orientation unidentifiable. 
% \end{itemize}

Fig.~\ref{fig:FIM-norm-Nb-Nm} numerically validates these scaling behaviors by plotting the exact (non-asymptotic) upper bounds in \eqref{eq:Fxy-upper}-\eqref{eq:Fpsi-upper}.
Specifically, Figs.~\ref{fig:FIM-norm-Nb-Nm}a-\ref{fig:FIM-norm-Nb-Nm}b confirm the linear growth of $\bar{F}_{x}$ and $\bar{F}_{y}$ with $N_{\rm b}$. For the off-axis case in Fig.~\ref{fig:FIM-norm-Nb-Nm}a, both $x$- and $y$-related components contribute significantly to the position information, whereas for the on-axis case in Fig.~\ref{fig:FIM-norm-Nb-Nm}b, the $y$-related component is negligible.  
Figs.~\ref{fig:FIM-norm-Nb-Nm}c-\ref{fig:FIM-norm-Nb-Nm}d confirms the quadratic growth of $\bar{F}_{\psi}$ with $N_{\rm m}$. 
When $\left|\sin(\theta-\psi)\right|=0$, e.g., $x=y$ with $\psi = \frac{\pi}{4}$ in Fig.~\ref{fig:FIM-norm-Nb-Nm}c or $y=0$ with $\psi = 0$ in Fig.~\ref{fig:FIM-norm-Nb-Nm}d, we see that $\bar{F}_{\psi}$ becomes very small. This agrees with the intuition that when the MS's effective aperture vanishes, the uplink observation carries negligible orientation information.

\vspace{-2mm}

\section{Predictive Analog Combining}\label{sec:AnalogCombiner}
 
In this section, we propose two low-complexity designs of analog combiners within the PAC-EKF framework. %, addressing the practical challenges outlined in Sec.~\ref{subsec:uplink}.
 
 \vspace{-2mm}
 
\subsection{FIM Based Predictive Analog Combining}\label{subsec:MFIM-PC}
 
A design principle for the analog combiner arises from the FIM perspective. 
Specifically, the \ac{FIM} $\mathbf{F}_{k} $ quantifies the information that the compressed observation $\mathbf{z}_k$ carries about the current pose $\mathbf{p}_{k}$, given the pilot $\mathbf{x}_k$. Therefore, $\mathbf{Q}_k$ can be designed to maximize the total Fisher information about the MS's position and orientation, which is measured by $F_{x,k}+F_{y,k}+F_{\psi,k}={\rm Tr}(\mathbf{F}_{k})$. From Proposition~\ref{prop:FIMdata}, this design objective can be rewritten as
\begin{align}\label{eq:trace}
         {\rm Tr}(\mathbf{F}_{k}) & =\frac{2}{\sigma_{\rm o}^2}{\rm Re}\{  {\rm Tr}\left(\mathbf{B}_k^{\mathsf{H}} \mathbf{P}_{\mathbf{Q}_k}\mathbf{B}_k\right) \}
         \overset{(a)}{=}\frac{2}{\sigma_{\rm o}^2}  {\rm Tr}\left(\mathbf{B}_k^{\mathsf{H}} \mathbf{P}_{\mathbf{Q}_k}\mathbf{B}_k\right) \nonumber\\& \overset{(b)}{=} \frac{2}{\sigma_{\rm o}^2} \left\| \mathbf{P}_{\mathbf{Q}_k}\mathbf{B}_k \right\| _{\mathsf{F}}^2 \overset{(c)}{\le} 
         \frac{2}{\sigma_{\rm o}^2} \left\| \mathbf{B}_k \right\| _{\mathsf{F}}^2,
\end{align}
where (a) holds since the trace of the Hermitian matrix $\mathbf{B}_k^{\mathsf{H}} \mathbf{P}_{\mathbf{Q}_k}\mathbf{B}_k$ is real, and (b)-(c) follow the derivation steps in \eqref{eq:trace_J}.
Accordingly, the FIM-based design can be formulated as 
\begin{align}\label{opt:mfim}
\begin{split}
&\max_{\mathbf{Q}_k \in \mathcal{U}^{N_{\rm rf}\times N_{\rm b}}} {\rm Tr}(\mathbf{F}_{k}) 
    = \max_{\mathbf{Q}_k \in \mathcal{U}^{N_{\rm rf}\times N_{\rm b}}} {\rm Tr}(\mathbf{B}_k^{\mathsf{H}} \mathbf{P}_{\mathbf{Q}_k}\mathbf{B}_k ),
\end{split}  
\end{align}
where $\mathbf{B}_k=\nabla_{\mathbf{s}_k} (\mathbf{H}_k \mathbf{x}_{k})|_{\mathbf{s}_{k|k-1}}$ is the observation Jacobian evaluated at $\mathbf{s}_{k|k-1}$. 
This predictive evaluation is a key feature of the PAC-EKF framework, which leverages the predicted state ${\mathbf{s}}_{k|k-1}$ as a surrogate for the true state $\mathbf{s}_k$ for the combiner design, compensating for the lack of instantaneous CSI.  
 
%%%%%%%%%%%

The problem \eqref{opt:mfim}, subject to the unit-modulus constraint, can be solved using manifold optimization~\cite{hybridMMSE,AltMin} but at a high computational cost.  
To gain further insight, we first consider the unconstrained case.
When $N_{\rm rf}< N_{\rm b}$ and the unit-modulus constraint is ignored, the upper bound in \eqref{eq:trace} can be achieved when $\mathbf{Q}_{k}$ spans the dominant left singular subspace of $\mathbf{B}_k$.
Since only the first three columns of $\mathbf{B}_k\in \mathbb{C}^{N_{\rm b}\times5}$ are non-zero as in \eqref{eq:Bk}, at most three singular vectors of $\mathbf{B}_k$ carry useful information.
Thus, the unconstrained optimal combiner is obtained via \ac{SVD} as
    \begin{align}\label{eq:Qsvd}
   \mathbf{Q}^{\rm svd}_{k} 
   = [\mathbf{u}_{B_k,1},...,\mathbf{u}_{B_k,\min\{N_{\rm rf},3 \}}]^{\mathsf{H}},  
\end{align}
where $\mathbf{u}_{B_k,i}$ is the $i$-th left singular vector of $\mathbf{B}_k$. 
 
Generally, $\mathbf{Q}^{\rm svd}_{k}$  fails to satisfy the unit-modulus constraint, since the singular vectors typically have varying magnitudes and phases across entries. 
A simple yet effective approximation is to retain only the entry-wise phases of $\mathbf{Q}^{\rm svd}_{k}$, i.e., 
\begin{align}
    \mathbf{Q}^{\rm fim}_{k}\triangleq\exp(\,{\rm j} \,\angle \mathbf{Q}^{\rm svd}_{k} ) .
\end{align} 
This procedure involves computing the SVD of a small $N_{\rm b} \times3$ matrix and extracting the corresponding entry-wise phases, which makes it computationally efficient. We abbreviate this procedure as SVD-PE and summarize it in Algorithm~\ref{algorthm:SVDphase}.

     	\begin{algorithm}[t!]
        
		\caption{\small FIM Based Predictive Analog Combining
		}
		\label{algorthm:SVDphase}
		\small
		\textbf{Input}: $\mathbf{s}_{k|k-1}$,  $\mathbf{B}_{k}$. 
		\textbf{Output}: $\mathbf{Q}^{\rm fim}_{k}$.
		\begin{algorithmic}[1]
        \STATE{SVD of $\mathbf{B}_{k}$;}
			\STATE{Obtain the unconstrained combiner $\mathbf{Q}^{\rm svd}_{k}$ in \eqref{eq:Qsvd};} 
			\STATE{Perform phase extraction $\mathbf{Q}^{\rm fim}_{k}=\exp({\rm j}\angle \mathbf{Q}^{\rm svd}_{k} )$.}
		\end{algorithmic}
		
	\end{algorithm}

\vspace{-2mm}

%%%%%%%%%%%%%%%%%%%%%%%%
\subsection{Geometry Inspired Predictive Analog Combining}\label{subsec:AFIM-PC}

% \subsection{Geometry Inspired Predictive Analog Combining}\label{subsec:AFIM-PC}

Another design is inspired by geometric intuition. In the downlink, the \ac{NF} beamfocusing effect can be used to construct an analog beamformer to deliver data effectively toward a specific point~\cite{NFtutorial,chen2025quasi}. Analogously, in the uplink, this effect can be exploited to construct an analog combiner that efficiently collects information from a desired point. 
In this regard, each MS antenna can be interpreted as a distinct ``viewpoint'' or {\em mode}, whose radiated signal impinges on the BS array with a unique phase profile, thereby contributing a distinct component to the composite spherical wavefront observed at the BS. Intuitively, modes that are physically close produce highly correlated observations and thus contribute little extra information. To construct an effective analog combiner with a limited number of RF chains, we therefore need to select a subset of well-separated modes that provide non-redundant information.

From a formal perspective, the average Fisher information $\bar{F}_{\mu,k}$  quantifies the information about each pose parameter $\mu \in \{x,y,\psi\}$ retained in the compressed observation $\mathbf{z}_k$, under the pilot statistics. This measure provides a principled criterion for the analog combiner design, as detailed below.
% which signal components are most informative, thereby providing a principled guide for selecting a subset of well-separated modes when designing the analog combiner.
 
\begin{remark}\label{reamrk:FIM}
    Proposition~\ref{prop:F_mu} reveals that, for each time instant $k\in[1,K]$, maximizing $\bar{F}_{\mu,k}$ requires $\mathbf{Q}_k$ to align with the dominant subspace spanned by the columns (i.e., the dominant left singular subspace) of the channel derivative $\mathbf{J}_{\mu,k}$, thereby capturing the most informative directions for estimating $\mu$. Proposition~\ref{prop:Jacobian} further shows that $\mathbf{J}_{\mu,k}$ can be approximated as a scaled or column-scaled version of the channel $\mathbf{H}_k$, implying that the left singular subspace of $\mathbf{J}_{\mu,k}$ is largely contained within that of $\mathbf{H}_k$.  
Therefore, an analog combiner aligned with the dominant channel subspace preserves most of the average Fisher information in the uplink observation.
\end{remark}
 
Motivated by the geometric principle of NF beamfocusing and the analytical insights into the average Fisher information, we adopt the \acp{QOM} proposed in our prior work~\cite{chen2025quasi} as the set of well-separated modes. Specifically, QOMs consist of resolvable MS antennas, whose channel vectors are nearly orthogonal and naturally aligned with the dominant channel subspace.
To show that the QOM-based analog combining effectively preserves average Fisher information $\bar{F}_{\mu,k}$, we proceed as follows. We first introduce how to construct QOMs and how QOMs relate to the subspace of channel $\mathbf{H}_{k}$, then extend this connection to the subspace of channel derivative $\mathbf{J}_{\mu,k}$, and finally link QOMs to $\bar{F}_{\mu,k}$. 
% For notational simplicity, the time index $k$ is omitted in Sec.~\ref{subsec:QOMs}–\ref{subsec:QOM-FIM}.
As we focus on the analog combiner design at a given time instant, the subscript $k$ is temporarily omitted from Sec.~\ref{subsec:QOMs} to Sec.\ref{subsec:QOM-FIM} for notational simplicity.

\begin{figure*}[t!]
% \vspace{-2mm}
    \subfloat[Aligning the subspace of $\mathbf{H}$ and $\mathbf{J}_\mu$ via QOM-based analog matrix $\mathbf{W}_{N_{\rm rf}}^{\rm cf}$.\label{fig:QOM-subspace}]
    {\begin{minipage}{.325\textwidth}
        \centering
        \includegraphics[width=1.1\linewidth]{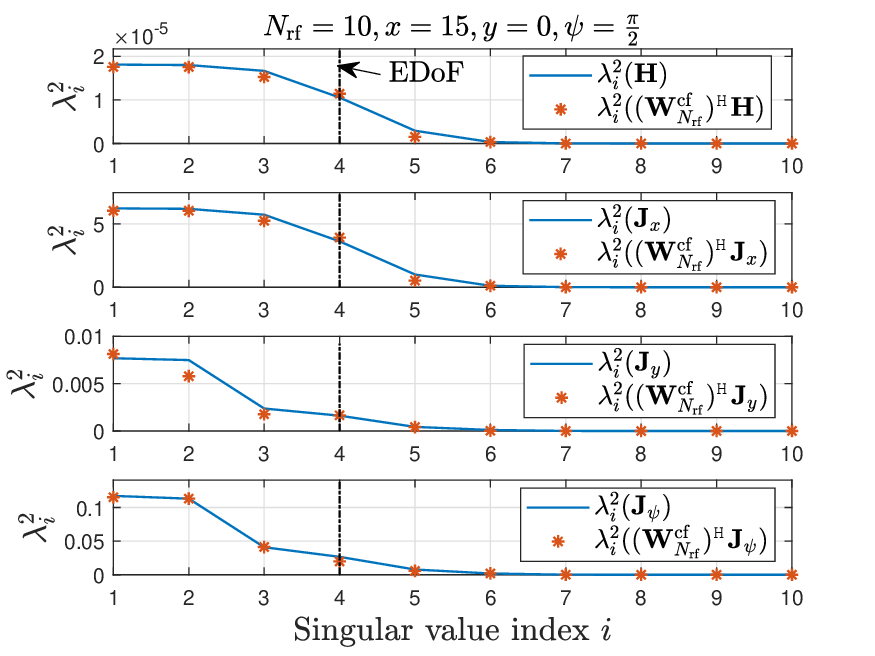}
     \end{minipage}}
    \hfill
    \subfloat[Mode energy: singular modes vs center-first ordered QOMs.\label{fig:QOMorder}]
    {\begin{minipage}{.325\textwidth}
        \centering
        \includegraphics[width=1.1\linewidth]{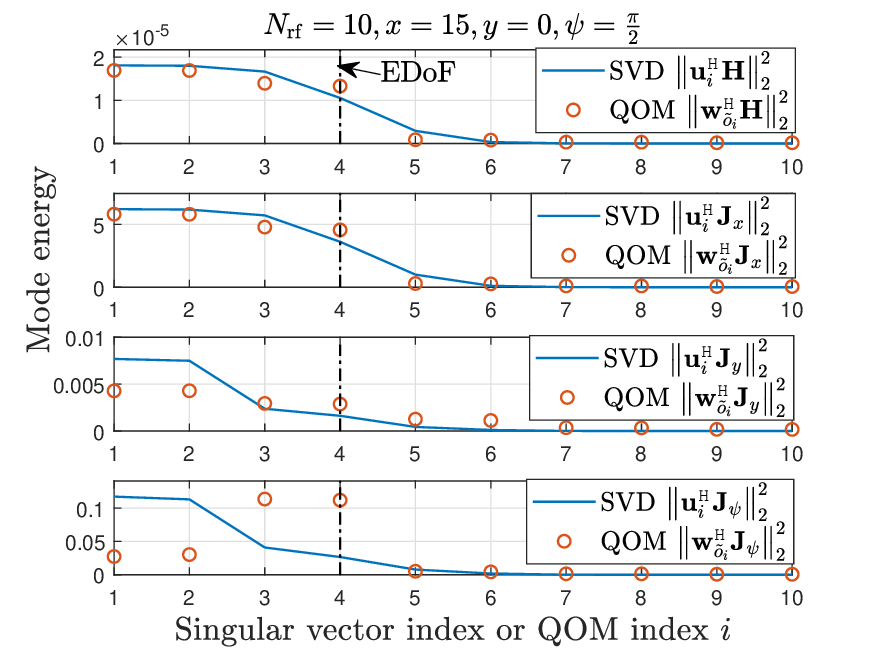}
     \end{minipage}}
    \hfill
    \subfloat[Impact of RF chain number $N_{\rm rf}$ and QOM ordering on average Fisher information $\bar{F}_{\mu}$.\label{fig:QOMorder-Nrf}]
    {\begin{minipage}{.325\textwidth}
        \centering
        \includegraphics[width=1.1\linewidth]{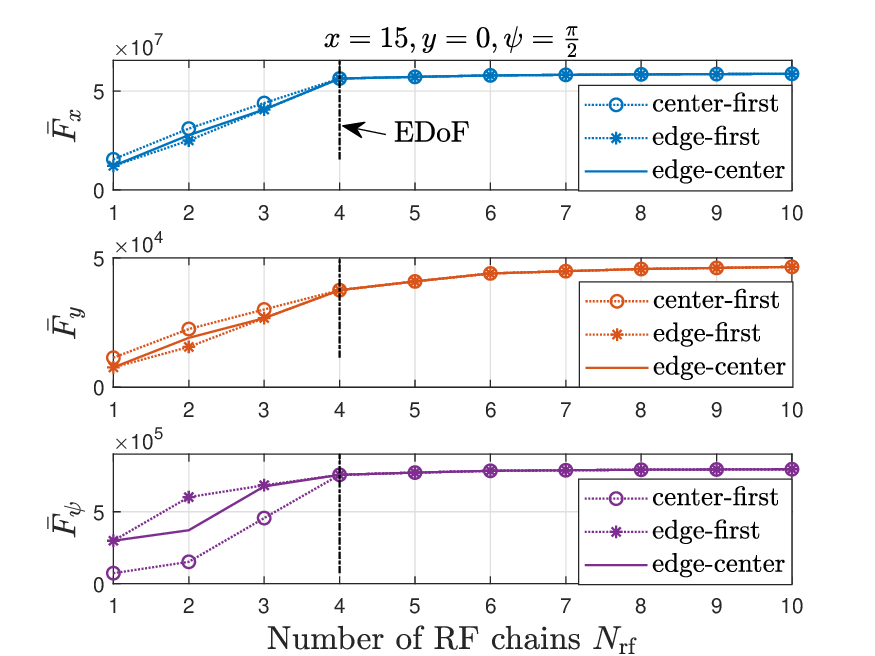}
     \end{minipage}}
     \vspace{-1mm}
    \caption{Connection between QOMs and average Fisher Information, where $f=28\,\rm GHz$, $N_{\rm b}=275$, $N_{\rm m}=75$, $P_{\rm m}=10\,\rm dBm$, and $\sigma^2_{\rm o}=-70\,\rm dBm$.} 
    \vspace{-2mm}
    \label{fig:QOM-comparison}
\end{figure*}

\subsubsection{QOM Construction}\label{subsec:QOMs}
 
The minimum spacing between the resolvable MS antenna indices, i.e., QOMs, is~\cite{chen2025quasi}  
\begin{align}\label{eq:delta}
				\Delta & =  \left \lceil \frac{\lambda r} {d_{\rm b}d_{\rm m} |\cos\theta \sin(\psi-\theta)| N_{\rm b} } \right \rceil \in \mathbb{Z},
			\end{align}
where the resolution $\Delta$ depends on the MS pose $\mathbf{p}=[x,y,\psi]^{\mathsf{T}}$ and its polar coordinate $[\theta,r]^{\mathsf{T}}$. 
Then, the QOM set is
\begin{align}\label{eq:QOM}
    \mathcal{Q}=\{\ell: \ell= i\Delta +\ell_0, i\in \mathbb{Z}, \ell_0,\ell\in\,<\!\!N_{\rm  m}\!\!>\},
\end{align}
where the reference index $\ell_0=-\bar N_{\rm m}+\left \lceil \frac{{\rm mod}(2\bar N_{\rm m},\Delta)}{2} \right \rceil$ ensures that QOMs are evenly distributed across the MS array.
The analog vector for selecting QOM $\ell\in \mathcal{Q}$ is
\begin{align} \label{eq:QO}
			% \begin{split}
			{\mathbf{w}}_{\ell}
			\!=\!\frac{1}{\sqrt{N_{\rm b}}}\left[ {\rm e}^{-{\rm j}\frac{2\pi}{\lambda}r_{\ell,-\bar N_{\rm b}}}  ,..., {\rm e}^{-{\rm j}\frac{2\pi}{\lambda}r_{\ell,\bar N_{\rm b}}}  \right]^{\mathsf{T}} \!\!\overset{(a)}{=} \!\frac{4\pi r }{\lambda \sqrt{N_{\rm b}} }  \mathbf{h}_{\ell},\!   
			% \end{split}
		\end{align}
        where ${\mathbf{w}}_{\ell}$ satisfies the constant-magnitude constraint, $\left\|{\mathbf{w}}_{\ell}\right\|_2^2=1$, and (a) holds under the uniform spherical wavefront model (A1).  
Moreover, for any distinct $\ell,\ell'\in\mathcal{Q}$, 
$ {\left|{\mathbf{w}}_{\ell'}^\mathsf{H} {\mathbf{w}}_{\ell}\right|}$ 
is close to zero, indicating their quasi-orthogonality~\cite{chen2025quasi}.   

The number of QOMs in $\mathcal{Q}\subseteq {< \!N_{\rm  m} \!>}$ is $N_{\rm e}=\left \lfloor \frac{2 \bar N_{\rm m}}{\Delta} \right \rfloor +1$. Importantly, $N_{\rm e}$ is related to the channel \ac{EDoF}, i.e., the number of dominant singular vectors of the channel, which preserves more than $95\%$ channel gain~\cite{chen2025quasi}.
Therefore, the QOMs lying inside the MS ULA, i.e., $\mathcal{Q}\subseteq{<\! N_{\rm  m}\! >}$, represents $N_{\rm e}$ dominant modes.
For limited RF chains with $N_{\rm rf}<N_{\rm e}$, we can select $N_{\rm rf}$ QOMs from these dominant modes as viewpoints. 
As a generalization, when $N_{\rm rf}>N_{\rm e}$, additional $N_{\rm rf}-N_{\rm e}$ non-dominant modes can be included by introducing virtual QOMs outside the MS ULA (sampled along the array axis with the resolution $\Delta$)~\cite{chen2025quasi}. 

% Given a limited number of RF chains, constructing the analog combiner requires selecting a subset of QOMs.
% To make this selection explicit, we introduce an ordering of QOMs.  

 \subsubsection{QOMs for Aligning With Channel Subspace}\label{subsec:QOM-channel}
Ordering QOMs by their distance to the MS array center yields the center-first ordered QOM set, denoted by
\begin{align}
  \mathcal{Q}^{\rm cf}_{N_{\rm rf}}=\{\tilde{o}_1,\tilde{o}_2,\dots,\tilde{o}_{N_{\rm rf}}\},  
\end{align} 
where $|\tilde{o}_1|\le...\le |\tilde{o}_{N_{\rm rf}}|$ and $\frac{{\tilde{o}}_{i}-\tilde{o}_{i'}}{\Delta}\in\mathbb{Z}$ for $1\le i,i' \le N_{\rm rf}$.
The corresponding analog matrix is $\mathbf{W}^{\rm cf}_{N_{\rm rf}}=[\mathbf{w}_{\tilde{o}_1},..., \mathbf{w}_{\tilde{o}_{N_{\rm rf}}}]$. 
As validated in~\cite{chen2025quasi}, for $N_{\rm e}$ dominant modes, $\mathbf{W}^{{\rm cf}^{*}}_{N_{\rm e}}$ nearly aligns with the dominant right singular subspace of the downlink channel.
%%%%%
The channel reciprocity in the TDD system 
implies that $(\mathbf{W}^{\rm cf}_{N_{\rm e}})^\mathsf{H}$ 
nearly aligns with the dominant left singular subspace of the uplink channel $\mathbf{H}$.
This is confirmed in Figs.~\ref{fig:QOM-subspace}-\ref{fig:QOMorder}, where $(\mathbf{W}^{\rm cf}_{N_{\rm e}})^\mathsf{H}$ preserves the energy and order of singular modes of $\mathbf{H}$.

\subsubsection{QOMs for Aligning With Channel Derivative Subspace}\label{subsec:QOM-Derivative}
 
From Proposition~\ref{prop:Jacobian}, under \ac{NF} assumptions, 
$\mathbf{J}_{x}$ and $\mathbf{J}_{y}$ 
are approximately scalar multiples of the uplink channel $\mathbf{H}$, so they share the same left singular vectors and ordering.
This suggests that the QOM-based analog vectors that approximately span the left singular subspace of $\mathbf{H}$ also span those of $\mathbf{J}_{x}$ and $\mathbf{J}_{y}$, as validated in Fig.~\ref{fig:QOM-subspace}.
Moreover, the center-first ordering of QOMs matches the ordering of singular modes of $\mathbf{H}$, $\mathbf{J}_{x}$, and $\mathbf{J}_{y}$, as validated in Fig.~\ref{fig:QOMorder}.  

In contrast, $\mathbf{J}_{\psi}$
can be approximated as a column-scaled version of $\mathbf{H}$, where each column is weighted proportionally to the MS antenna index $n_{\rm m} \in \{-\bar{N}_{\rm m}, ..., 0,..., \bar{N}_{\rm m}\}$. 
This scaling preserves the column space (and hence the left singular subspace) of $\mathbf{H}$ but modifies the energy distribution across columns.
Consequently, the QOM-based analog vectors approximately span the left singular subspaces of $\mathbf{H}$ and $\mathbf{J}_{\psi}$, as validated in Fig.~\ref{fig:QOM-subspace}. 
However, the column weight shifts the dominant modes toward the array edges. 
As shown in Fig.~\ref{fig:QOMorder}, for the center-first ordered dominant QOMs $\{\tilde{o}_i:i\le N_{\rm e}\}$, the $4$-th QOM (near the MS array edge) exhibits stronger energy than the $1$-st QOM (near the MS array center). To align with the left singular subspace of $\mathbf{J}_\psi$, we consider an edge-first ordered QOM set, denoted by
% for the dominant QOMs while keeping the center-first order for the remaining non-dominant QOMs.
\begin{align}
   \mathcal{Q}^{\rm ef}_{N_{\rm rf}}=\{\dot{o}_1,\dot{o}_2,\dots,\dot{o}_{N_{\rm rf}}\}, 
\end{align} 
where dominant QOMs are ordered from the edges inward, i.e., 
$|\dot{o}_i|\ge |\dot{o}_{i+1}|$ for $1\le i<N_{\rm e}$, while non-dominant QOMs retain center-first ordering, i.e., $|\dot{o}_i|\le |\dot{o}_{i+1}|$ for $ i > N_{\rm e}$, and $\frac{\dot{o}_{i}-\dot{o}_{i'}}{\Delta}\in\mathbb{Z}$ for $1\le i,i' \le N_{\rm rf}$. 
The corresponding analog matrix is 
% Correspondingly, 
$\mathbf{W}^{\rm ef}_{N_{\rm rf}}=[\mathbf{w}_{\dot{o}_1},...,\mathbf{w}_{\dot{o}_{N_{\rm rf}}}]$.

\subsubsection{QOMs for Preserving Average Fisher Information}\label{subsec:QOM-FIM}
 
From Proposition~\ref{prop:F_mu} and the above analysis, we can conclude that position information $\bar{F}_{x}+\bar{F}_{y}$ is more sensitive to the dominant QOMs near the center of the MS array, while orientation information $\bar{F}_{\psi}$ is more sensitive to those near the endpoints of the MS array. A geometric intuition linking QOMs to position and orientation information is detailed below.

\begin{remark} 
Although QOMs inside the MS ULA correspond to dominant modes, their contributions to different pose parameter estimation are inherently asymmetric. 
An intuitive analogy is to compare the MS array to a ruler: its midpoint is most informative for determining position, whereas its endpoints are more sensitive to rotation. 
This asymmetry originates from the underlying signal physics: the position is encoded in the global curvature of the wavefront, while the orientation is encoded in differential phase shifts across the antenna array.
Hence, QOMs near the MS array \textbf{center} primarily capture information about the MS \textbf{position}, whereas QOMs near the \textbf{edges} primarily capture information about the MS \textbf{orientation}.  
\end{remark}

Given that different QOMs contribute unequally to the preservation of position and orientation information, while only a subset of QOMs can be selected under the RF-chain constraint, it becomes necessary to prioritize them according to their information relevance. 
To this end, we adopt a mixed edge-center ordering of QOMs, where the analog combiner is formed by selecting the first $N_{\rm rf}$ modes, thereby balancing the preservation of position and orientation information.
Specifically, the mixed edge-center ordered QOM set is denoted as
\begin{align}\label{eq:QOMmix}
    \mathcal{Q}_{N_{\rm rf}}^{\rm ec}= \{{o}_1,...,{o}_{N_{\rm rf}}\},
\end{align}
where dominant QOMs alternate between edges and center, i.e., $|{o}_{2i-1}|\ge|{o}_{2i+1}|$, $|{o}_{2i}|\le|{o}_{2i+2}|$, and $|{o}_{2i-1}|>|{o}_{2i}|$ for $1\le i< N_{\rm e}/2$, while non-dominant QOMs follow center-first ordering, i.e., $|{o}_{i}|\le|{o}_{i+1}|$ for $i> N_{\rm e}$, and $\frac{{o}_{i}-{o}_{i'}}{\Delta}\in\mathbb{Z}$ for $1\le i,i'\le N_{\rm rf}$. 
% The corresponding analog matrix for selecting $N_{\rm rf}$ ordered QOMs as effective viewpoints is $\mathbf{W}^{\rm ec}_{N_{\rm rf}}=[\mathbf{w}_{{o}_1},...,\mathbf{w}_{{o}_{N_{\rm rf}}}]$. 
The corresponding analog matrix is $\mathbf{W}^{\rm ec}_{N_{\rm rf}}=[\mathbf{w}_{{o}_1},...,\mathbf{w}_{{o}_{N_{\rm rf}}}]$. 
%%%%%
Fig.~\ref{fig:QOMorder-Nrf} shows the impact of $N_{\rm rf}$ and QOM ordering on $\bar{F}_{\mu}$. Specifically, the analog combiner $\mathbf{Q}$ is set as $ {\sqrt{N_{\rm b}}} ({\mathbf{W}^{{\rm cf}}_{N_{\rm rf}}})^{\mathsf{H}}$, ${\sqrt{N_{\rm b}}} ({\mathbf{W}^{{\rm ef}}_{N_{\rm rf}}})^{\mathsf{H}} $, and ${\sqrt{N_{\rm b}}}({\mathbf{W}^{{\rm ec}}_{N_{\rm rf}}})^{\mathsf{H}}$, respectively, where ${\sqrt{N_{\rm b}}}$ ensures each entry of $\mathbf{Q}$ has a unit modulus. When $N_{\rm rf}<N_{\rm e}$, only a subset of dominant modes can be captured.  
We see that the center-first ordering preserves more position information, while the edge-first ordering preserves more orientation information. 
When $N_{\rm rf}\ge N_{\rm e}$, all dominant modes are included, so position/orientation information becomes identical across ordering strategies.  
% Overall, the mixed ordering balances the preservation of both types of information.  

\subsubsection{QOMs Based Predictive Analog Combining}\label{subsec:QOM-combiner}

     	\begin{algorithm}[t!]
        
		\caption{\small  Geometry Inspired Predictive Analog Combining
		}
		\label{algorthm:QOM}
		% \hspace*{\algorithmicindent}
		\small
		\textbf{Input}:  $\mathbf{s}_{k|k-1}$.
		\textbf{Output}: $\mathbf{Q}^{\rm qom}_{k}$.
		\begin{algorithmic}[1]
        \STATE{Set $\mathbf{p}=[x,y,\psi]^{\mathsf{T}}$ as the first three elements of $\mathbf{s}_{k|k-1}$;}
			\STATE{Calculate the resolution $\Delta$ in \eqref{eq:delta} based on $\mathbf{p}$;} 
			\STATE{Generate QOMs based on $\Delta$ and order QOMs following \eqref{eq:QOMmix};}
			\STATE{Generate analog matrix $\mathbf{W}_{N_{\rm rf}}^{\rm ec}$;} % 
            \STATE{Return $\mathbf{Q}^{\rm qom}_{k}=\sqrt{N_{\rm b}}(\mathbf{W}^{{\rm ec}}_{N_{\rm rf}})^{\mathsf{H}}$.}
		\end{algorithmic}
		
	\end{algorithm}
    
Note from \eqref{eq:delta} that constructing QOMs requires knowledge of the MS pose. Under the PAC-EKF framework, we leverage ${\mathbf{s}}_{k|k-1}$ as a surrogate for $\mathbf{s}_k$, and thus the geometry-inspired predictive analog combining is 
\begin{align}
    \mathbf{Q}^{\rm qom}_k=\sqrt{N_{\rm b}}(\mathbf{W}^{{\rm ec}}_{N_{\rm rf},k})^{\mathsf{H}},
\end{align}
where $\mathbf{W}^{{\rm ec}}_{N_{\rm rf},k}$ specifies the design of $\mathbf{W}^{\rm ec}_{N_{\rm rf}}$ at time instant $k$. The procedure is summarized in Algorithm~\ref{algorthm:QOM}.
Since QOMs are determined by the closed-form resolution $\Delta$, the overall design is computationally efficient.

\vspace{-2mm}

\subsection{Design Comparison}\label{subsec:compare}

%%%%%%%
Compared with the \ac{MMSE}-based analog combiner design (introduced in Sec.~\ref{subsec:PAC-EKF}) with iterative manifold optimization at each time instant, Algorithms~\ref{algorthm:SVDphase} and~\ref{algorthm:QOM} are computationally lightweight, making them particularly attractive in scenarios with strict real-time constraints.  
Moreover, Algorithm~\ref{algorthm:SVDphase} exploits the FIM dependent on pilot realization at each time instant, whereas Algorithm~\ref{algorthm:QOM} relies on the average Fisher information over the pilot distribution. Thus, Algorithm~\ref{algorthm:SVDphase} may achieve slightly better tracking performance. 
A potential advantage of QOM-based design in Algorithm~\ref{algorthm:QOM} is that it naturally extends to downlink beamforming~\cite{chen2025quasi}. This enables a unified configuration of analog phase shifters across uplink tracking and downlink transmission in the NF MIMO system with a hybrid ELAA architecture.

% \vspace{-2mm}

\section{Results and Discussions}\label{sec:results}

This section evaluates the performance of the proposed predictive analog combiners under the PAC-EKF framework. The acronyms of different combiners are defined as follows.
\begin{itemize}
    \item \textbf{FD}: Fully digital combiner with $\mathbf{Q}_{k}=\mathbf{Q}^{\rm fd}=\mathbf{I}$ and $N_{\rm rf}=N_{\rm b}$. 
    \item \textbf{Rand}: Random binary combiner with entries in $\{\pm 1\}$. % $\mathbf{Q}_{k}^{{\rm rand}}$
    \item \textbf{SVD-PE} (Algorithm~\ref{algorthm:SVDphase}): Phase-extracted SVD combiner. % $\mathbf{Q}_{k}^{\rm fim}$. %,  $\mathbf{Q}_{k}^{\rm fim}=\exp(j\angle \mathbf{Q}_{k}^{\rm svd})$, where $\mathbf{Q}_{k}^{\rm svd}$ is obtained from the right singular vectors of the Jacobian $\mathbf{B}_k$
    \item \textbf{QOM} (Algorithm~\ref{algorthm:QOM}): QOM-based combiner. % $\mathbf{Q}_{k}^{\rm qom}$.  
    \item \textbf{MO(Init)}: \ac{MMSE}-based analog combiner via manifold optimization, as introduced in Sec.~\ref{subsec:PAC-EKF}, initialized by random, SVD-PE, and QOM-based combiners, referred to as \textbf{MO(Rand)}, \textbf{MO(SVD-PE)}, \textbf{MO(QOM)}, respectively.
\end{itemize}
Note that only the FD scheme uses $N_{\rm rf} = N_{\rm{b}}$; all other schemes operate under $N_{\rm{rf}} \ll N_{\rm{b}}$. 
The FD scheme therefore represents the best possible tracking performance.

\vspace{-2mm}

% \begin{figure}
% \vspace{-4mm}
%     \centering
%     \includegraphics[width=0.8\linewidth]{Figure/MC1_Traj_10dBm_20dBm.eps}
%     \vspace{-2mm}
%     \caption{Estimated trajectories under the PAC-EKF framework with the transmit power $P_{\rm m}$ being $10 ~\rm dBm$ and $20 ~\rm dBm$.}
%     \label{fig:Trajectory}
% \end{figure}

\begin{figure*}[t!]
\vspace{-2mm}
    \subfloat[Estimation error of PAC-EKF at time instant $k$ with low-complexity designs.\label{fig:RMSE_time_k}]
    {\begin{minipage}{.325\textwidth}
        \centering
        \includegraphics[width=1.1\linewidth]{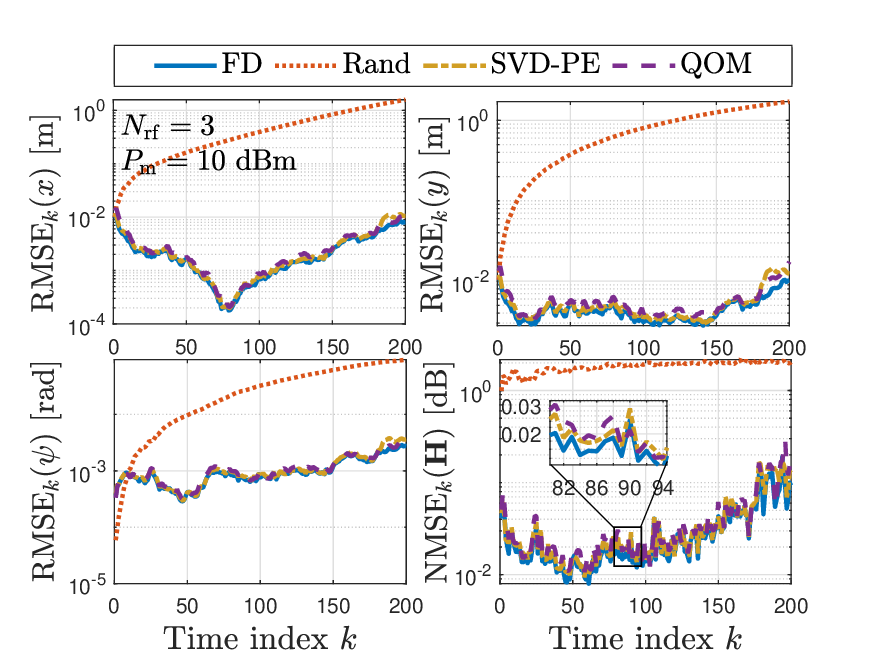}
     \end{minipage}}
    \hfill
    \subfloat[Time-averaged estimation error of PAC-EKF under different $P_{\rm m}$.\label{fig:RMSE}]
    {\begin{minipage}{.325\textwidth}
        \centering
        \includegraphics[width=1.1\linewidth]{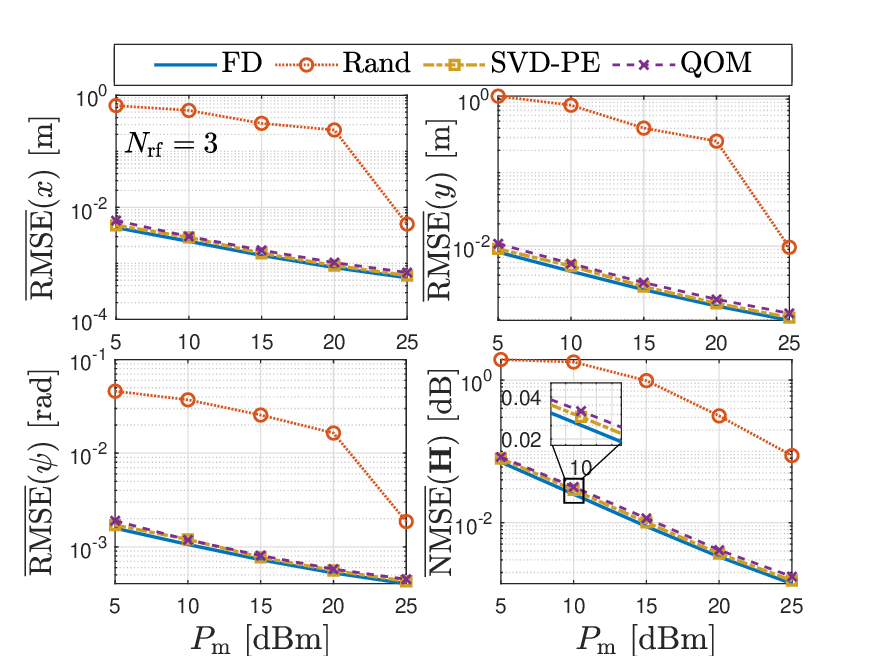}
     \end{minipage}}
    \hfill
    \subfloat[MMSE-based PAC-EKF with low-complexity combiners as initialization.\label{fig:MO_NRF}]
    {\begin{minipage}{.325\textwidth}
        \centering
        \includegraphics[width=1.1\linewidth]{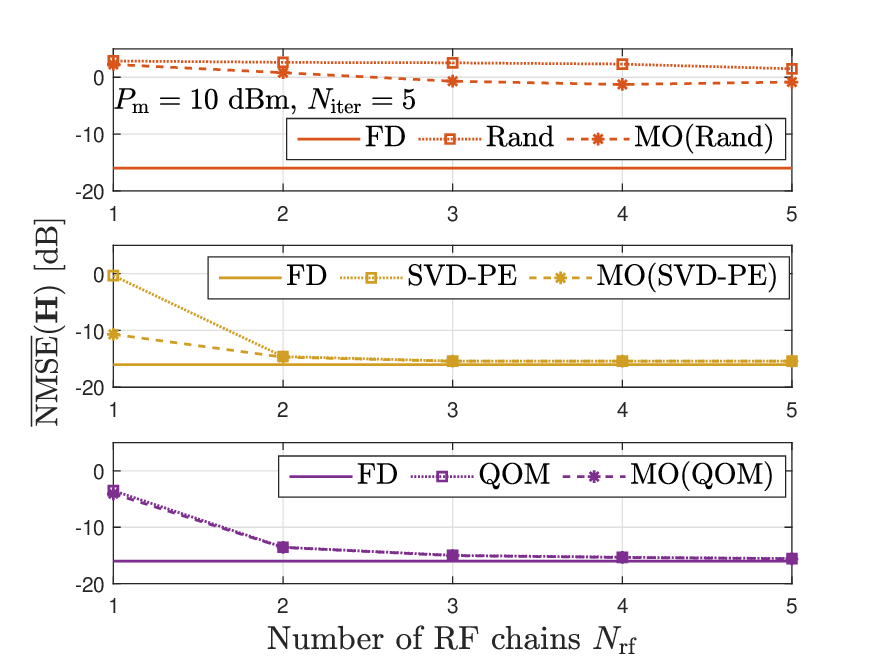}
     \end{minipage}}
     \vspace{-1mm}
    \caption{Pose estimation accuracy and channel reconstruction accuracy of the proposed PAC-EKF.}
    \label{fig:PAC-EKF-performance}
    % \vspace{-2mm}
\end{figure*}

\subsection{Simulation Setting}\label{subsec:simu}

We set the carrier frequency as $f=28~\rm GHz$, the numbers of antennas as $N_{\rm b}=275$ and $N_{\rm m}=75$, and the channel noise power as $\sigma_{\rm o}^2=-70~\rm dBm$. 
% the MS transmit power as $P_{\rm m}=10~\rm dBm$
The initial MS state is $\mathbf{s}_0=[15~\rm m,-15 ~\rm m,3\pi/8~\rm rad, 10 ~\rm m/ s, 0.1~\rm rad/s]^{\mathsf{T}}$ with $\sigma_v=2 ~\rm m/ s^2$ and $\sigma_\omega=0.1~\rm rad/s^2$. For EKF initialization~\cite{GDN2016tracking,LiuISAC2}, we set ${\mathbf{s}}_{0|0}=\mathbf{s}_0$ and $\mathbf{P}_{0|0}=\operatorname{diag}([0.05^2,0.05^2,0.001^2,\frac{v_0^2}{100},\frac{\omega_0^2}{100}])$.
%\ac{5G-NR}. 
% In practice, the interval $\tau$ is influenced by several factors, such as the degree of nonlinearity in the state transition or observation model, the process and channel noise levels, and the maximum expected rate of change of the tracked parameters. \ac{5G-NR} supports much shorter reference signal periodicity, making update intervals of less than $1~\rm ms$ feasible when finer-grained tracking is required~\cite{clock2,GDN2016tracking,LiuISAC2}.

% We conduct simulations based on the ray-tracing channel model in \eqref{eq:Hexact}.
% Simulation is repeated over $N_{\rm mc}=50$ Monte Carlo trials. 
\textcolor{black}{The tracking time interval is $\tau=20~\rm ms$, equal to the duration of two radio frames in \ac{5G-NR} systems. Simulations use the ray-tracing channel model in \eqref{eq:Hexact}, repeated over $N_{\rm mc}=50$ Monte Carlo trials. 
Each trial consists of $K=200$ time instants and the corresponding total tracking duration is $4~\rm s$. 
This simulated duration is used to illustrate the temporal evolution of the tracking performance under the considered setting, rather than to indicate a universal limit for reliable tracking. In practice, the reliable tracking duration is fundamentally determined by when significant error propagation occurs in the EKF-based tracking process.
% This is influenced by several factors, e.g., the degree of nonlinearity in the state transition and observation models, the process and observation noise levels, and the tracking update interval $\tau$. 
A more detailed discussion on the factors affecting error propagation will be provided in Sec.~\ref{subsec:factors}.
}

% Simulations use the ray-tracing channel model in \eqref{eq:Hexact}, repeated over $N_{\rm mc}=50$ Monte Carlo trials.
For each Monte Carlo trial $t\in\{1,...,N_{\rm mc}\}$, a single pilot vector is generated once from $\mathcal{CN}(\mathbf{0},\frac{P_{\rm m}}{N_{\rm m}}\mathbf{I})$ and reused across all tracking time instants $\{1,...,K\}$ within the same trial. Similarly, a random binary combiner with entries from $\{\pm 1\}$ is generated once per trial and kept fixed for all $k$.
Moreover, for each trial $t$, independent realizations of the state transition noise and uplink observation noise are generated as $ \{\mathbf{n}_{{\rm s},k}^{(t)}\}_{k=1}^{K}$ and $ \{\mathbf{n}_{{\rm o},k}^{(t)}\}_{k=1}^{K}$ with $\mathbf{n}_{{\rm s},k}^{(t)}\sim \mathcal{N}(\mathbf{0},\mathbf{N}_{\rm s})$ and $\mathbf{n}_{{\rm o},k}^{(t)}\sim \mathcal{CN}(\mathbf{0},\mathbf{N}_{\rm o})$, respectively. 
All combiner schemes are tested under identical pilot and noise realizations in the same trial to enable fair comparison.

% We evaluate both pose estimation accuracy and channel reconstruction accuracy. 
The position/orientation estimation accuracy at time instant $k$ is measured by root mean-square error (RMSE) as 
\begin{align}
\!{\rm RMSE}_k(\mu) \!=\! \sqrt{\frac{\sum_{i=1}^{N_{\rm mc}} |\mu_{k|k}^{(i)} - \mu_{k}^{(i)}|^2}{N_{\rm mc}}},\mu\!\in\!\{x,y,\psi\},
\end{align}
where $\{x_{k|k}, y_{k|k}, \psi_{k|k}\}$ are estimates of the MS's position and orientation, corresponding to the first three elements of ${\mathbf{s}}_{k|k}$, and the superscript $(i)$ denotes the $i$-th Monte Carlo trial.

The channel reconstruction accuracy at time instant $k$ is measured by normalized mean-square error (NMSE) as 
\begin{align}
{\rm NMSE}_k(\mathbf{H}) = \frac{ \frac{1}{N_{\rm mc}}\sum_{i=1}^{N_{\rm mc}} \|\mathbf{H}(\mathbf{p}_{k|k}^{(i)}) - \mathbf{H}(\mathbf{p}_{k}^{(i)})\|_\mathsf{F}^2} { \frac{1}{N_{\rm mc}}\sum_{i=1}^{N_{\rm mc}} \|\mathbf{H}(\mathbf{p}_{k}^{(i)})\|_\mathsf{F}^2}.
\end{align}
Besides, $\overline{\rm RMSE}(\mu) \!=  \frac{1}{K}\sum_{k=1}^{K} {\rm RMSE}_k(\mu)$ and $\overline{\rm NMSE}(\mathbf{H}) \!= \frac{1}{K}\sum_{k=1}^{K}  {\rm NMSE}_k(\mathbf{H})$  are time-averaged metrics. 

 \vspace{-2mm}

\subsection{Tracking Performance Evaluation}  

% \subsubsection{Estimated Trajectories}
 
% Fig.~\ref{fig:Trajectory} presents the true and estimated trajectories in one Monte-Carlo trial under the PAC-EKF framework for different MS transmit power $P_{\rm m}$. At both $P_{\rm m}=10~\rm dBm$ and $P_{\rm m}=20~\rm dBm$, we see that the FD, SVD-PE, and QOM combiners yield trajectory estimates that closely follow the ground truth. 
% In contrast, the Rand combiner performs reasonably well only at $P_{\rm m}=20~\rm dBm$. % while its accuracy degrades significantly at $10~\rm dBm$. 
% These suggest that even with a limited number of RF chains ($N_{\rm rf}=3$), the SVD-PE and QOM combiners can accurately track the MS pose under moderate transmit power, while the Rand combiner requires higher power. % to maintain tracking accuracy.

Fig.~\ref{fig:RMSE_time_k} compares the low-complexity analog combiners at each time instant $k$ over $50$ Monte Carlo trials. We see that both the SVD-PE and QOM combiners achieve nearly the same accuracy as the FD combiner, confirming their effectiveness in capturing Fisher information from the uplink observation with only a few RF chains.
%%%%%%%%%%
% An error trend is observed across time: estimation errors first decrease and then increase. This behavior is explained by the MS trajectory in Fig.~\ref{fig:Trajectory} and the scaling behavior of Fisher information in Sec.~\ref{subsec:FIMapprox}, where the MS initially moves closer to the BS at the origin (reducing path loss and thus increasing Fisher information), but then departs (leading to reduced received power and diminished Fisher information).
%%%%%%%%%%
However, the Rand combiner consistently suffers from higher errors. % due to its inability to capture informative signal components in the observation.
% which lowers the observable Fisher information.

Fig.~\ref{fig:RMSE} evaluates the time-averaged performance for different MS transmit power. As expected, higher transmit power improves tracking accuracy across all analog combiners by enhancing the \ac{SNR}. However, the gap between well-designed and random combiners remains: the SVD-PE and QOM combiners nearly match the performance of the FD combiner, whereas the Rand combiner is noticeably inferior. 
Specifically, the performance of the proposed combiners at $P_{\rm m}=5~\rm dBm$ is comparable to the Rand combiner at $P_{\rm m}=25~\rm dBm$, indicating a transmit power saving of up to $20~\rm dBm$.
Moreover, the SVD-PE combiner slightly outperforms the QOM combiner, as the SVD-PE combiner leverages the FIM associated with the pilot realization at each trial, while the QOM combiner considers the pilot statistics, as discussed in Sec.~\ref{subsec:compare}.
These results highlight the analog combiner design in reducing the RF chain number and transmit power while maintaining tracking accuracy.
 
In Fig.~\ref{fig:MO_NRF}, we investigate the performance of MMSE-based analog combiners obtained via iterative manifold optimization, where the total number of iterations at each time instant is limited to $N_{\rm iter}=5$ to keep the computational load compatible with the short tracking interval $\tau$.
% Fig.~\ref{fig:MO_NRF} shows that the MMSE-based design strongly depends on the initialization. 
With Rand initialization, although manifold optimization significantly improves the Rand combiner using 5 iterations, the performance gap to the FD benchmark remains large. 
% This shows that while optimization can partially correct uninformative initializations, it cannot fully close the gap within a few iterations.
%%%%%%%%%%%
For SVD-PE initialization, manifold optimization provides a gain only when $N_{\rm rf}=1$, while for $N_{\rm rf}\ge2$ the result essentially coincides with the SVD-PE baseline, suggesting that the SVD-PE combiner is nearly a local minimum of \eqref{opt:mmse}.
Moreover, the QOM combiner is nearly a local minimum of \eqref{opt:mmse} as manifold optimization offers negligible improvement.
%%%%%%%%%
These findings provide two key insights: In a general RF-limited case with $N_{\rm rf}\ge2$, low-complexity designs such as SVD-PE and QOM are already near-optimal, making iterative optimization largely unnecessary; in an extreme case with $N_{\rm rf}=1$, manifold optimization can offer additional gain, but at the expense of computational overhead that may not be practical for fast tracking.

\vspace{-2mm}

\color{black}
 
\subsection{Finite Resolution of Analog Phase Shifters}\label{subsec:Quantize}

\begin{figure}
% \vspace{-3mm}
    \centering
    \includegraphics[width=0.8\linewidth]{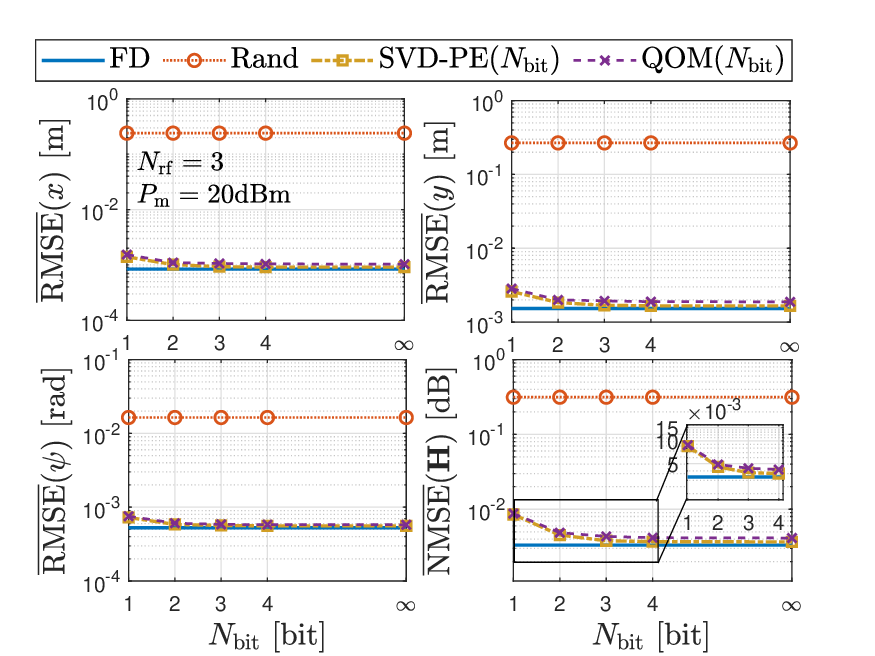}
    \vspace{-1mm}
    \caption{Impact of finite-resolution analog phase shifters.}
    \label{fig:Nbit}
\end{figure}

We further incorporate the finite-resolution constraint of practical phase shifters into the proposed analog combiner design.
Specifically, for a phase shifter with $N_{\rm bit}$-bit resolution, only $2^{N_{\rm bit}}$ discrete phase values can be realized.
We denote the corresponding feasible phase set by
$\mathcal{P}_{N_{\rm bit}}=\{ n_{\rm p} \frac{2\pi}{2^{N_{\rm bit}}}: n_{\rm p} = 0,1,\ldots,2^{N_{\rm bit}}-1\}$.
Accordingly, since each entry of the analog combining matrix must take its phase value from $\mathcal{P}_{N_{\rm bit}}$, we quantize the phase of each entry of the proposed analog combiner to its nearest value in $\mathcal{P}_{N_{\rm bit}}$~\cite{chen2025quasi}. 
 
To evaluate the resulting performance loss, Fig.~\ref{fig:Nbit} shows the tracking accuracy achieved by the proposed SVD-PE and QOM combiners under different phase resolutions. In particular, $N_{\rm bit}=\infty$ corresponds to the ideal continuous-phase implementation without quantization error. The results show that $1$-bit phase quantization causes noticeable degradation compared to the ideal infinite-resolution case. Notably, even with $1$-bit phase shifters, both the SVD-PE and QOM combiners still outperform the Rand combiner. As the phase resolution increases, the tracking accuracy of both SVD-PE and QOM combiners improves, and with $4$-bit phase shifters, the quantization loss becomes negligible.
These results indicate that the proposed designs remain practically effective under phase quantization and can approach ideal performance with moderate phase resolution.

\color{black}

% \vspace{-2mm}

\color{black}

\section{Impact of Prediction Error in PAC-EKF}
\label{sec:ErrorProp}

As illustrated in Fig.~\ref{fig:PAC-EKF}, the predicted state plays a central role in the proposed PAC-EKF framework. 
It is used not only for the EKF update but also for the analog combiner design. 
Importantly, both operations rely on the \emph{observation Jacobian} evaluated at the predicted state.  
As a result, prediction error affects both the acquisition of the compressed observation and the subsequent EKF update, potentially leading to error propagation in the recursive tracking process.  
In this section, we discuss the underlying mechanisms of this effect and the factors that influence its severity. To facilitate the discussion, we take the SVD-PE combiner as an illustrative example, and distinguish between the observation Jacobian evaluated at the true state and that evaluated at the predicted state, denoted by $\mathbf{B}_k^{\rm true}$ and $\mathbf{B}_k^{\rm pred}$, respectively.  

\vspace{-2mm}

\subsection{Fundamental Impact of Prediction Error}\label{subsec:impact}

Prediction error introduces a mismatch between the predicted and true observation Jacobians, i.e., $\mathbf{B}_k^{\rm pred} \neq \mathbf{B}_k^{\rm true}$. This mismatch affects the tracking process through two coupled aspects, as described below.

\subsubsection{Information Loss due to Analog Combiner Misalignment}

In the PAC-EKF, the analog combiner $\mathbf{Q}_k$ can be designed based on $\mathbf{B}_k^{\rm pred}$ to capture the dominant observation subspace. 
However, the informative observation subspace is determined by $\mathbf{B}_k^{\rm true}$. 
When $\mathbf{B}_k^{\rm pred} \neq \mathbf{B}_k^{\rm true}$, the designed combiner $\mathbf{Q}_k$ becomes misaligned with the true informative subspace, leading to a loss of Fisher information in the compressed observation. 
Consequently, the measurement fed into the EKF update becomes less informative for state estimation. 

\subsubsection{Observation Model Mismatch in EKF Linearization}
In the EKF update stage, the nonlinear observation model is locally linearized around the predicted state ${\mathbf{s}}_{k|k-1}$ using the Jacobian $\mathbf{B}_k^{\rm pred}$, which defines the linear observation model assumed in the update.  
When $\mathbf{B}_k^{\rm pred} \neq \mathbf{B}_k^{\rm true}$, this linearization becomes inaccurate, resulting in a mismatch between the assumed and true observation models. 
Consequently, the EKF update may fail to effectively correct the prediction error.

 \vspace{-2mm}
 
\subsection{Behavior in Small- and Large-Error Regimes}\label{subsec:Small-Large}

\begin{figure}
    \vspace{-3mm}
    \centering \includegraphics[width=0.8\columnwidth]{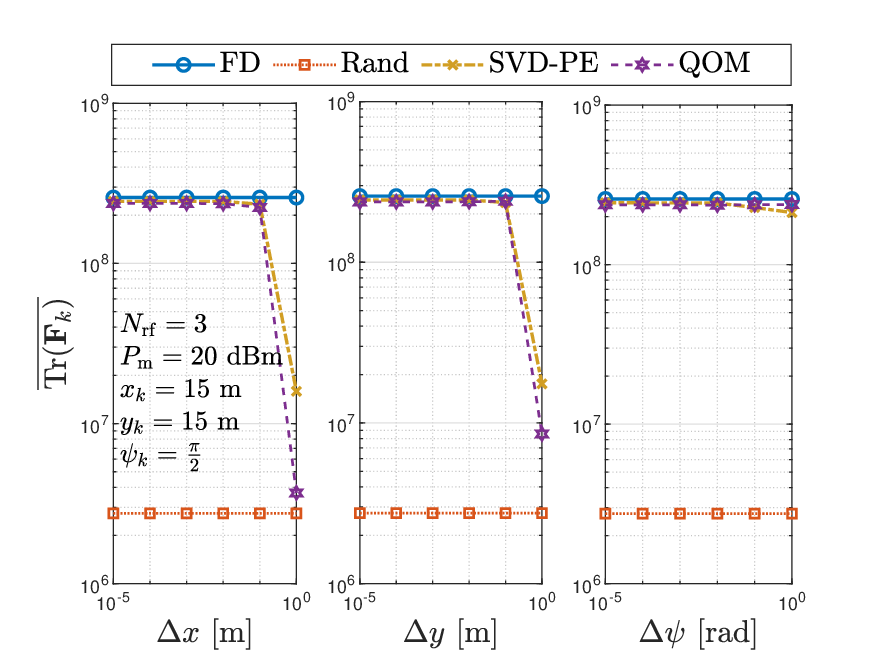}
    % \vspace{-2mm}
     \caption{The impact of prediction error on the Fisher information of the compressed observation. Here, $\mathbf{B}_k$ is evaluated at the true pose $\mathbf{p}_k$, and $\mathbf{Q}_k$ is designed based on the predicted pose $\mathbf{p}_k+\Delta \mathbf{p}$, where the prediction error is set as $\Delta \mathbf{p}=[\Delta x, 0,0]^\mathsf{T}$, $\Delta \mathbf{p}=[0, \Delta y,0]^\mathsf{T}$, and $\Delta \mathbf{p}=[0, 0,\Delta \psi]^\mathsf{T}$, respectively. The metric $\overline{{\rm Tr}(\mathbf{F}_k)}$ denotes the average trace of $\mathbf{F}_k$ over $50$ Monte Carlo realizations of the uplink pilot signals. The default simulation setting follows Sec.~\ref{subsec:simu}, unless otherwise specified.} 
    \label{fig:PredError_true}
\end{figure}

As discussed in Sec.~\ref{subsec:impact}, the impact of the prediction error is fundamentally governed by the mismatch between $\mathbf{B}_k^{\rm pred}$ and $\mathbf{B}_k^{\rm true}$. 
This mismatch determines (i) the accuracy of the locally linearized observation model used in the EKF update and (ii) the alignment between the analog combiner and the true informative observation subspace.  
Based on this, the system behavior can be characterized under two regimes.

\subsubsection{Small Prediction Error (Valid Local Regime)}\label{subsec:SmallError}

When the prediction error is small, the mismatch between $\mathbf{B}_k^{\rm pred}$ and $\mathbf{B}_k^{\rm true}$ is limited. 
In this regime, the locally linearized observation model remains accurate, and the dominant subspaces associated with $\mathbf{B}_k^{\rm pred}$ and $\mathbf{B}_k^{\rm true}$ are well aligned. 
%%%%%%%%%%%%%
As a result, the predictive analog combiners (e.g., SVD-PE and QOM) can effectively capture the informative observation subspace, and the compressed observation retains most of the Fisher information. 
This is consistent with Fig.~\ref{fig:PredError_true}, where a small prediction error in $x_k$, $y_k$, or $\psi_k$ (e.g., $\Delta x\le 0.1~\rm m$, $\Delta y\le 0.1~\rm m$, or $\Delta \psi\le 0.1~\rm rad$) leads to negligible information loss.
Therefore, the EKF update can reliably exploit the measurement to correct the prediction error, thereby maintaining bounded tracking error over time. 

\subsubsection{Large Prediction Error (Error Propagation Regime)}\label{subsec:LargeError}

When the prediction error becomes large, the mismatch between $\mathbf{B}_k^{\rm pred}$ and $\mathbf{B}_k^{\rm true}$ becomes significant, leading to both information loss and model mismatch. 

First, the analog combiner becomes misaligned with the true informative observation subspace, resulting in substantial information loss in the compressed observation. 
As shown in Fig.~\ref{fig:PredError_true}, a large prediction error in $x_k$, $y_k$, or $\psi_k$ (e.g., on the order of $1~\rm m$ or $1~\rm rad$) leads to severe degradation in the Fisher information.
Second, the local linearization of the observation model becomes inaccurate, resulting in a mismatch between the linearized and true observation models in the EKF update. 
Importantly, this limitation is intrinsic to the EKF: even with full-dimensional observations (i.e., without RF-domain compression), an inaccurate local observation model prevents effective correction of the prediction error.  
As a consequence, the residual estimation error is not properly corrected and propagates to subsequent time steps, leading to progressively degraded predictions and updates, i.e., error propagation.

% Notably, these two effects are inherently coupled: the loss of measurement information further limits the ability of the EKF update to correct the prediction error, 
% while an inaccurate local observation model prevents effective utilization of the available measurement information.

This error propagation mechanism also highlights the role of analog combiner design. 
When the prediction error is small, the proposed predictive combiners preserve most of the measurement information, whereas uninformative combiners (e.g., random design) provide weak measurements even at the initial stage, making error propagation more likely.  
This effect can be seen in Fig.~\ref{fig:RMSE_time_k}, where the Rand combiner exhibits progressively increasing estimation error over time, while the proposed designs maintain stable performance.

 \vspace{-2mm}
 
\subsection{Factors Affecting Error Propagation}\label{subsec:factors}

Based on the above discussion, the severity of error propagation depends on whether the EKF operates within the valid local linearization region and how much observation information is preserved under analog combining.

%%%%%%
The impact of error propagation is exacerbated by:
(i) larger tracking interval $\tau$, 
especially under strongly nonlinear state transition or observation models, which reduces the validity region of the local model;
(ii) lower SNR, which reduces the Fisher information of the observation; 
(iii) uninformative analog combining, such as random compression or insufficient RF chains; 
(iv) larger process noise, which increases the prediction uncertainty.

Correspondingly, the risk of error propagation can be mitigated by:
(i) using a sufficiently small tracking interval to limit the prediction error, which is feasible in \ac{5G-NR} systems due to their support for short reference-signal periodicities;
(ii) increasing the pilot transmit power to improve the SNR; 
(iii) employing informative analog combiners such as the proposed SVD-PE and QOM combiners or increasing the number of RF chains; 
(iv) periodically or adaptively triggering a higher-accuracy re-estimation stage when the predicted error becomes too large. In practice, such re-estimation can be implemented by leveraging flexible uplink resource allocation in \ac{5G-NR} systems to obtain multiple-pilot-aided observations.

\vspace{-2mm}

\subsection{Operating Conditions and Practical Implications}

The analysis in Sec.~\ref{subsec:Small-Large} reveals that error propagation is fundamentally driven by the breakdown of the locally linearized observation model in the EKF, while the analog combiner determines how much observation information is preserved when the EKF operates within its valid local regime.  
In particular, the predictive analog combiner cannot compensate for an inaccurate local model, but it can effectively retain the informative observation subspace when the EKF's linearized observation model remains valid. 

In practical communication systems with sufficiently small tracking intervals (e.g., millisecond-level updates), the state variation between consecutive tracking time instants is typically limited, so that the EKF operates within its valid local regime. 
Under such conditions, the proposed PAC-EKF can nearly achieve fully digital performance while significantly reducing the number of RF chains.

\color{black}

\vspace{-2mm}

\section{Conclusions}\label{sec:conclusion}

We studied NF position and orientation tracking for a multi-antenna MS served by a multi-antenna BS with limited RF chains, where the analog combiner compresses uplink observations. To preserve informative observations, we proposed a PAC-EKF framework, designing the analog combiner in the EKF prediction stage, before uplink pilot reception. This predictive approach leverages the temporal correlation in the MS pose trajectory to provide the EKF update stage with more informative compressed observations. 
By analyzing the fundamental performance limits of the NF pose tracking, we revealed how the NF channel structure and the analog combiner design influence the pose-relevant information contained in the uplink observation, and how this information scales with the array size, SNR, and MS pose.
Guided by these insights, we developed two low-complexity combiner designs under the unit-modulus constraint, namely a FIM-based SVD-PE scheme and a geometry-inspired QOM scheme.
% (i) a \ac{FIM}-based design that captures the informative signal components by extracting phases from the SVD of the observation Jacobian; (ii) a geometry-inspired design that captures non-redundant spatial information by selecting \acp{QOM} (i.e., well-separated viewpoints) based on the NF beamfocusing effect. 
\textcolor{black}{Numerical results show that the proposed designs nearly achieve the fully digital tracking accuracy with 3 RF chains, outperform the random combiner with up to $20~\rm dBm$ lower transmit power, and remain effective under 4-bit phase quantization. 
We further show that prediction error affects both the analog combiner design and the EKF update in the PAC-EKF framework, while reliable performance is maintained when the EKF operates within its valid local linearization regime.}

% We also show that prediction error affects both the analog combiner design and the EKF update through Jacobian mismatch, potentially leading to error propagation when the local linearization becomes invalid.

\textcolor{black}{
For future work, the proposed predictive information-preserving design can be extended beyond the current receiver-side implementation by incorporating transmit beamforming as an additional degree of freedom. In particular, predictive transmit beamforming can be leveraged to actively shape the observation, leading to a joint transmit-receive design for enhancing the resulting Fisher information in tracking.
Moreover, extending the framework from \ac{LoS} scenarios to \ac{NLoS} scenarios is another important direction.  
Techniques such as \ac{RIS} to create a cascaded \ac{LoS} link~\cite{NF-RIS}, or wavefront engineering (e.g., Airy beams) to enable curved propagation paths~\cite{Airy}, can be leveraged to support tracking when the direct LoS channel is blocked.
}

\appendices

\vspace{-2mm} 

\section{Proof of Proposition~\ref{prop:FIMdata}}\label{app:dataFIM}
Since the following analysis applies to any time instant $k \in [1,K]$, we omit the subscript $k$ for notational simplicity.
% As in Appendix~\ref{app:jacobian}, we omit the subscript $k$ for notational simplicity. 
Then, the observation model in \eqref{eq:zk-observe} becomes
\begin{align}\label{eq:z_observe}
    \mathbf{z}
	=  \mathbf{Q}\mathbf{b}(\mathbf{s})+\mathbf{Q}\mathbf{n}_{\rm o},
\end{align}
where $\mathbf{Qn}_{\rm o}\sim\mathcal{CN}(0,\mathbf{Q}\mathbf{N}_{\rm o}\mathbf{Q}^{\mathsf{H}})$ and $\mathbf{N}_{\rm o}=\sigma_{\rm o}^2\mathbf{I}$.
The corresponding log-likelihood function is
 \begin{align}\label{eq:logp(z|s)}
 \begin{split}
       &\log p(\mathbf{z}|\mathbf{s})
		= -\log \left({\pi^{N_{\rm rf}} {\rm det}(\mathbf{M}^{-1}) }\right)  - \mathbf{r}^{\mathsf{H}} \mathbf{M} \mathbf{r},
 \end{split}
\end{align}
where $\mathbf{r} \triangleq \mathbf{z}-\mathbf{Q}\mathbf{b}(\mathbf{s})$ and $\mathbf{M} \triangleq (\mathbf{Q}\mathbf{N}_{\rm o}\mathbf{Q}^{\mathsf{H}})^{-1}$. 
With \eqref{eq:logp(z|s)}, the gradient of the log-likelihood function w.r.t. $\mathbf{s}$ is 
\begin{align}\label{eq:score1}
        \mathbf{g}=\nabla_{\mathbf{s}}\log p(\mathbf{z}|\mathbf{s}) &= -\frac{\partial \mathbf{r}^{\mathsf{H}} \mathbf{M}\mathbf{r}}{ \partial \mathbf{s}} 
        \overset{(a)}{=} -2 {\rm Re} \left\{  \left(\frac{\partial \mathbf{r}}{ \partial \mathbf{s}^{\mathsf{T}} } \right)^{\mathsf{H}} \mathbf{M} \mathbf{r}  \right\}
 \nonumber\\& \overset{(b)}{=} 2 {\rm Re} \left\{ \mathbf{B}^{\mathsf{H}} \mathbf{Q}^{\mathsf{H}} \mathbf{M}   \mathbf{r}\right\},
 % \overset{(b)}{=} 2 {\rm Re} \left\{ \mathbf{B}^{\mathsf{H}} \mathbf{Q}^{\mathsf{H}} \mathbf{M}   (\mathbf{z}-\mathbf{Q}\mathbf{b}(\mathbf{s}))\right\},
\end{align}
where $\mathbf{B}\triangleq\nabla_{\mathbf{s}} \mathbf{b}(\mathbf{s})$, (a) is from 
% $\frac{\partial\mathbf{r}^{\mathsf{H}}\mathbf{M}\mathbf{r}}{\partial \mathbf{s}}=\left[ \frac{\partial\mathbf{r}^{\mathsf{H}}\mathbf{M}\mathbf{r}}{\partial s_1},...,\frac{\partial \mathbf{r}^{\mathsf{H}}\mathbf{M}\mathbf{r} }{\partial s_5}\right]^{\mathsf{T}}$ for $\mathbf{s}=[s_1,...,s_5]^{\mathsf{T}}$ and 
$\frac{\partial \mathbf{r}^{\mathsf{H}}\mathbf{M}\mathbf{r}}{\partial s_i}= \frac{\partial \mathbf{r}^{\mathsf{H}}}{\partial s_i} \mathbf{M}\mathbf{r} + \mathbf{r}^{\mathsf{H}}\mathbf{M}\frac{\partial \mathbf{r}}{\partial s_i} = 2 {\rm Re} \left\{ \frac{\partial \mathbf{r}^{\mathsf{H}}}{\partial s_i} \mathbf{M} \mathbf{r}\right\}$ for $\mathbf{s}=[s_1,...,s_5]^{\mathsf{T}}$ and $i\in\{1,...,5\}$, and (b) is from $\frac{\partial \mathbf{r} }{ \partial \mathbf{s}^{\mathsf{T}} }=\frac{\partial (\mathbf{z}-\mathbf{Q}\mathbf{b}(\mathbf{s}) ) }{ \partial \mathbf{s}^{\mathsf{T}} }
        = -\mathbf{Q}\frac{\partial \mathbf{b}(\mathbf{s}) }{ \partial \mathbf{s}^{\mathsf{T}} }= - \mathbf{Q} \mathbf{B}$.
% where $\mathbf{B}=\nabla_{\mathbf{s}} \mathbf{b}(\mathbf{s})$ is given in Appendix~\ref{app:Hjacobian}.
With $\mathbf{b}(\mathbf{s})=\mathbf{H}(\mathbf{p}) \mathbf{x}$, 
we obtain $\nabla_{\mathbf{s}} \mathbf{b}(\mathbf{s})
= \left[ \frac{\partial \mathbf{H}(\mathbf{p})}{\partial x}\mathbf{x}, 
 \frac{\partial \mathbf{H}(\mathbf{p})}{\partial y} \mathbf{x}, 
 \frac{\partial \mathbf{H}(\mathbf{p})}{\partial \psi} \mathbf{x}, 
\mathbf{0}, \mathbf{0}\right]$.

Following the definition in \eqref{eq:Fdata1}, the \ac{FIM} $\mathbf{F}=\mathbb{E}_{\mathbf{z}|\mathbf{s}}[ \mathbf{g}\mathbf{g}^{\mathsf{T}} ]$ is
\begin{align}\label{eq:Fd}
\mathbf{F}&=\mathbb{E}_{\mathbf{z}|\mathbf{s}}\left[\left(2 {\rm Re} \left\{  \mathbf{B}^{\mathsf{H}}  \mathbf{Q}^{\mathsf{H}}  \mathbf{M}  \mathbf{r}\right\}\right) \left(2 {\rm Re} \left\{ \mathbf{B}^{\mathsf{H}}  \mathbf{Q}^{\mathsf{H}}  \mathbf{M}  \mathbf{r} \right\}\right)^{\mathsf{T}} \right]
\nonumber\\&\overset{(a)}{=} 2{\rm Re}\{\mathbb{E}_{\mathbf{z}|\mathbf{s}}\left[
        \mathbf{B}^{\mathsf{H}} \mathbf{Q}^{\mathsf{H}} \mathbf{M} \mathbf{r} \mathbf{r} ^{\mathsf{H}}  \mathbf{M}\mathbf{Q}\mathbf{B}
 \right]\}
\nonumber\\&\overset{(b)}{=} 2{\rm Re}\left\{
        \mathbf{B}^{\mathsf{H}} \mathbf{Q}^{\mathsf{H}} \mathbf{M} \mathbf{Q}\mathbf{N}_{\rm o} \mathbf{Q}^{\mathsf{H}} \mathbf{M}\mathbf{Q}\mathbf{B}
 \right\}
 \nonumber\\&\overset{(c)}{=}\frac{2}{\sigma_{\rm o}^2}{\rm Re}\{ \mathbf{B}^{\mathsf{H}} \mathbf{Q}^{\mathsf{H}} (\mathbf{Q}\mathbf{Q}^{\mathsf{H}})^{-1} \mathbf{Q} \mathbf{B} \},
\end{align}
where (a) is from $\mathbb{E}[{\rm Re}\{\mathbf{u}\}({\rm Re}\{\mathbf{u}\})^\mathsf{T}]=\frac{1}{2}{\rm Re}\{\mathbb{E}[\mathbf{u}\mathbf{u}^\mathsf{H}]\}$ with $\mathbf{u}= \mathbf{B}^\mathsf{H} \mathbf{Q}^\mathsf{H}  \mathbf{M} \mathbf{r} $ and $\mathbf{r} \sim \mathcal{CN} (\mathbf{0},\mathbf{Q}\mathbf{N}_{\rm o} \mathbf{Q}^\mathsf{H})$, (b) is from $\mathbb{E}[\mathbf{r}\mathbf{r}^\mathsf{H}]=\mathbf{Q}\mathbf{N}_{\rm o} \mathbf{Q}^\mathsf{H}=\mathbf{M}^{-1}$, and (c) is from $\mathbf{N}_{\rm o}=\sigma_{\rm o}^2 \mathbf{I}$. 

\color{black}

\vspace{-2mm}

\section{Conjugate Gradient}\label{app:conjugate}

We adopt a differential approach to derive $\frac{\partial{\rm Tr}( \mathbf{D}{\mathbf{P}}_{k|k})}{\partial \mathbf{Q}_k^*}$ in this appendix.
For notational simplicity, define $\mathbf{R} \triangleq {\mathbf{P}}_{k|k}^{-1}$, $\mathbf{S} \triangleq {\mathbf{P}}_{k|k-1}^{-1}$, 
$\mathbf{F}\triangleq \mathbf{F}_{k}$,
$\mathbf{B} \triangleq \mathbf{B}_{k|k-1}$, and $\mathbf{Q} \triangleq \mathbf{Q}_k$.
% According to some basic differentiation rules for complex-value matrices [32], the differential of J(VRF) can be expressed as
Following \cite[Lemma 1]{hybridMMSE}, the differential of ${\rm Tr}( \mathbf{D}\mathbf{R}^{-1})$ w.r.t. $\mathbf{Q}^*$ is
\begin{align}\label{eq:derivative1}
    {\rm d}({\rm Tr}( \mathbf{D}\mathbf{R}^{-1}))={\rm Tr} \left( \frac{\partial {\rm Tr}( \mathbf{D}\mathbf{R}^{-1})}{\partial \mathbf{Q}^*} {\rm d}(\mathbf{Q}^{\mathsf{H}}) \right),
\end{align}
where ${\rm d}(\cdot)$ denotes the differential w.r.t. $\mathbf{Q}^*$ while treating $\mathbf{Q}$ as constant.
% taking $\mathbf{Q}$ as a constant matrix during the derivation of $\frac{\partial{\rm Tr}( \mathbf{D}\mathbf{R}^{-1})}{\partial \mathbf{Q}^*}$.
We next explicitly derive ${\rm d}(\operatorname{Tr}(\mathbf{D}\mathbf{R}^{-1}))$ and express it in a form that is linear in ${\rm d}(\mathbf{Q}^{\mathsf{H}})$, based on some basic differentiation rules for complex-valued matrices.  

% we next derive ${\rm d}({\rm Tr}( \mathbf{D}\mathbf{R}^{-1}))$ 
Using the identity ${\rm d}(\mathbf{R}^{-1})=-\mathbf{R}^{-1}{\rm d}(\mathbf{R})\mathbf{R}^{-1}$, we obtain 
\begin{align}\label{eq:tr(DW)0}
   {\rm d}({\rm Tr}( \mathbf{D}\mathbf{R}^{-1}))=-{\rm Tr}(\mathbf{R}^{-1} \mathbf{D}\mathbf{R}^{-1}{\rm d}(\mathbf{R}) ).
\end{align}
From \eqref{eq:P(k|k)}, $ \mathbf{R}=\mathbf{S}+\mathbf{F}$, which can be further expressed as 
\begin{align}\label{eq:W}
    \mathbf{R}&\overset{(a)}{=}\mathbf{S}+\frac{2}{\sigma_{\rm o}^2}{\rm Re}\{ \mathbf{B}^{\mathsf{H}} \mathbf{P}_{\mathbf{Q}}\mathbf{B} \}
    \overset{(b)}{=}\mathbf{S}+\frac{1}{\sigma_{\rm o}^2}(\mathbf{G}+\mathbf{G}^*),
\end{align} 
where $\mathbf{P}_{\mathbf{Q}}=\mathbf{Q}^{\mathsf{H}} (\mathbf{Q}\mathbf{Q}^{\mathsf{H}})^{-1} \mathbf{Q}$, (a) is from \eqref{eq:Fdata1}, and (b) is from $\mathbf{G} \triangleq \mathbf{B}^{\mathsf{H}} \mathbf{P}_{\mathbf{Q}} \mathbf{B}$.
Since $\mathbf{S}$ is independent of $\mathbf{Q}$,  we have 
\begin{align}
    {\rm d}(\mathbf{R})=\frac{1}{\sigma_{\rm o}^2}{\rm d}(\mathbf{G})+\frac{1}{\sigma_{\rm o}^2}{\rm d}(\mathbf{G}^*).
\end{align}
%%%%%%%%%%%%%%%%%%%
Substituting ${\rm d}(\mathbf{R})$ into  \eqref{eq:tr(DW)0} yields
\begin{align}\label{eq:tr(DW)1}
    {\rm d}({\rm Tr}( \mathbf{D}\mathbf{R}^{-1}))&=-\frac{1}{\sigma_{\rm o}^2}\big({\rm Tr}( \mathbf{R}^{-1}\mathbf{D}\mathbf{R}^{-1}  {\rm d}(\mathbf{G})) \nonumber\\& \qquad +{\rm Tr}( \mathbf{R}^{-1}\mathbf{D}\mathbf{R}^{-1}  {\rm d}(\mathbf{G}^*))\big).
\end{align}
% Since a matrix and its transpose have the same trace, we have
To express the second term in \eqref{eq:tr(DW)1} in a form involving ${\rm d}(\mathbf{G})$, we rewrite it using basic trace identities as
\begin{align}\label{eq:dG*}
    & {\rm Tr}( \mathbf{R}^{-1}\mathbf{D}\mathbf{R}^{-1}  {\rm d}(\mathbf{G}^*))\!=\!{\rm Tr}(   ({\rm d}(\mathbf{G}^*))^{\mathsf{T}} (\mathbf{R}^{-1}\mathbf{D}\mathbf{R}^{-1})^{\mathsf{T}})
    \nonumber\\ & \overset{(a)}{=}{\rm Tr}(   {\rm d}(\mathbf{G}^{\mathsf{H}}) \mathbf{R}^{-1}\mathbf{D}\mathbf{R}^{-1})
   \overset{(b)}{=}\!{\rm Tr}(   \mathbf{R}^{-1}\mathbf{D}\mathbf{R}^{-1}  {\rm d}(\mathbf{G}^{\mathsf{H}})), 
\end{align}
where (a) is from $({\rm d}(\mathbf{G}^*))^{\mathsf{T}}= {\rm d}(\mathbf{G}^{\mathsf{H}})$, $\mathbf{R}=\mathbf{R}^{\mathsf{T}}$, and $\mathbf{D}=\mathbf{D}^{\mathsf{T}}$, and (b) is from the cyclic property of the trace.
With \eqref{eq:dG*}, we can further express \eqref{eq:tr(DW)1} as
\begin{align}\label{eq:tr(DW)2}
    &{\rm d}({\rm Tr}( \mathbf{D}\mathbf{R}^{-1}))\nonumber \\&=-\frac{1}{\sigma_{\rm o}^2}{\rm Tr}( \mathbf{R}^{-1}\mathbf{D}\mathbf{R}^{-1}  {\rm d}(\mathbf{G})) -\frac{1}{\sigma_{\rm o}^2} {\rm Tr}(   \mathbf{R}^{-1}\mathbf{D}\mathbf{R}^{-1} {\rm d}(\mathbf{G}^{\mathsf{H}}))
    \nonumber\\&\overset{(a)}{=}-\frac{2}{\sigma_{\rm o}^2}{\rm Tr}( \mathbf{R}^{-1}\mathbf{D}\mathbf{R}^{-1}  {\rm d}(\mathbf{G})),
\end{align} 
where (a) follows from the Hermitian property of $\mathbf{G}$, i.e., ${\rm d}(\mathbf{G}^{\mathsf{H}}) = {\rm d}(\mathbf{G})$. 
%%%%%%%%%%
% Recall $\mathbf{G})=  \mathbf{B}^{\mathsf{H}} \mathbf{Q}^{\mathsf{H}} (\mathbf{Q}\mathbf{Q}^{\mathsf{H}})^{-1} \mathbf{Q}\mathbf{B}$
Then, taking $\mathbf{Q}$ as constant, the differential of $\mathbf{G}=  \mathbf{B}^{\mathsf{H}} \mathbf{Q}^{\mathsf{H}} (\mathbf{Q}\mathbf{Q}^{\mathsf{H}})^{-1} \mathbf{Q}\mathbf{B}$ w.r.t. $\mathbf{Q}^*$ is given by
\begin{align}\label{eq:dG}
     {\rm d}(\mathbf{G})
    % = {\rm d}(\mathbf{B}^{\mathsf{H}} \mathbf{Q}^{\mathsf{H}} (\mathbf{Q}\mathbf{Q}^{\mathsf{H}})^{-1} \mathbf{Q}\mathbf{B})
    % \nonumber\\&
    &=\mathbf{B}^{\mathsf{H}}{\rm d}(\mathbf{Q}^{\mathsf{H}}) (\mathbf{Q}\mathbf{Q}^{\mathsf{H}})^{-1} \mathbf{Q}\mathbf{B} +  \mathbf{B}^{\mathsf{H}} \mathbf{Q}^{\mathsf{H}}  {\rm d} ((\mathbf{Q}\mathbf{Q}^{\mathsf{H}})^{-1} ) \mathbf{Q}\mathbf{B}
    % \nonumber\\&\overset{(a)}{=}
    % \mathbf{B}^{\mathsf{H}}{\rm d}(\mathbf{Q}^{\mathsf{H}}) (\mathbf{Q}\mathbf{Q}^{\mathsf{H}})^{-1} \mathbf{Q}\mathbf{B} -
    % \mathbf{B}^{\mathsf{H}} \mathbf{Q}^{\mathsf{H}}  (\mathbf{Q}\mathbf{Q}^{\mathsf{H}})^{-1} \mathbf{Q} {\rm d} (\mathbf{Q}^{\mathsf{H}}) (\mathbf{Q}\mathbf{Q}^{\mathsf{H}})^{-1} \mathbf{Q}\mathbf{B}
    \nonumber\\&\overset{(a)}{=} \mathbf{B}^{\mathsf{H}} (\mathbf{I}-\mathbf{P}_{\mathbf{Q}}) {\rm d} (\mathbf{Q}^{\mathsf{H}}) (\mathbf{Q}\mathbf{Q}^{\mathsf{H}})^{-1} \mathbf{Q}\mathbf{B},
\end{align}
where (a) is from 
\begin{align}
    {\rm d} ((\mathbf{Q}\mathbf{Q}^{\mathsf{H}})^{-1} )&=-(\mathbf{Q}\mathbf{Q}^{\mathsf{H}})^{-1} {\rm d} (\mathbf{Q}\mathbf{Q}^{\mathsf{H}}) (\mathbf{Q}\mathbf{Q}^{\mathsf{H}})^{-1} 
    \nonumber\\&= -(\mathbf{Q}\mathbf{Q}^{\mathsf{H}})^{-1} \mathbf{Q}{\rm d} (\mathbf{Q}^{\mathsf{H}}) (\mathbf{Q}\mathbf{Q}^{\mathsf{H}})^{-1}.
\end{align}
Substituting \eqref{eq:dG} into \eqref{eq:tr(DW)2}, we obtain
\begin{align}\label{eq:tr(DW)3}
    &{\rm d}({\rm Tr}( \mathbf{D}\mathbf{R}^{-1})) 
    \\&= \frac{2}{\sigma_{\rm o}^2}{\rm Tr}\left( \mathbf{R}^{-1}\mathbf{D}\mathbf{R}^{-1}  \mathbf{B}^{\mathsf{H}} (\mathbf{P}_{\mathbf{Q}}-\mathbf{I}) {\rm d} (\mathbf{Q}^{\mathsf{H}}) (\mathbf{Q}\mathbf{Q}^{\mathsf{H}})^{-1} \mathbf{Q}\mathbf{B}\right)
    \nonumber\\&= {\rm Tr}\left( \frac{2}{\sigma_{\rm o}^2} (\mathbf{Q}\mathbf{Q}^{\mathsf{H}})^{-1} \mathbf{Q}\mathbf{B}\mathbf{R}^{-1}\mathbf{D}\mathbf{R}^{-1}  \mathbf{B}^{\mathsf{H}} (\mathbf{P}_{\mathbf{Q}}-\mathbf{I}) {\rm d} (\mathbf{Q}^{\mathsf{H}}) \right).\nonumber
\end{align}

% \vspace{-2mm}

Finally, by comparing \eqref{eq:derivative1} with \eqref{eq:tr(DW)3}, we obtain 
\begin{align} 
    &\frac{\partial {\rm Tr}( \mathbf{D}\mathbf{R}^{-1})}{\partial \mathbf{Q}^*}=\frac{2}{\sigma_{\rm o}^2} (\mathbf{Q}\mathbf{Q}^{\mathsf{H}})^{-1} \mathbf{Q}\mathbf{B}\mathbf{R}^{-1}\mathbf{D}\mathbf{R}^{-1}  \mathbf{B}^{\mathsf{H}} (\mathbf{P}_{\mathbf{Q}}-\mathbf{I}).\nonumber
\end{align}
% \vspace{-2mm}
% \textcolor{red}{which completes the derivation.} 
This completes the derivation. 
\color{black}
 
\vspace{-2mm}

\section{Proof of Proposition~\ref{prop:Jacobian}}\label{app:SVD_Hjacobian}
 
% \textcolor{red}{As in Appendix~\ref{app:dataFIM}, we omit the subscript $k$ for notational simplicity.}
% Moreover, the following derivations are conducted under NF assumptions (A1)-(A3.1).
From \eqref{eq:Hexact}, the $(\bar{n}_{\rm b},\bar{n}_{\rm m})$-th element of $\frac{\partial \mathbf{H} (\mathbf{p}) }{\partial \mu}$ is
\begin{align}\label{eq:dH}
    \left[ \frac{\partial \mathbf{H} (\mathbf{p}) }{\partial \mu} \right]_{\bar{n}_{\rm b},\bar{n}_{\rm m}}= 
\frac{\partial h_{n_{\rm b},n_{\rm m}}}{\partial \mu} = \frac{\partial h_{n_{\rm b},n_{\rm m}}}{\partial r_{n_{\rm b},n_{\rm m}} } \frac{\partial r_{n_{\rm b},n_{\rm m}}}{\partial \mu},
\end{align}
where $\mu\in\{x,y,\psi\}$,  $\bar{n}_{\rm m}\triangleq n_{\rm m}+\bar N_{\rm m}+1$, $\bar{n}_{\rm b}\triangleq n_{\rm b}+\bar N_{\rm b}+1$, $h_{n_{\rm b},n_{\rm m}}\!\!=\left[  \mathbf{H} \right]_{\bar{n}_{\rm b},\bar{n}_{\rm m}}$, and
 \begin{align} %\label{eq:dh}
  \nonumber\frac{\partial h_{n_{\rm b},n_{\rm m}}}{\partial r_{n_{\rm b},n_{\rm m}} } &= -\frac{\lambda}{4\pi r^2_{n_{\rm b},n_{\rm m}}} \left(1+{\rm j} \frac{2\pi}{\lambda} r_{n_{\rm b},n_{\rm m}} \right) e^{-{\rm j}\frac{2\pi}{\lambda} r_{n_{\rm b},n_{\rm m}} }.
\end{align}  
From \eqref{eq:r(nb,nm)}, the partial derivative of $r_{n_{\rm b},n_{\rm m}}$ w.r.t. $\mu$ is 
 \begin{align}\label{eq:dr}
    \frac{\partial r_{n_{\rm b},n_{\rm m}}}{\partial x} &= \frac{x
+n_{\rm m} d_{\rm m}\cos\psi} { r_{n_{\rm b},n_{\rm m}} },\nonumber\\
\frac{\partial r_{n_{\rm b},n_{\rm m}}}{\partial y} &= \frac{y
+n_{\rm m} d_{\rm m} \sin\psi -n_{\rm b} d_{\rm b}} { r_{n_{\rm b},n_{\rm m}} }, \\
        \frac{\partial r_{n_{\rm b},n_{\rm m}}}{\partial \psi} & 
=-\frac{\partial r_{n_{\rm b},n_{\rm m}}}{\partial x}n_{\rm m} d_{\rm m}\sin\psi+\frac{\partial r_{n_{\rm b},n_{\rm m}}}{\partial y}n_{\rm m} d_{\rm m}\cos\psi. \nonumber
\end{align}
% Substituting \eqref{eq:dh} and \eqref{eq:dr} into \eqref{eq:dH}, we obtain $\nabla_{\mathbf{s}} \mathbf{b}(\mathbf{s})$.

\subsubsection{Analysis of $\mathbf{J}_x=\frac{\partial \mathbf{H} }{\partial x}$}
With \eqref{eq:dH}-\eqref{eq:dr}, we obtain
% the $(\bar{n}_{\rm b}, \bar{n}_{\rm m})$-th element of $\mathbf{J}_x$ can be written as 
\begin{align}\label{eq:Jx-rn}
      \left[  \mathbf{J}_x \right]_{\bar{n}_{\rm b},\bar{n}_{\rm m}} & 
      =\frac{a_{n_{\rm b},n_{\rm m}} \lambda}{4\pi r_{n_{\rm b},n_{\rm m}}  } e^{-{\rm j}\frac{2\pi}{\lambda} r_{n_{\rm b},n_{\rm m}} },
\end{align}
where $a_{n_{\rm b},n_{\rm m}} = - (\frac{1}{r_{n_{\rm b},n_{\rm m}} } + {\rm j} \frac{2\pi}{\lambda}){\frac{x
+n_{\rm m} d_{\rm m}\cos\psi} { r_{n_{\rm b},n_{\rm m}}  }}$,  and $\frac{a_{n_{\rm b},n_{\rm m}}\lambda}{4\pi r_{n_{\rm b},n_{\rm m}}  }$ is the amplitude of $\left[\mathbf{J}_x \right]_{\bar{n}_{\rm b},\bar{n}_{\rm m}}$.
Based on the uniform power model in (A1), where the amplitude variation across antenna elements is negligible, 
$r_{n_{\rm b},n_{\rm m}}$ involved in the amplitude of $\left[\mathbf{J}_x \right]_{\bar{n}_{\rm b},\bar{n}_{\rm m}}$ can be replaced by the MS-BS distance $r$, i.e., 
\begin{align}\label{eq:Jx-r}
        \left[  \mathbf{J}_x \right]_{\bar{n}_{\rm b},\bar{n}_{\rm m}} &= - \left(\frac{1}{r}+{\rm j} \frac{2\pi}{\lambda}  \right)  \frac{x
+n_{\rm m} d_{\rm m}\cos\psi} { r } h_{n_{\rm b},n_{\rm m}} 
\nonumber\\& \overset{(a)}{=} \eta  \frac{x
} { r } \left(1+ \varepsilon_{x}\right) h_{n_{\rm b},n_{\rm m}}, 
\end{align}
where $\eta \triangleq -\left(\frac{1}{r}+{\rm j} \frac{2\pi}{\lambda}  \right) $,  $h_{n_{\rm b},n_{\rm m}}=\frac{\lambda}{4\pi r} e^{-{\rm j}\frac{2\pi}{\lambda} r_{n_{\rm b},n_{\rm m}} }$, and (a) is from $\varepsilon_{x}  \triangleq \frac{
n_{\rm m} d_{\rm m}\cos\psi} { x }$ for $x\neq0$. Since $n_{\rm m}\in[-\bar N_{\rm m},\bar N_{\rm m}]$, $\left|\varepsilon_{x}\right|$ satisfies
\begin{align}\label{eq:x/r0}
        \left|\varepsilon_{x}\right| & =\left|\frac{
n_{\rm m} d_{\rm m}\cos\psi} { x }\right| \le  \left|\frac{\bar N_{\rm m} d_{\rm m} \cos\psi }{x}\right|= \frac{D_{\rm m}}{2 |x|} \!\overset{(a)}{<}\! \frac{D_{\rm b}}{2 |x|},\!
\end{align}
where $D_{\rm m}=2\bar N_{\rm m} d_{\rm m}$ is the MS array aperture, (a) is from $D_{\rm m}<D_{\rm b}$ under assumption (A2).
% that the MS array is smaller than the BS array with $D_{\rm m}<D_{\rm b}$,
% Under assumptions (A2) and (A3.1), 
$\left|\varepsilon_{x}\right|$ is further bounded by
\begin{align}\label{eq:x/r}
         \left|\varepsilon_{x}\right|  
\!\overset{(a)}{<}\! \frac{D_{\rm b}}{1.24\sqrt{\frac{D_{\rm b}^3}{\lambda}}}   \overset{(b)}{=} \! \frac{ 1 }{\sqrt{0.7688  (N_{\rm b}\!-\!1) }}  
\!=\mathcal{O}\left(\frac{1}{\sqrt{N_{\rm b}}}\right),
\end{align}
where (a) is from $|x|>d_{\rm Fn}$ and $d_{\rm Fn}=0.62\sqrt{\frac{D_{\rm b}^3}{\lambda}}$ under assumption (A3.1),
% the MS lies in the radiative \ac{NF} region of the BS ULA with $|x|>d_{\rm Fn}$, where $d_{\rm Fn}=0.62\sqrt{\frac{D_{\rm b}^3}{\lambda}}$ is Fresnel distance, 
and (b) is from $d_{\rm b}=\frac{\lambda}{2}$ and $D_{\rm b} =(N_{\rm b}-1) d_{\rm b}$. 
% Since $\left|\varepsilon_{x}\right|$ in \eqref{eq:x/r} decays as $N_{\rm b}^{-\frac{1}{2}}$, 
Eqs.~\eqref{eq:Jx-r} and \eqref{eq:x/r} indicate that $\lim_{N_{\rm b}\to \infty} \left[  \mathbf{J}_x \right]_{\bar{n}_{\rm b},\bar{n}_{\rm m}} = \eta\frac{x
} { r }  h_{n_{\rm b},n_{\rm m}}$.
Hence,
$\mathbf{J}_x$ can be decomposed into a leading term  $\tilde{\mathbf{J}}_x$ and a residual error term $\mathbf{E}_x$ as 
\begin{align}\label{eq:Jx-H}
    \begin{split}
        \mathbf{J}_x = \eta  \frac{x
} { r } \mathbf{H} + \mathbf{E}_x \triangleq \tilde{\mathbf{J}}_x + \mathbf{E}_x,
    \end{split}
\end{align}
where $ [  \tilde{\mathbf{J}}_x ]_{\bar{n}_{\rm b},\bar{n}_{\rm m}} =  \eta  \frac{x
} { r }  h_{n_{\rm b},n_{\rm m}}$, and $ \left[  \mathbf{E}_x \right]_{\bar{n}_{\rm b},\bar{n}_{\rm m}} = \varepsilon_{x} [\tilde{\mathbf{J}}_x ]_{\bar{n}_{\rm b},\bar{n}_{\rm m}} $. This shows that $\mathbf{J}_x$ is asymptotically a scaled version of $\mathbf{H}$, with vanishing relative error $\frac{\|\mathbf{E}_x\|_{\mathsf{F}}^2}{\|\tilde{\mathbf{J}}_x\|_{\mathsf{F}}^2}=\mathcal{O}(N_{\rm b}^{-1})$.

\subsubsection{Analysis of $\mathbf{J}_y=\frac{\partial \mathbf{H} }{\partial y}$}
Following the steps \eqref{eq:Jx-rn}-\eqref{eq:Jx-r}, 
 \begin{align}\label{eq:Jy-r}
      \left[ \mathbf{J}_y  \right]_{\bar{n}_{\rm b},\bar{n}_{\rm m}} 
& = - \left(\frac{1}{r}+{\rm j} \frac{2\pi}{\lambda}  \right)  \frac{y
+n_{\rm m} d_{\rm m}\sin\psi-n_{\rm b} d_{\rm b}} { r } h_{n_{\rm b},n_{\rm m}} 
\nonumber\\&
\overset{(a)}{=}
\eta \frac{y}{r} (1+ \varepsilon_{y} )h_{n_{\rm b},n_{\rm m}},
\end{align} 
where (a) is from $\varepsilon_{y}\triangleq\frac{n_{\rm m} d_{\rm m} \sin\psi -n_{\rm b} d_{\rm b}} { y}$ for $y\neq0$.
Under assumption (A3.1), when $|y|>d_{\rm Fn}$,   $\left|\varepsilon_{y}\right| $ is bounded by  
\begin{align}\label{eq:y/r}
         \left| \varepsilon_{y} \right|  
         % \le   \frac{\bar N_{\rm m} d_{\rm m}\!+\! \bar N_{\rm b} d_{\rm b}}{|y|} 
        {<} \frac{D_{\rm b}}{|y|}  
 \!= \! \frac{1}{\!\! \sqrt{0.1922(\!N_{\rm b}\!-\!1)}} \! = \!\mathcal{O}\left(\frac{1}{\sqrt{N_{\rm b}}}\right), 
\end{align}
which follows the steps in \eqref{eq:x/r0}-\eqref{eq:x/r}.
With \eqref{eq:Jy-r} and \eqref{eq:y/r}, we can conclude that under assumptions (A1)-(A3.1), $\mathbf{J}_y$ becomes 
\begin{align}\label{eq:Jy-H}
        \mathbf{J}_y = \eta \frac{y
} { r } \mathbf{H} + \mathbf{E}_y \triangleq \tilde{\mathbf{J}}_y + \mathbf{E}_y,
\end{align}
where $\tilde{\mathbf{J}}_y$ is a scaled version of $\mathbf{H}$,  $ [  \tilde{\mathbf{J}}_y ]_{\bar{n}_{\rm b},\bar{n}_{\rm m}} =  \eta  \frac{y
} { r }  h_{n_{\rm b},n_{\rm m}}$,  $ \left[  \mathbf{E}_y \right]_{\bar{n}_{\rm b},\bar{n}_{\rm m}} = \varepsilon_{y}[  \tilde{\mathbf{J}}_y ]_{\bar{n}_{\rm b},\bar{n}_{\rm m}} $, and 
$\frac{\|\mathbf{E}_y\|_{\mathsf{F}}^2}{\|\tilde{\mathbf{J}}_y\|_{\mathsf{F}}^2}=\mathcal{O}(N_{\rm b}^{-1})$.  

\subsubsection{Analysis of $\mathbf{J}_\psi= \frac{\partial \mathbf{H} }{\partial \psi}$}
From \eqref{eq:dH}-\eqref{eq:dr}, we obtain
% the $(\bar{n}_{\rm b}, \bar{n}_{\rm m})$-th element of $\mathbf{J}_\psi$ can be expressed as
\begin{align}\nonumber%\label{eq:Jpsi-n}
      \!\!\!\left[  \mathbf{J}_\psi \right]_{\bar{n}_{\rm b},\bar{n}_{\rm m}}  
      \!\!\!\!=\!-\left[  \mathbf{J}_x \right]_{\bar{n}_{\rm b},\bar{n}_{\rm m}}   \!\!n_{\rm m} d_{\rm m}\sin\psi \!+ \!\left[  \mathbf{J}_y \right]_{\bar{n}_{\rm b},\bar{n}_{\rm m}} \!\! n_{\rm m} d_{\rm m}\cos\psi .\!\!
\end{align} 
Let $\mathbf{D}_{N_{\rm m}}\triangleq{\rm diag}[-\bar N_{\rm m}, ..., 0,..., \bar N_{\rm m}]$. We can express 
$\mathbf{J}_\psi$ as
\begin{align}\label{eq:Jpsi=J+E}
        \!\mathbf{J}_\psi \!& = \!-d_{\rm m}\sin\psi \mathbf{J}_x \mathbf{D}_{N_{\rm m}} \!+\! d_{\rm m}\cos\psi \mathbf{J}_y \mathbf{D}_{N_{\rm m}} \!\!\overset{(a)}{=}  \tilde{\mathbf{J}}_\psi  \!+\!  \mathbf{E}_\psi,\!
\end{align}
where  $\tilde{\mathbf{J}}_\psi\triangleq(-\sin\psi \tilde{\mathbf{J}}_x + \cos\psi\tilde{\mathbf{J}}_y)d_{\rm m}\mathbf{D}_{N_{\rm m}}$, $\mathbf{E}_\psi\triangleq(-\sin\psi \mathbf{E}_x+\cos\psi\mathbf{E}_y)d_{\rm m}\mathbf{D}_{N_{\rm m}}$, and (a) is from \eqref{eq:Jx-H} and \eqref{eq:Jy-H}, under assumptions (A1)-(A3.1).
% $\mathbf{J}_\psi=-d_{\rm m}\sin\psi \mathbf{J}_x \mathbf{D}_{N_{\rm m}} + d_{\rm m}\cos\psi \mathbf{J}_y \mathbf{D}_{N_{\rm m}}$. From \eqref{eq:Jx-H} and \eqref{eq:Jy-H}, under assumptions (A1)-(A3.1), $\mathbf{J}_\psi$ can be written as 
% \begin{align}\label{eq:Jpsi=J+E}
%         \mathbf{J}_\psi & =  \underbrace{(-\sin\psi \tilde{\mathbf{J}}_x + \cos\psi\tilde{\mathbf{J}}_y)d_{\rm m}\mathbf{D}_{N_{\rm m}}}_{\tilde{\mathbf{J}}_\psi}   +  \nonumber\\& \qquad \underbrace{(-\sin\psi \mathbf{E}_x+\cos\psi\mathbf{E}_y)d_{\rm m}\mathbf{D}_{N_{\rm m}} }_{\mathbf{E}_\psi},
% \end{align}
% where $\mathbf{D}_{N_{\rm m}}={\rm diag}[-\bar N_{\rm m}, ..., 0,..., \bar N_{\rm m}]$.
% Moreover,
$\tilde{\mathbf{J}}_\psi$ can be rewritten as
\begin{align}\label{eq:J_psi_approx}
    \tilde{\mathbf{J}}_\psi & \overset{(a)}{=} \eta d_{\rm m} (-\sin\psi   \frac{x
} { r } +  \cos\psi   \frac{y
} { r } )  \mathbf{H} \mathbf{D}_{N_{\rm m}}
\overset{(b)}{=} \tilde{\eta} \mathbf{H} \mathbf{D}_{N_{\rm m}},
\end{align}
where 
$\tilde{\eta} = \eta d_{\rm m}\sin(\theta-\psi)$, (a) is from \eqref{eq:Jx-H} and \eqref{eq:Jy-H}, and (b) relates polar coordinate $[\theta,r]^{\mathsf{T}}$ to Cartesian coordinate $[x,y]^{\mathsf{T}}$.  
This indicates that $\tilde{\mathbf{J}}_\psi$ is a column-scaled version of $\mathbf{H}$, i.e.,
\begin{align}\label{eq:J_psi_n}
    [  \tilde{\mathbf{J}}_\psi ]_{\bar{n}_{\rm b},\bar{n}_{\rm m}} = \sin(\theta-\psi) \eta d_{\rm m} h_{n_{\rm b},n_{\rm m}} n_{\rm m}.
\end{align}
From \eqref{eq:Jx-H} and \eqref{eq:Jy-H}, $\mathbf{E}_\psi$ can be element-wise expressed as
\begin{align}\label{eq:E_psi}
    \left[  \mathbf{E}_\psi \right]_{\bar{n}_{\rm b},\bar{n}_{\rm m}} & = 
\varepsilon_\psi \eta d_{\rm m}h_{n_{\rm b},n_{\rm m}} n_{\rm m},
\end{align}
where $\varepsilon_\psi \triangleq -  \sin\psi  \frac{x \varepsilon_{x} } { r }+ \cos\psi  \frac{y \varepsilon_{y} 
} { r }=\frac{ -n_{\rm b} d_{\rm b} \cos\psi} { r}$. 
Since $r=\sqrt{x^2+y^2}>d_{\rm Fn}$, following the steps in \eqref{eq:x/r0}-\eqref{eq:x/r}, we obtain
% $|\varepsilon_{\psi}|$ is bounded by
\begin{align}\label{eq:e_psi}
   |\varepsilon_\psi| & < \frac{ 1 }{\sqrt{ 0.7688  (N_{\rm b}-1) }}  = \mathcal{O}\left(\frac{1}{\sqrt{N_{\rm b}}}\right).
\end{align}
Eqs. \eqref{eq:J_psi_n}-\eqref{eq:e_psi} indicate that when $|\sin(\theta-\psi)|>c$ for some non-zero constant $c$,  
$\frac{\|\mathbf{E}_\psi\|_{\mathsf{F}}^2}{\|\tilde{\mathbf{J}}_\psi\|_{\mathsf{F}}^2}=\mathcal{O}\left(\frac{N_{\rm b}^{-1}}{|\sin(\theta-\psi)|^2}\right)=\mathcal{O}\left(N_{\rm b}^{-1}\right)$. 
 
 %%%%%%%%%%%%%%%%%%%%%%%%%
 
\vspace{-2mm}

\section{Proof of Proposition~\ref{prop:Fnorm}} 
\label{app:Hjacobian_norm}
 
% \textcolor{red}{As in Appendix~\ref{app:dataFIM}, we omit the subscript $k$ for notational simplicity.}
% \color{black}
% The following derivations are conducted under NF assumptions (A1)-(A3), including the {\em off-axis} case (A3.1), where the MS lies away from the axis perpendicular to the BS ULA with $|x| > d_{\rm Fn}$ and $|y| > d_{\rm Fn}$, and the {\em on-axis} case (A3.2),  where the MS lies on this axis with $|x| > d_{\rm Fn}$ and $y=0$. 

\subsubsection{Position Information}\label{app:position}

From Proposition~\ref{prop:F_mu}, the average Fisher information about the MS position is bounded by
\begin{align}\label{eq:Jx+Jy0}
  \bar{F}_x+\bar{F}_y & 
  \le  \frac{2  P_{\rm m}}{\sigma_{\rm o}^2 N_{\rm m}} \left(  \left\|\mathbf{J}_{x}\right\|_{\mathsf{F}}^2 + \left\| \mathbf{J}_{y}\right\|_{\mathsf{F}}^2\right).
\end{align}
To analyze this bound, we next evaluate $\left\| \mathbf{J}_{\mu}\right\|_{\mathsf{F}}^2$,  
$\mu\in\{x,y\}$, under the {\em off-axis} case (A3.1), where the MS lies away from the axis perpendicular to the BS ULA with $|x| > d_{\rm Fn}$ and $|y| > d_{\rm Fn}$, and the {\em on-axis} case (A3.2) with $|x| > d_{\rm Fn}$ and $y=0$, respectively.

\paragraph{Off-Axis Case} 
% To analyze the bound in \eqref{eq:Jx+Jy0} asymptotically, we first evaluate $\left\| \mathbf{J}_{\mu}\right\|_{\mathsf{F}}^2$,  
% $\mu\in\{x,y\}$, when $|x| > d_{\rm Fn}$ and $|y| > d_{\rm Fn}$.
% \color{black}
Under assumptions (A1)-(A3.1), we obtain from \eqref{eq:Jx-r} and \eqref{eq:Jy-r} that $| \left[ \mathbf{J}_\mu  \right]_{\bar{n}_{\rm b},\bar{n}_{\rm m}} |^2$, $\mu\in\{x,y\}$, is given by
\begin{align}\label{eq:Jmu:n}
    | \left[ \mathbf{J}_\mu  \right]_{\bar{n}_{\rm b},\bar{n}_{\rm m}} |^2 & = \left|\eta \frac{\mu} { r } (1+ \varepsilon_{\mu} ) h_{n_{\rm b},n_{\rm m}} \right|^2
    \nonumber\\& =|\eta h_{n_{\rm b},n_{\rm m}}|^2 \left( \frac{\mu} { r }\right)^2 (1+2\varepsilon_{\mu}+\varepsilon_{\mu}^2).
\end{align}
Summing over $n_{\rm b}$ and $n_{\rm m}$ yields 
\begin{align}\label{eq:J_mu_Fnorm0}
    \left\|\mathbf{J}_{\mu}\right\|_{\mathsf{F}}^2&  \overset{(a)}{=}\sum_{n_{\rm b}=-\bar N_{\rm b}}^{\bar N_{\rm b}} \sum_{n_{\rm m}=-\bar N_{\rm m}}^{\bar N_{\rm m}} |\eta|^2 \left( \frac{\mu} { r }\right)^2 |h_{n_{\rm b},n_{\rm m}}|^2 (1+\varepsilon_{\mu}^2) \nonumber\\& \overset{(b)}{=}\|\tilde{\mathbf{J}}_{\mu} \|_{\mathsf{F}}+\left\|\mathbf{E}_{\mu}\right\|_{\mathsf{F}}^2
\overset{(c)}{=} \|\tilde{\mathbf{J}}_{\mu} \|_{\mathsf{F}}^2\left(1+\mathcal{O}(N_{\rm b}^{-1})\right),
\end{align}
where (a) is from $|h_{n_{\rm b},n_{\rm m}}|=|\frac{\lambda}{4\pi r} e^{-{\rm j}\frac{2\pi}{\lambda} r_{n_{\rm b},n_{\rm m}} }|=\frac{\lambda}{4\pi r}$ and from $\sum_{n_{\rm b}=-\bar N_{\rm b}}^{\bar N_{\rm b}}  \sum_{n_{\rm m}=-\bar N_{\rm m}}^{\bar N_{\rm m}} \varepsilon_{\mu} =0$ for both $\varepsilon_{x}$ in \eqref{eq:Jx-r} and $\varepsilon_{y}$ in \eqref{eq:Jy-r},  (b) follows the decomposition in \eqref{eq:Jx-H} and \eqref{eq:Jy-H}, and  (c) is from  $\varepsilon_{\mu}^2=\mathcal{O}(N_{\rm b}^{-1})$.
Furthermore,  
% from \eqref{eq:Jx-H} and \eqref{eq:Jy-H}, 
$\|\tilde{\mathbf{J}}_{\mu} \|_{\mathsf{F}}^2$ is given by
\begin{align}\label{eq:Jxy-Fnorm} \|\tilde{\mathbf{J}}_{\mu} \|_{\mathsf{F}}^2 & = \left\| \eta \frac{\mu}{r} \mathbf{H}\right\|_{\mathsf{F}}^2 
\overset{(a)}{=}  |\eta|^2 
 \frac{\mu^2}{r^2}  \left(\frac{\lambda}{4\pi r}\right)^2 N_{\rm b} N_{\rm m},
\end{align}
where (a) is from $\left\|\mathbf{H}\right\|_{\mathsf{F}}^2=\left(\frac{\lambda}{4\pi r}\right)^2 N_{\rm b} N_{\rm m}$ under assumption (A1).
Moreover, $|\eta|^2$ can be expanded as
\begin{align}\label{eq:eta^2}
    |\eta|^2&=\left|\frac{1}{r}+{\rm j}\frac{2\pi}{\lambda} \right|^2
    =\frac{1}{r^2}+\frac{4\pi^2}{\lambda^2}=\frac{4\pi^2}{\lambda^2}\left(1+\frac{\lambda^2}{4\pi^2r^2}\right)
    \nonumber\\& \overset{(a)}{\le} \!\frac{4\pi^2}{\lambda^2}\left(\!1\!+\!\frac{1}{0.1922\pi^2 N_{\rm b}^3}\!\right)\!=\! \frac{4\pi^2}{\lambda^2}\left(1\!+ \!\mathcal{O}(N_{\rm b}^{-3})\right),
\end{align}
where (a) uses $r=\sqrt{x^2+y^2}>d_{\rm Fn}$ from (A3.1).
% where (a) is from assumption (A3.1), i.e., $r=\sqrt{x^2+y^2}>d_{\rm Fn}=0.62\sqrt{\frac{D_{\rm b}^3}{\lambda}}=0.62\sqrt{\frac{(N_{\rm b}-1)^3d_{\rm b}^3}{\lambda}}=0.62\lambda\sqrt{\frac{(N_{\rm b}-1)^3}{8}}$.
Substituting \eqref{eq:Jxy-Fnorm} and \eqref{eq:eta^2} into \eqref{eq:J_mu_Fnorm0}, we obtain the asymptotic form
\begin{align}\label{eq:J_mu_Fnorm1}
\left\|\mathbf{J}_{\mu}\right\|_{\mathsf{F}}^2 &=
\frac{\mu^2}{4 r^4} N_{\rm b} N_{\rm m} (1+ \mathcal{O}(N_{\rm b}^{-3}))\left(1+\mathcal{O}(N_{\rm b}^{-1})\right)
\nonumber\\&=
\frac{\mu^2}{4 r^4} N_{\rm b} N_{\rm m} \left(1+\mathcal{O}(N_{\rm b}^{-1})\right).
\end{align}
Plugging this result into \eqref{eq:Jx+Jy0}, we obtain \eqref{eq:Fxy-upper}.
% \begin{align}\label{eq:Jx+Jy1}
%   \bar{F}_x+\bar{F}_y & \le
%   \frac{2  P_{\rm m}}{\sigma_{\rm o}^2 N_{\rm m}} \left(\frac{x^2}{4r^4}+\frac{y^2}{4r^4}\right) N_{\rm b} N_{\rm m} (1+\mathcal{O}(N_{\rm b}^{-1}))
%   \nonumber \\&= \frac{ P_{\rm m} N_{\rm b} }{2\sigma_{\rm o}^2r^2} \left(1+ \mathcal{O}(N_{\rm b}^{-1})\right).
% \end{align}

\paragraph{On-Axis Case}\label{app:Broadside} 
Under assumption (A3.2), $|x| > d_{\rm Fn}$ and $|y| = 0$. 
The analysis of $\left\|\mathbf{J}_{x}\right\|_{\mathsf{F}}^2$ remains identical to the off-axis case due to $|x| > d_{\rm Fn}$. Specifically, the distance simplifies to $r=\sqrt{x^2+y^2}=|x|>d_{\rm Fn}$. When $r=|x|$, \eqref{eq:J_mu_Fnorm0} becomes
\begin{align}\label{eq:Jx:n}
    \left\|\mathbf{J}_{x}\right\|_{\mathsf{F}}^2 &= \sum_{n_{\rm b}=-\bar N_{\rm b}}^{\bar N_{\rm b}} \sum_{n_{\rm m}=-\bar N_{\rm m}}^{\bar N_{\rm m}}  |\eta h_{n_{\rm b},n_{\rm m}}|^2 (1+\varepsilon_{x}^2).
\end{align}
However, 
the expression of $\left\|\mathbf{J}_{y}\right\|_{\mathsf{F}}^2$ differs due to $y=0$. Under assumption (A3.2), $\left[ \mathbf{J}_y  \right]_{\bar{n}_{\rm b},\bar{n}_{\rm m}} $ in \eqref{eq:Jy-r} satisfies 
% $\left[ \mathbf{J}_y  \right]_{\bar{n}_{\rm b},\bar{n}_{\rm m}}=\eta h_{n_{\rm b},n_{\rm m}} \frac{y+n_{\rm m} d_{\rm m} \sin\psi  -n_{\rm b} d_{\rm b}} { r }=\eta h_{n_{\rm b},n_{\rm m}} \frac{n_{\rm m} d_{\rm m} \sin\psi  -n_{\rm b} d_{\rm b}} { r }$. Therefore,
\begin{align}\label{eq:Jy=0}
    \left|  \left[ \mathbf{J}_y  \right]_{\bar{n}_{\rm b},\bar{n}_{\rm m}} \right|^2 
&= |\eta h_{n_{\rm b},n_{\rm m}}|^2 \left( \frac{y+n_{\rm m} d_{\rm m} \sin\psi  -n_{\rm b} d_{\rm b}} { r } \right)^2 
\nonumber\\&\overset{(a)}{=} |\eta h_{n_{\rm b},n_{\rm m}}|^2 \varepsilon_{r}^2,
\end{align}
% where $\varepsilon_{r}\triangleq\frac{y+n_{\rm m} d_{\rm m} \sin\psi  -n_{\rm b} d_{\rm b}} { r }=\frac{n_{\rm m} d_{\rm m} \sin\psi -n_{\rm b} d_{\rm b}} { r } $ is from $y=0$.
where $\varepsilon_{r}\triangleq \frac{n_{\rm m} d_{\rm m} \sin\psi -n_{\rm b} d_{\rm b}} { r } $ and (a) is from $y=0$.
Given $r>d_{\rm Fn}$,  it follows from the steps in \eqref{eq:y/r} that 
\begin{align}\label{eq:r/r}
            \left| \varepsilon_{r} \right|  \le   \frac{1}{\sqrt{0.1922(N_{\rm b}-1)}} = \mathcal{O}\left(\frac{1}{\sqrt{N_{\rm b}}}\right).
\end{align}
Using \eqref{eq:Jx:n}-\eqref{eq:r/r}, in the on-axis case (A3.2), we obtain
\begin{align}\label{eq:Jx+Jy|y=0}
&\left\|\mathbf{J}_{x}\right\|_{\mathsf{F}}^2+\left\|\mathbf{J}_{y}\right\|_{\mathsf{F}}^2 = \sum_{n_{\rm b}=-\bar N_{\rm b}}^{\bar N_{\rm b}} \sum_{n_{\rm m}=-\bar N_{\rm m}}^{\bar N_{\rm m}} |\eta h_{n_{\rm b},n_{\rm m}}|^2 (1+ \varepsilon_{x}^2+\varepsilon_{r}^2)
\nonumber\\&\overset{(a)}{=} \|\tilde{\mathbf{J}}_{x} \|_{\mathsf{F}}^2\left(1+\mathcal{O}(N_{\rm b}^{-1}) \right)=\frac{N_{\rm b} N_{\rm m}}{4 r^2}  \left(1+\mathcal{O}(N_{\rm b}^{-1})\right),\!
\end{align} 
where (a) follows the steps in \eqref{eq:J_mu_Fnorm0}-\eqref{eq:J_mu_Fnorm1}.
%%%%%%%%% 
Substituting \eqref{eq:Jx+Jy|y=0} into \eqref{eq:Jx+Jy0} yields the same asymptotic bound of the position information as the off-axis case. 
% The difference lies in the source of information: 
However, 
Eq.~\eqref{eq:J_mu_Fnorm1} shows that in the off-axis case, both the $x$ and $y$ related components contribute significantly; Eq.~\eqref{eq:Jx+Jy|y=0} shows that in the on-axis case, the contribution from $\mathbf{J}_{y}$ is negligible.

\subsubsection{Orientation Information}\label{app:orientation}
Following a similar approach to Appendix~\ref{app:position}, we begin by analyzing $\left\| \mathbf{J}_{\psi}\right\|_{\mathsf{F}}^2$. 
\paragraph{Off-Axis Case} Under assumptions (A1)-(A3.1), we obtain from \eqref{eq:Jpsi=J+E}-\eqref{eq:E_psi} that
\begin{align}
    | \left[  \mathbf{J}_\psi \right]_{\bar{n}_{\rm b},\bar{n}_{\rm m}}  |^2 & = | [  \tilde{\mathbf{J}}_\psi  ]_{\bar{n}_{\rm b},\bar{n}_{\rm m}}+ [  \mathbf{E}_\psi  ]_{\bar{n}_{\rm b},\bar{n}_{\rm m}}|^2
    \nonumber\\& = | (\sin(\theta-\psi)+\varepsilon_\psi) \eta d_{\rm m}h_{n_{\rm b},n_{\rm m}} n_{\rm m} |^2,
\end{align}
where $\varepsilon_{\psi}=\frac{ -n_{\rm b} d_{\rm b} \cos\psi} { r}$ is given in \eqref{eq:E_psi}. Therefore,
% Then, $\left\|\mathbf{J}_{\psi}\right\|_{\mathsf{F}}^2$ becomes
\begin{align}\label{eq:J_psi_Fnorm0}
    \left\|\mathbf{J}_{\psi}\right\|_{\mathsf{F}}^2&  \overset{(a)}{=}\!\!\!\sum_{n_{\rm b}=-\bar N_{\rm b}}^{\bar N_{\rm b}} \sum_{n_{\rm m}
    =-\bar N_{\rm m}}^{\bar N_{\rm m}} \!\!\!|\eta d_{\rm m} h_{n_{\rm b},n_{\rm m}} n_{\rm m} |^2 ( \sin^2(\theta-\psi)+ \varepsilon_\psi^2  ) 
     \nonumber\\& =\|\tilde{\mathbf{J}}_{\psi} \|_{\mathsf{F}}^2+\left\|\mathbf{E}_{\psi}\right\|_{\mathsf{F}}^2 \overset{(b)}{=} \|\tilde{\mathbf{J}}_{\psi}\|_{\mathsf{F}}^2\left(1+\mathcal{O}(N_{\rm b}^{-1})\right),
\end{align}
where (a) is because the summation over the cross term $ 2|\eta d_{\rm m} h_{n_{\rm b},n_{\rm m}} n_{\rm m} |^2\sin(\theta-\psi)\varepsilon_\psi$ is $0$,
and (b) is from \eqref{eq:J_psi_n}-\eqref{eq:e_psi} for $|\sin(\theta-\psi)|>c>0$.
Moreover, Eqs.~\eqref{eq:J_psi_approx}-\eqref{eq:J_psi_n} give
% $\|\tilde{\mathbf{J}}_{\psi} \|_{\mathsf{F}}^2$ is calculated as
\begin{align}   \label{eq:J_psi_Fnorm_approx}
\|\tilde{\mathbf{J}}_{\psi} \|_{\mathsf{F}}^2 & 
\overset{(a)}{=} |\eta d_{\rm m} |^2 \sin^2(\theta-\psi) \frac{\lambda^2}{(4\pi r)^2} \sum_{n_{\rm b}=-\bar N_{\rm b}}^{\bar N_{\rm b}} \sum_{n_{\rm m}=-\bar N_{\rm m}}^{\bar N_{\rm m}}  n_{\rm m}^2
\nonumber\\&  \overset{(b)}{=}  \frac{\lambda^2 |\eta|^2 N_{\rm b} }{(4\pi r)^2}   \sin^2(\theta-\psi) d_{\rm m}^2\frac{\bar N_{\rm m} (\bar N_{\rm m}+1) (2\bar N_{\rm m}+1)}{3} 
 \nonumber\\&  \overset{(c)}{=}  \frac{\lambda^2 |\eta|^2 N_{\rm b} N_{\rm m}}{(4\pi r)^2}    \frac{ D_{\rm m} (D_{\rm m}+2 d_{\rm m})  }{12} \sin^2(\theta-\psi)
\nonumber\\& \overset{(d)}{=}\frac{N_{\rm b} N_{\rm m}}{48 r^2} D_{\rm m} (D_{\rm m}\!+\!2 d_{\rm m})\sin^2(\theta-\psi)  (1\!+\!\mathcal{O}(N_{\rm b}^{-3}))
\nonumber\\& =\frac{N_{\rm b} N_{\rm m} (D_{\rm m}^{\rm eff})^2}{48 r^2}  \!\left(\!1\!+\!\frac{2}{N_{\rm m}\!-1}\right)\!\left(1\!+\!\mathcal{O}(N_{\rm b}^{-3})\right),\!
 \end{align}
where (a) is from $h_{n_{\rm b},n_{\rm m}}=\frac{\lambda}{4\pi r} e^{-{\rm j}\frac{2\pi}{\lambda} r_{n_{\rm b},n_{\rm m}} }$, (b) is from $\sum_{n=-\bar N}^{\bar N} n^2\!=\!2\sum_{n=1}^{\bar N} n^2\!=\!\frac{\bar N  (\bar N +1) (2\bar N +1)}{3}$, (c) is from $D_{\rm m} =2\bar N_{\rm m}d_{\rm m}$ and $N_{\rm m}=2\bar N_{\rm m}+1$, (d) is from \eqref{eq:eta^2},  and $D_{\rm m}^{\rm eff}\!\triangleq \!D_{\rm m}\left|\sin(\theta-\psi)\right|$.
% Following the steps in \eqref{eq:J_mu_Fnorm1}-\eqref{eq:Jx+Jy1}, under assumptions (A1)-(A3.1), $\bar{F}_\psi$ is bounded by
% % the average Fisher information about the MS orientation is bounded by
% \begin{align}\label{eq:Fpsi-upper0}
%   \bar{F}_\psi &
%  \overset{(a)}{\le}  \frac{2  P_{\rm m}}{\sigma_{\rm o}^2 N_{\rm m}}   \left\|\mathbf{J}_{\psi}\right\|_{\mathsf{F}}^2  
%  \nonumber\\&\overset{(b)}{=} \frac{ P_{\rm m} N_{\rm b} (D_{\rm m}^{\rm eff})^2}{24\sigma_{\rm o}^2r^2}\!\left(1+\frac{2}{N_{\rm m}-1}\right)\!\left(1+\mathcal{O}(N_{\rm b}^{-1})\right),
% \end{align}
% where (a) is from~\eqref{eq:barFmu}, (b) 
% substitutes \eqref{eq:J_psi_Fnorm_approx} into \eqref{eq:J_psi_Fnorm0}, and (b) holds when $|\sin(\theta-\psi)|>c>0$.
Substituting \eqref{eq:J_psi_Fnorm_approx} into \eqref{eq:J_psi_Fnorm0} yields
\eqref{eq:Fpsi-upper}.

\paragraph{On-Axis Case} 
The analysis for the on-axis case (A3.2) proceeds analogously to Appendix~\ref{app:Broadside}.
% and yields the same asymptotic expression as \eqref{eq:Fpsi-upper0},  
% which is omitted here for brevity.
 
\vspace{-2mm}

%%%%%%%%%%%%%%%%
%%%%%%%%%%%%%%%%%
\bibliographystyle{IEEEtran}
\bibliography{reference}

\begin{thebibliography}{10}
\providecommand{\url}[1]{#1}
\csname url@samestyle\endcsname
\providecommand{\newblock}{\relax}
\providecommand{\bibinfo}[2]{#2}
\providecommand{\BIBentrySTDinterwordspacing}{\spaceskip=0pt\relax}
\providecommand{\BIBentryALTinterwordstretchfactor}{4}
\providecommand{\BIBentryALTinterwordspacing}{\spaceskip=\fontdimen2\font plus
\BIBentryALTinterwordstretchfactor\fontdimen3\font minus
  \fontdimen4\font\relax}
\providecommand{\BIBforeignlanguage}[2]{{%
\expandafter\ifx\csname l@#1\endcsname\relax
\typeout{** WARNING: IEEEtran.bst: No hyphenation pattern has been}%
\typeout{** loaded for the language `#1'. Using the pattern for}%
\typeout{** the default language instead.}%
\else
\language=\csname l@#1\endcsname
\fi
#2}}
\providecommand{\BIBdecl}{\relax}
\BIBdecl

\bibitem{6G_freq_antenna}
S.~Chen, Y.-C. Liang, S.~Sun, S.~Kang, W.~Cheng, and M.~Peng, ``Vision,
  requirements, and technology trend of {6G}: How to tackle the challenges of
  system coverage, capacity, user data-rate and movement speed,'' \emph{IEEE
  Wirel. Commun.}, vol.~27, no.~2, pp. 218--228, 2020.

\bibitem{6G_freq}
W.~Saad, M.~Bennis, and M.~Chen, ``A vision of {6G} wireless systems:
  Applications, trends, technologies, and open research problems,'' \emph{IEEE
  Netw.}, vol.~34, no.~3, pp. 134--142, 2019.

\bibitem{NFtutorial_Dai}
M.~Cui, Z.~Wu, Y.~Lu, X.~Wei, and L.~Dai, ``Near-field {MIMO} communications
  for {6G}: Fundamentals, challenges, potentials, and future directions,''
  \emph{IEEE Commun. Mag.}, vol.~61, no.~1, pp. 40--46, 2023.

\bibitem{selvan2017fraunhofer}
K.~T. Selvan and R.~Janaswamy, ``{Fraunhofer} and {Fresnel distances}: Unified
  derivation for aperture antennas,'' \emph{IEEE Antennas Propag. Mag.},
  vol.~59, no.~4, pp. 12--15, 2017.

\bibitem{NFtutorial}
Y.~Liu, Z.~Wang, J.~Xu, C.~Ouyang, X.~Mu, and R.~Schober, ``Near-field
  communications: A tutorial review,'' \emph{IEEE Open J. Commun. Soc.},
  vol.~4, pp. 1999--2049, 2023.

\bibitem{mismatch}
A.~Elzanaty, J.~Liu, A.~Guerra, F.~Guidi, Y.~Ma, and R.~Tafazolli, ``Near and
  far field model mismatch: Implications on {6G} communications, localization,
  and sensing,'' \emph{IEEE Internet Thing Mag.}, vol.~7, no.~5, pp. 120--126,
  2024.

\bibitem{NFlocalization}
Z.~Wang, P.~Ramezani, Y.~Liu, and E.~Bj{\"o}rnson, ``Near-field localization
  and sensing with large-aperture arrays: From signal modeling to processing,''
  \emph{IEEE Signal Process. Mag.}, vol.~42, no.~1, pp. 74--87, 2025.

\bibitem{yuan2024scalable}
X.~Yuan, M.~Zhang, Y.~Zheng, B.~Teng, and W.~Jiang, ``Scalable near-field
  localization based on partitioned large-scale antenna array,'' \emph{IEEE
  Trans. Wirel. Commun.}, vol.~24, no.~3, pp. 2203--2217, 2025.

\bibitem{yuan-MIMO}
\BIBentryALTinterwordspacing
M.~Zhang, X.~Yuan, B.~Teng, and L.~Wang, ``Array partitioning based near-field
  attitude and location estimation,'' 2025. [Online]. Available:
  \url{https://arxiv.org/abs/2504.16800}
\BIBentrySTDinterwordspacing

\bibitem{guerra2021near}
A.~Guerra, F.~Guidi, D.~Dardari, and P.~M. Djuri{\'c}, ``Near-field tracking
  with large antenna arrays: Fundamental limits and practical algorithms,''
  \emph{IEEE Trans. Signal Process.}, vol.~69, pp. 5723--5738, 2021.

\bibitem{AltMin}
X.~Yu, J.-C. Shen, J.~Zhang, and K.~B. Letaief, ``Alternating minimization
  algorithms for hybrid precoding in millimeter wave {MIMO} systems,''
  \emph{IEEE J. Sel. Top. Signal Process.}, vol.~10, no.~3, pp. 485--500, 2016.

\bibitem{hybridMMSE}
T.~Lin, J.~Cong, Y.~Zhu, J.~Zhang, and K.~Ben~Letaief, ``Hybrid beamforming for
  millimeter wave systems using the {MMSE} criterion,'' \emph{IEEE Trans.
  Commun.}, vol.~67, no.~5, pp. 3693--3708, 2019.

\bibitem{el2014spatially}
O.~El~Ayach, S.~Rajagopal, S.~Abu-Surra, Z.~Pi, and R.~W. Heath, ``Spatially
  sparse precoding in millimeter wave {MIMO} systems,'' \emph{IEEE Trans.
  Wirel. Commun.}, vol.~13, no.~3, pp. 1499--1513, 2014.

\bibitem{mendez2015channel}
R.~M{\'e}ndez-Rial, C.~Rusu, A.~Alkhateeb, N.~Gonz{\'a}lez-Prelcic, and R.~W.
  Heath, ``Channel estimation and hybrid combining for {mmWave}: Phase shifters
  or switches?'' in \emph{Proc. 2015 Information Theory and Applications
  Workshop (ITA)}.\hskip 1em plus 0.5em minus 0.4em\relax San Diego, CA, USA:
  IEEE, Feb. 2015, pp. 90--97.

\bibitem{simon2006optimal}
D.~Simon, \emph{Optimal state estimation: Kalman, H infinity, and nonlinear
  approaches}.\hskip 1em plus 0.5em minus 0.4em\relax John Wiley \& Sons, 2006.

\bibitem{salmi2008detection}
J.~Salmi, A.~Richter, and V.~Koivunen, ``Detection and tracking of {MIMO}
  propagation path parameters using state-space approach,'' \emph{IEEE Trans.
  Signal Process.}, vol.~57, no.~4, pp. 1538--1550, 2008.

\bibitem{clock1}
M.~Koivisto, M.~Costa, J.~Werner, K.~Heiska, J.~Talvitie, K.~Lepp{\"a}nen,
  V.~Koivunen, and M.~Valkama, ``Joint device positioning and clock
  synchronization in {5G} ultra-dense networks,'' \emph{IEEE Trans. Wirel.
  Commun.}, vol.~16, no.~5, pp. 2866--2881, 2017.

\bibitem{clock2}
J.~Talvitie, M.~S{\"a}ily, and M.~Valkama, ``Orientation and location tracking
  of {XR} devices: {5G} carrier phase-based methods,'' \emph{IEEE J. Sel. Top.
  Signal Process.}, vol.~17, no.~5, pp. 919--934, 2023.

\bibitem{koivisto2021channel}
M.~Koivisto, J.~Talvitie, E.~Rastorgueva-Foi, Y.~Lu, and M.~Valkama, ``Channel
  parameter estimation and {TX} positioning with multi-beam fusion in {5G}
  mmwave networks,'' \emph{IEEE Trans. Wirel. Commun.}, vol.~21, no.~5, pp.
  3192--3207, 2021.

\bibitem{LiuISAC2}
\BIBentryALTinterwordspacing
H.~Jiang, Z.~Wang, and Y.~Liu, ``Near-field sensing enabled predictive
  beamforming: From estimation to tracking,'' 2024. [Online]. Available:
  \url{https://arxiv.org/abs/2408.02027}
\BIBentrySTDinterwordspacing

\bibitem{dai2025attitude}
X.~Dai, M.~Zhang, B.~Teng, X.~Yuan, and X.~Wang, ``Attitude estimation assisted
  short-range {UAV} localization and tracking based on extremely large antenna
  array,'' \emph{IEEE Trans. Wirel. Commun.}, vol.~24, no.~12, pp.
  10\,391--10\,407, 2025.

\bibitem{cui2022channel}
M.~Cui and L.~Dai, ``Channel estimation for extremely large-scale {MIMO}:
  Far-field or near-field?'' \emph{IEEE Trans. Commun.}, vol.~70, no.~4, pp.
  2663--2677, 2022.

\bibitem{mixed}
Y.~Lu and L.~Dai, ``Near-field channel estimation in mixed {LoS/NLoS}
  environments for extremely large-scale {MIMO} systems,'' \emph{IEEE Trans.
  Commun.}, vol.~71, no.~6, pp. 3694--3707, 2023.

\bibitem{kosasih2024finite}
A.~Kosasih and E.~Bj{\"o}rnson, ``Finite beam depth analysis for large
  arrays,'' \emph{IEEE Trans. Wirel. Commun.}, vol.~23, no.~8, pp.
  10\,015--10\,029, 2024.

\bibitem{tang2023line}
A.~Tang, J.-B. Wang, Y.~Pan, W.~Zhang, Y.~Chen, H.~Yu, and R.~C. de~Lamare,
  ``Line-of-sight extra-large {MIMO} systems with angular-domain processing:
  Channel representation and transceiver architecture,'' \emph{IEEE Trans.
  Commun.}, vol.~72, no.~1, pp. 570--584, 2023.

\bibitem{indoorSM}
P.~Liu, M.~Di~Renzo, and A.~Springer, ``Line-of-sight spatial modulation for
  indoor {mmWave} communication at 60 {GHz},'' \emph{IEEE Trans. Wirel.
  Commun.}, vol.~15, no.~11, pp. 7373--7389, 2016.

\bibitem{chen2025quasi}
L.~Chen, X.~Yuan, and Y.-J.~A. Zhang, ``Quasi-orthogonal beamforming in
  near-field line-of-sight {MIMO} channel,'' \emph{IEEE Trans. Wirel. Commun.},
  vol.~25, pp. 3766--3784, 2026.

\bibitem{CTRV+CV}
R.~Schubert, E.~Richter, and G.~Wanielik, ``Comparison and evaluation of
  advanced motion models for vehicle tracking,'' in \emph{Proc. 2008 11th
  International Conference on Information Fusion}, Cologne, Germany, June 2008,
  pp. 1--6.

\bibitem{CTRV+diag}
H.~Ragab, S.~Khamis, and S.~A. Napoleon, ``Advanced object tracking in
  self-driving cars: {EKF} and {UKF} performance evaluation,'' in \emph{Proc.
  2024 6th Novel Intelligent and Leading Emerging Sciences Conference
  (NILES)}.\hskip 1em plus 0.5em minus 0.4em\relax Giza, Egypt: IEEE, Oct.
  2024, pp. 25--29.

\bibitem{GDN2016tracking}
C.~Zhang, D.~Guo, and P.~Fan, ``Tracking angles of departure and arrival in a
  mobile millimeter wave channel,'' in \emph{Proc. 2016 IEEE international
  conference on communications (ICC)}.\hskip 1em plus 0.5em minus 0.4em\relax
  Kuala Lumpur, Malaysia: IEEE, May 2016, pp. 1--6.

\bibitem{tichavsky1998posterior}
P.~Tichavsky, C.~H. Muravchik, and A.~Nehorai, ``Posterior {Cram{\'e}r-Rao}
  bounds for discrete-time nonlinear filtering,'' \emph{IEEE Trans. Signal
  Process.}, vol.~46, no.~5, pp. 1386--1396, 1998.

\bibitem{koohifar2018autonomous}
F.~Koohifar, I.~Guvenc, and M.~L. Sichitiu, ``Autonomous tracking of
  intermittent {RF} source using a {UAV} swarm,'' \emph{IEEE Access}, vol.~6,
  pp. 15\,884--15\,897, 2018.

\bibitem{NF-RIS}
L.~Xia, R.~Wang, X.~Yuan, B.~Teng, and J.~Rodr{\'\i}guez-Pi{\~n}eiro,
  ``Near-field localization for reconfigurable intelligent surface aided
  {XL-MIMO} systems harnessing the {NLoS} components,'' \emph{IEEE Trans.
  Wirel. Commun.}, vol.~25, pp. 1857--1874, 2026.

\bibitem{Airy}
A.~Singh, V.~Petrov, H.~Guerboukha, I.~V. Reddy, E.~W. Knightly, D.~M.
  Mittleman, and J.~M. Jornet, ``Wavefront engineering: Realizing efficient
  terahertz band communications in {6G} and beyond,'' \emph{IEEE Wirel.
  Commun.}, vol.~31, no.~3, pp. 133--139, 2024.

\end{thebibliography}
\end{document}